\documentclass[11pt,prd,aps,letterpaper,amsmath,amssymb,superscriptaddress]{revtex4-2}
\usepackage{geometry}[letter,11pt]
\usepackage{graphicx}
\usepackage{xcolor}
\usepackage{gensymb}
\usepackage{textcomp}
\usepackage[T1]{fontenc}
\usepackage[hyperindex,breaklinks]{hyperref}
\hypersetup{
    colorlinks=true,
    linkcolor=red,
    citecolor=blue,
    urlcolor=blue
}

\newcommand \theia{\textsc{Theia}{}}
\newcommand \eos{\textsc{Eos}{}}
\newcommand \mimir{\textsc{Mimir}{}}
\newcommand \snd{\textsc{SeedNDestroy{}}}
\newcommand \hitman{\textsc{HITMAN{}}}
\newcommand \geant{\textsc{Geant4{}}}
\newcommand{\AERO}{\textsc{Aero{}}}
\newcommand{\pp}[1]{\left(#1\right)}
\newcommand{\tz}{t_{0}}
\newcommand{\rp}{\textsc{RAT-PAC-2}}
\newcommand{\np}{$N^{\mathrm{p}}_\mathrm{hit}${}}
\newcommand{\nhit}{$N_\mathrm{hit}$}
\newcommand{\chindf}{$\chi^{2}/\rm{NDF}${} }

\begin{document}
\title{Performance of the \eos{} detector with water}

\author{S.~Arora}\affiliation{\ucb}\affiliation{\lbnl}
\author{M.~Askins}\affiliation{\ucb}\affiliation{\lbnl}
\author{A.~J.~Bacon}\affiliation{\penn}
\author{Z.~Bagdasarian}\affiliation{\ucb}\affiliation{\lbnl}
\author{A.~Baldoni}\affiliation{\psu}\affiliation{\westpoint}
\author{L.~Bartoszek}\affiliation{\bart}
\author{M.~Bergevin}\affiliation{\llnl}
\author{Y.~Bezawada}\affiliation{\ucb}\affiliation{\lbnl}
\author{E.~Blucher}\affiliation{\chic}
\author{J.~Boissevain}\affiliation{\bart}\author{R.~Bonventre}\affiliation{\lbnl}
\author{E.~J.~Callaghan}\affiliation{\ucb}\affiliation{\lbnl}
\author{D.~F.~Cowen}\affiliation{\psu} 
\author{K.~DeHolton}\affiliation{\psu} 
\author{M.~Diwan}\affiliation{\bnl} 
\author{M.~Dubnowski}\affiliation{\penn}
\author{P.~Englezos}\affiliation{\ucb}\affiliation{\lbnl}
\author{S.~Gadamsetty}\affiliation{\ucb}\affiliation{\lbnl}
\author{C.~Grant}\affiliation{\bu}
\author{B.~Harris}\affiliation{\penn}
\author{M.~R.~Hebert}\affiliation{\ucb}\affiliation{\lbnl}
\author{S.~Jeon}\affiliation{\bu}
\author{T.~Kaptanoglu}\affiliation{\ucb}\affiliation{\lbnl} 
\author{A.~Katt}\affiliation{\bu}
\author{J.~R.~Klein}\affiliation{\penn}
\author{T.~Kroupov\'a}\affiliation{\penn} 
\author{L.~Lebanowski}\affiliation{\ucb}\affiliation{\lbnl}
\author{A.~Mastbaum}\affiliation{\rut}
\author{C.~Mauger}\affiliation{\penn}
\author{G.~Mayers}\affiliation{\penn} 
\author{M.~Miller}\affiliation{\iow}
\author{J.~Nachtman}\affiliation{\iow}
\author{S.~Naugle}\affiliation{\penn}
\author{J. Newby}\affiliation{\ornl}
\author{M.~Newcomer}\affiliation{\penn}
\author{A.~Nikolica}\affiliation{\penn} 
\author{G.~D.~Orebi~Gann}\affiliation{\ucb}\affiliation{\lbnl}
\author{A.~Phipps}\affiliation{\cseb}
\author{L.~Pickard}\affiliation{\ucb}\affiliation{\lbnl}
\author{R.~C.~Pitelka}\affiliation{\penn}
\author{L.~Ren}\affiliation{\boul} 
\author{A.~Rincon}\affiliation{\ucb}\affiliation{\lbnl} 
\author{R.~Rosero}\affiliation{\bnl}
\author{N.~Rowe}\affiliation{\ucb}\affiliation{\lbnl} 
\author{H.~J.~Ryoo}\affiliation{\ucb}\affiliation{\lbnl} 
\author{J.~Ryshkewitch}\affiliation{\bu}
\author{J.~Saba}\affiliation{\lbnl} 
\author{S.~Schoppmann}\affiliation{\ucb}\affiliation{\lbnl}
\author{J.~Shen}\affiliation{\penn}
\author{M.~Smiley}\affiliation{\ucb}\affiliation{\lbnl}
\author{H.~Song}\affiliation{\ucb}\affiliation{\lbnl}\affiliation{\bu}
\author{H.~Steiger}\affiliation{\mainz}\affiliation{\mun}
\author{B. Tam}\affiliation{\oxford}
\author{E.~Tiras}\affiliation{\erc}\affiliation{\iow}
\author{W.~H.~To}\affiliation{\css}
\author{M.~R.~Vagins}\affiliation{\irvine}
\author{R.~Van~Berg}\affiliation{\bart}
\author{J.~Wallig}\affiliation{\lbnl}
\author{G.~Wendel}\affiliation{\psu}
\author{M.~Wetstein}\affiliation{\iowa}
\author{M.~Wurm}\affiliation{\mainz}
\author{G.~Yang}\affiliation{\ucb}\affiliation{\lbnl}
\author{M.~Yeh}\affiliation{\bnl}
\author{E.~D.~Zimmerman}\affiliation{\boul}
\author{A.~Zummo}\affiliation{\rut}

\newcommand{\oxford}{Department of Physics, University of Oxford, Parks Rd, Oxford OX1 3PU, United Kingdom}

\newcommand{\ornl}{Oak Ridge National Laboratory, 1 Bethel Valley Road, Oak Ridge, TN 37830}

\newcommand{\ucb}{Physics Department, University of California at Berkeley, Berkeley, CA 94720-7300}

\newcommand{\lbnl}{
Lawrence Berkeley National Laboratory, 1 Cyclotron Road, Berkeley, CA 94720-8153, USA}

\newcommand{\penn}{Department of Physics and Astronomy, University of Pennsylvania, Philadelphia, PA 19104-6396}

\newcommand{\chic}{The Enrico Fermi Institute and Department of Physics, The University of Chicago, Chicago, IL 60637, USA}

\newcommand{\bnl}{Brookhaven National Laboratory, Upton, New York 11973, USA}

\newcommand{\iowa}{Department of Physics and Astronomy, Iowa State University, Ames, IA 50011, USA}

\newcommand{\bu}{Boston University, Department of Physics, Boston, MA 02215, USA}

\newcommand{\mainzprisma}{Johannes Gutenberg-Universit{\"a}t, Institute of Physics and EC PRISMA$^{++}$, Mainz, 55099 Mainz, Germany}

\newcommand{\mainz}{Johannes Gutenberg-Universit{\"a}t, Institute of Physics and EC PRISMA$^{++}$, Mainz, 55099 Mainz, Germany}

\newcommand{\ham}{Institut f{\"u}r Experimentalphysik, Universit{\"a}t Hamburg, 22761 Hamburg, Germany}

\newcommand{\rut}{Department of Physics and Astronomy, Rutgers, The State University of New Jersey, 136 Frelinghuysen Road, Piscataway, NJ 08854-8019 USA}

\newcommand{\boul}{University of Colorado at Boulder, Department of Physics, Boulder, CO 80309, USA}

\newcommand{\mun}{Physics Department, Technische Universit{\"a}t M{\"u}nchen, 85748 Garching, Germany}

\newcommand{\llnl}{Lawrence Livermore National Laboratory, Livermore, CA 94550, USA}

\newcommand{\erc}{Department of Physics, Erciyes University, 38030, Kayseri, Turkey}

\newcommand{\iow}{Department of Physics and Astronomy, The University of Iowa, Iowa City, IA 52242, USA}

\newcommand{\psu}{Department of Physics, Pennsylvania State University, University Park, PA 16802, USA}

\newcommand{\westpoint}{United States Military Academy, Department of Physics and Nuclear Engineering, West Point, NY 10996}

\newcommand{\bart}{Bartoszek Engineering, Aurora, IL 60506, USA}

\newcommand{\css}{California State University, Stanislaus, Department of Physics, Turlock, CA 95382, USA}

\newcommand{\cseb}{California State University, East Bay, Department of Physics, 25800 Carlos Bee Blvd, Hayward, CA 94542}

\newcommand{\irvine}{Department of Physics and Astronomy, University of California, Irvine, Irvine, CA 92697-4575, USA}

\begin{abstract}
In this manuscript we present the first results from \eos{}, a four tonne optical detector located at the University of California, Berkeley. The primary goal of \eos{} is to demonstrate the performance capabilities of scintillation-based, `hybrid' detector technology for future neutrino detectors. The data presented were collected while both the inner target vessel and the outer buffer vessel were filled with water. The water target acts as a well-understood medium that produces only Cherenkov light, which can be used to calibrate and develop the detector model and reconstruction algorithms prior to the deployment of scintillating material. Using deployed optical and radioactive calibration sources, a series of detailed detector calibrations are performed. These enable a suite of tests for various reconstruction algorithms. Simulations that use calibrated models are compared with the data across a variety of different types of calibration sources, source positions, and rotations. 

\end{abstract}

\maketitle 

\clearpage

\section{Introduction}

The study of neutrinos is a tremendously broad field of science, with insights into the fusion processes in our sun, the explosion mechanism of distant supernovae, monitoring of terrestrial nuclear activity, and potentially the matter-antimatter asymmetry in the early Universe. The detection of neutrinos is a rich field, and historically significant progress has been made utilizing deep-underground detectors, such as Super-Kamiokande and SNO~\cite{Super-Kamiokande:1998kpq,SNO:2002tuh}, filled with water. In these experiments, neutrino interactions produce Cherenkov light, which can be utilized to reconstruct the direction of the incoming neutrino. In more modern experiments, such as Borexino and SNO+~\cite{Bonventre:2018hyd,SNO:2015wyx}, liquid scintillator is chosen as the target material, which produces far more light than water and can be more easily cleaned of the radioactive contaminants that produce backgrounds. However, detecting the relatively small amount of Cherenkov light in a liquid scintillator detector is challenging, and has only been achieved statistically across many events~\cite{BOREXINO:2021efb} or at relatively high energy~\cite{SNO:2023cnz}, in large-scale neutrino detectors.

Future detectors such as \theia~\cite{Askins:2019oqj} hope to utilize groundbreaking technology in order to simultaneously detect both Cherenkov and scintillation photons. This `hybrid' technique would allow for low thresholds, excellent energy and vertex resolution, event-by-event directionality, and particle identification that leverages the ratio of Cherenkov to scintillation photons. \theia{} offers world-leading sensitivity to a broad range of physics, including measurements of neutrinoless double beta decay, low-energy solar neutrinos, and the diffuse supernova neutrino background~\cite{TheiaDSNB}. Additionally, hybrid detectors are being investigated for far-field detection of antineutrinos for the purpose of nuclear nonproliferation~\cite{Askins:2015bmb,Bat:2021jyq}. 

There are several technology developments that are being explored for hybrid event detection. Novel scintillators can emphasize the Cherenkov signal, either by reducing the scintillation light yield~\cite{YehWbLS} or by slowing the scintillation emission timing~\cite{Biller:2020uoi,GUO201933,Steiger:2024nes}. Red-sensitive and/or fast photodetectors~\cite{Lyashenko:2019tdj,Kaptanoglu:2021prv} can enhance the Cherenkov signal, particularly in combination with spectral photon sorting provided by devices like the dichroicon~\cite{Kaptanoglu:2018sus,Kaptanoglu:2019gtg}. Advances in fast-electronics, waveform processing algorithms, and sophisticated reconstruction algorithms~\cite{Elagin:2016zgp} will aid in the identification of the Cherenkov photons. Other ideas include unique approaches to photon collection  to isolate the Cherenkov component in liquid argon neutrino detectors~\cite{CCM:2025kal}.

The \eos{} detector, described in detail in Sec.~\ref{sec:detector}, is a multi-tonne-scale optical detector that utilizes many of these advanced technologies. The primary goal of \eos{} is to demonstrate the performance capabilities of the hybrid detector technology, and to test Monte Carlo (MC) simulation model predictions. This includes a demonstration of full event reconstruction, ultimately utilizing both Cherenkov and scintillation light. These objectives largely drove the detector design and the selection of hardware and electronics for \eos{}. In the future, \eos{} plans to probe the dependence of the reconstruction resolution on detector configurations, such as the properties of the target material, the photocoverage, photon detector time resolution, and the addition of dichroicons for spectral sorting. MC models will be calibrated and validated against data, which will support predictions for next-generation experiments. More details about the detector goals are provided in Ref.~\cite{Anderson:2022lbb}.

\eos{} operated with the inner vessel filled with water in various time periods between May 2024 and April 2026 (the full water data collection period is referred to as the water phase). During this period, data were collected with deployed optical and radioactive calibration sources to evaluate the performance of the detector prior to introducing scintillation light, to calibrate the behavior of the various detector components, and to develop the infrastructure and analysis tools necessary for achieving the experimental goals. The relatively simple optical behavior of the water target makes this period critical for evaluating the baseline performance of \eos{} and provides a benchmark to which future phases can be compared. The future phases involve the deployment of both water-based liquid scintillator and `pure' liquid scintillator (e.g., LAB+PPO) as the primary target material. In this manuscript, we describe the detector in Sec.~\ref{sec:detector}, outline the simulation, processing, data collection, analysis tools, and event selection in Secs.~\ref{sec:detector-modeling}, ~\ref{sec:data-processing}, ~\ref{sec:data-collection}, ~\ref{sec:analysis}, and ~\ref{sec:event-selection} respectively, present the full calibration of the detector in Sec.~\ref{sec:detector-calibration}, and finally detail the systematics, model verification, and water phase reconstruction performance results in Secs.~\ref{sec:systematics}, \ref{sec:model-validation}, and \ref{sec:results}.



\section{The \eos{} Detector}\label{sec:detector}

\begin{figure}[ht!]
    \centering
    \includegraphics[width=0.8\textwidth]{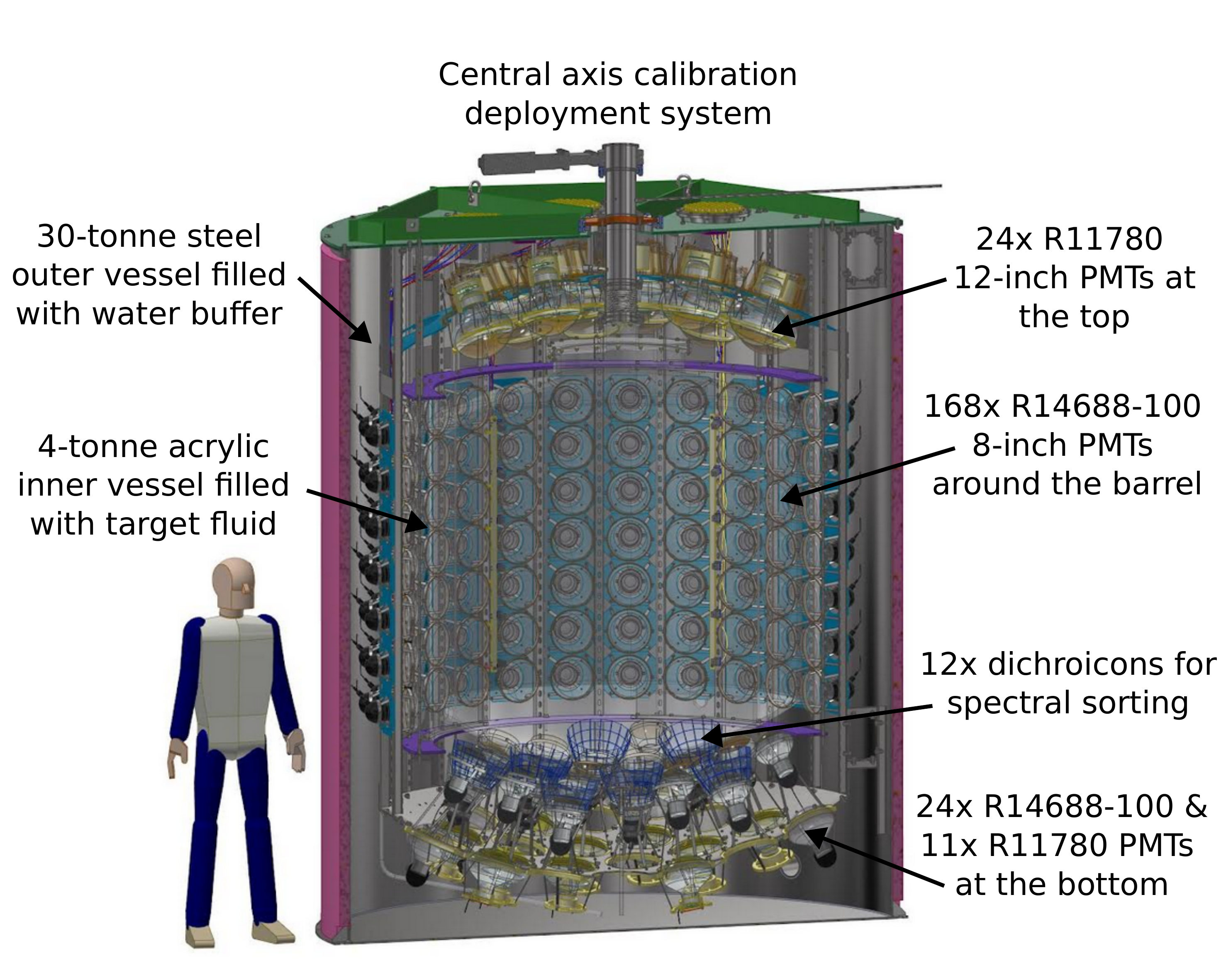}
    \caption{A schematic of the \eos{} detector showing the various detector components.}
    \label{fig:eos-detector}
\end{figure}

\begin{figure}[ht!]
    \centering    \includegraphics[width=0.8\textwidth]{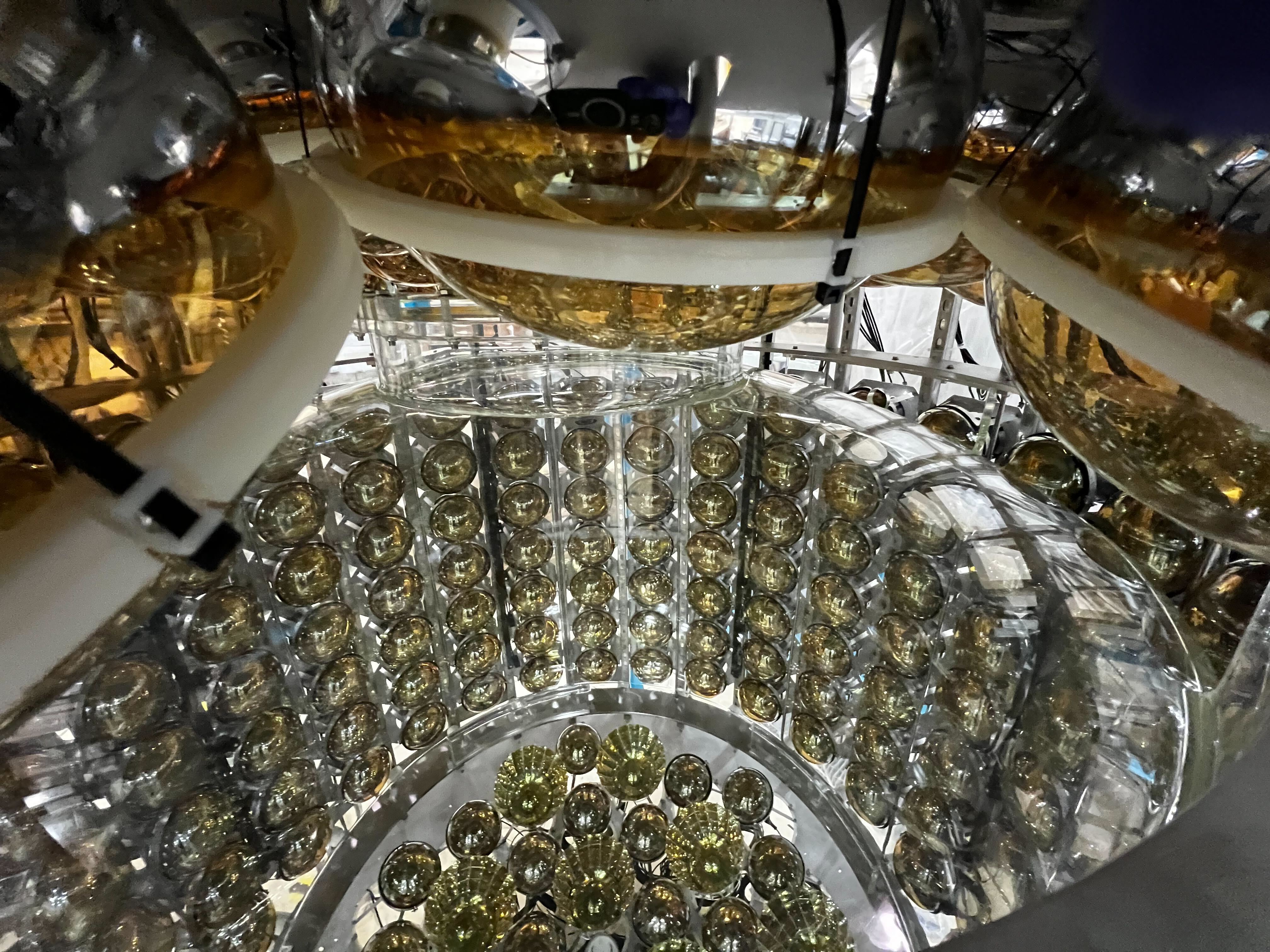}
    \caption{A picture of the \eos{} detector taken during the construction, while the IV was filled air and prior to moving the detector components into the OV. This image is looking down through the acrylic vessel towards the eight-inch R14688-100 PMTs along the barrel as well as some of the dichroicons at the bottom of the detector. The twelve-inch R11780 PMTs above the acrylic vessel are seen at the top. The top surface and neck of the inner acrylic vessel are also visible. More images of the detector are shown in Ref.~\cite{detector_paper}.}
    \label{fig:eos-picture}
\end{figure}

The \eos{} detector, shown in Fig.~\ref{fig:eos-detector}, has a four-tonne capacity acrylic inner vessel (IV), which can contain either a water or scintillator target. The IV is roughly cylindrical in shape, with ellipsoidal end caps and a 91~cm radius at the center. The IV is viewed by 241 PMTs, which detect the light created from interactions within the detector. \eos{} utilizes three different types of PMTs to study the impact of varying photon detector configurations. 204 Hamamatsu R14688-100 PMTs \cite{ham_datasheet_r14688} are placed around the barrel and at the bottom of the detector. These eight-inch diameter PMTs have excellent single photoelectron (SPE) timing resolution, less than one~ns FWHM, and high quantum efficiencies ($>$ 25\% at 400~nm)~\cite{Kaptanoglu:2023ayz}. Twenty-four Hamamatsu twelve-inch R11780 PMTs~\cite{Brack:2012ig} are placed at the top of the detector to increase the total light collection. An array of twelve dichroicons~\cite{Kaptanoglu:2019gtg} are placed in front of twelve of the eight-inch R14688-100 PMTs at the bottom of the detector to provide spectral sorting of the Cherenkov and scintillation light~\cite{Kaptanoglu:2019gtg}. Thirteen Hamamatsu ten-inch R7081 PMTs~\cite{ham_datasheet_r7081} are placed 35~cm behind the dichroicons to collect the light that passes through the dichroic filters. The PMTs and IV are contained within a 30-tonne capacity stainless steel outer vessel (OV) filled with water. A neck that runs from the IV to the top lid of the detector allows the calibration sources to be deployed down the central axis. Fig.~\ref{fig:eos-picture} shows a picture of the detector, taken during construction, that shows many of the components, including the various PMT types, the dichroicons, and the IV. The detector is surrounded by a muon veto system consisting of optically isolated plastic scintillator panels coupled to PMTs. For the data collected in this manuscript, the muon veto system was not yet operational. Ref.~\cite{detector_paper} provides more details about the detector design, configuration and construction of the \eos{} detector. 

The size of the \eos{} outer vessel was designed with practical limitations in mind, such as the size of the space available at UC Berkeley, the weight limitation of the overhead crane, and the width of the door into the space. With these limitations, the size of the detector was maximized, such that timing-based reconstruction algorithms could be tested. The choice to utilize the 12-inch and 10-inch PMTs (all repurposed from previous experiments) was made to maximize photocathode coverage, including coverage behind the dichroicons. The size of the inner vessel was limited by the space needed to fit the PMTs and, at the bottom, the dichroicons as well. The shape of the inner vessel was based on structural and practical needs, in coordination with the supplier, Reynolds Polymer Technology. The R14688 PMTs outperform the other PMTs, and 20 additional R14688 PMTs have been purchased since \eos{} has begun operating. These additional PMTs could be used to replace some or all of the other types of PMTs in the future. More details about how detector design choices were made is provided in Ref.~\cite{detector_paper}.

Based on the mechanical tolerances of the detector and calibration deployment system as well as measurements performed while constructing \eos{}, the vertical position ($z$) of the calibration source is known to within 1.5~cm. Additionally, a tilt of the calibration deployment system relative to the IV is known to be small ($<$ 5 degrees), but is determined much more precisely using the laserball source described in Sec.~\ref{sec:detector-calibration}.

The PMT cables are plugged into custom high voltage boards (HVSSs) that extract the PMT pulses and pass them to a set of CAEN V1730 digitizers \cite{caen_v1730}. The HVSSs are also responsible for producing an approximately 16~ns wide, 20~mV square pulse for each PMT pulse that crosses a programmable threshold. These square pulses are used for triggering the detector. The HVSSs each perform an analog sum of the square pulses across sixteen channels, and pass the sum to a central analog summing board (CASB). The CASB performs a global analog sum of the trigger signals and compares the sum against a configurable global threshold. The CASB provides five thresholds across two gain paths with different dynamic ranges. If any of the thresholds are crossed, a trigger is emitted and sent to the central trigger board (CTB). The CTB hosts an FPGA, which is used to apply programmable trigger logic to the trigger inputs. For example, the CTB allows one to trigger on the coincidence between any of the five thresholds within a certain time-window. If the trigger condition on the CTB is met, a final `global trigger' is emitted and sent to the CAEN boards to inform them to read out their data to disk. In addition to the CAEN data, trigger information from the CTB is also written out. The CTB provides a clock that is fanned out to the CAEN boards in order to keep the separate electronic components synchronized. 

All 241 PMT signals are digitized using CAEN V1730 waveform digitizers at a sampling rate of 500~MS/s. The data is readout over optical link and stored in \texttt{hdf5} files for processing. A description of the complete downstream processing chain can be found in Sec.~\ref{sec:data-processing}.

\section{Detector Modeling}\label{sec:detector-modeling}

\begin{figure}[ht!]
    \centering
    \includegraphics[width=0.5\textwidth]{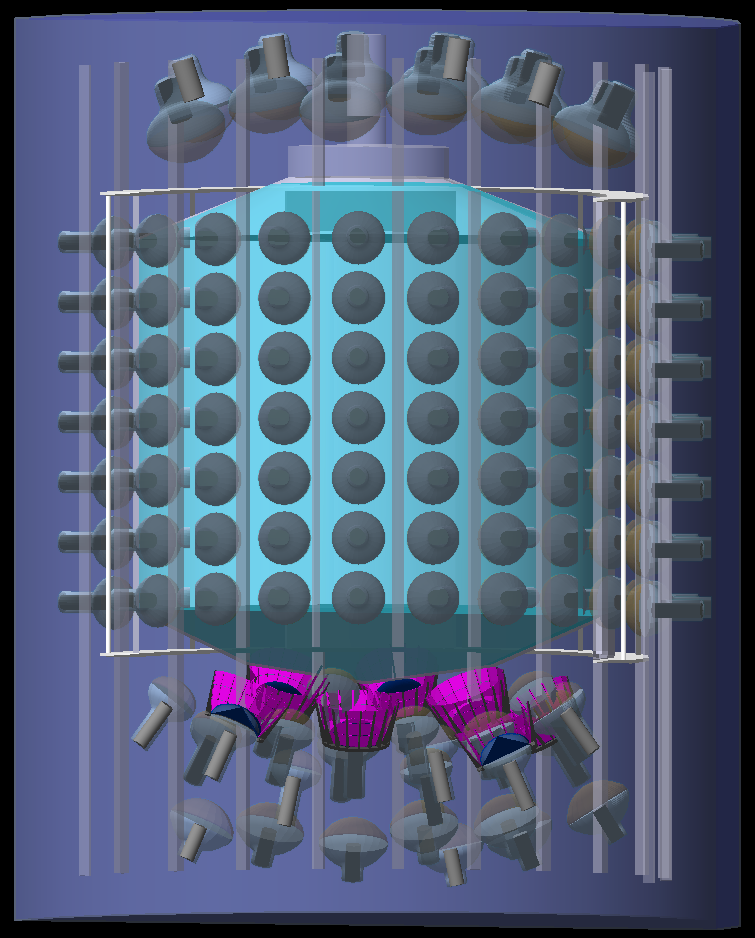}
    \caption{The \eos{} detector geometry model as implemented in the \rp{} software. The various geometry features include the three different types of PMTs, the dichroicons (pink), the inner IV (cyan), the acrylic neck, and the steel support structures.}
    \label{fig:simulated-detector}
\end{figure}

The \eos{} detector is modeled using the open-source, publicly-available \rp{} simulation and analysis package \cite{ratpac-two}. \rp{} contains \geant{}-based \cite{GEANT4:2002zbu} detector geometry descriptions, custom optical photon generation and propagation, individual 3D PMT modeling, a flexible database that holds run-dependent parameters, and customizable data-acquisition simulations. In \eos{} we simulate a detailed detector geometry, as visualized in Fig.~\ref{fig:simulated-detector}, that includes all of the PMTs, dichroicons, and material within the detector. The calibration sources are also modeled in similar detail. The physics interactions are handled by \geant{} and are initiated according to the details of each \eos{} calibration source. \geant{} version v11.1.2 is used for the simulation. Standard particle interactions are simulated using the "Shielding" reference physics list in \geant{}~\cite{g4guide}, which is optimized for low energy interactions and neutron transport. Optical photons are generated and propagated through the detector using custom optical physics modules, which are based on standard \geant{} processes but include improvements to the handling of Scintillation photon generation, absorption, and Rayleigh scattering.

The PMT response model is based on ex-situ and in-situ measurements of the quantum efficiency, transit time distribution, pulse shape, and pulse charge. The cable delays and set of offline PMTs are monitored on a regular basis and included into the \rp{} database. Further details about how the PMT parameters are extracted are provided in Sec.~\ref{sec:detector-calibration}.

A simplified simulation of the \eos{} trigger performs an analog sum of the square pulses created for each PMT hit. This trigger decides when an event has crossed threshold and flags \rp{} to write the data to disk. Typically, the threshold is selected to be sufficiently low such that we trigger on more events in the simulation than in the data. This allows for the flexibility to study below-threshold events. An analysis-level threshold is applied to both the data and simulation, as described further in Sec.~\ref{sec:detector-calibration}.

For triggered events, each of the above-threshold pulses are passed to a model of the CAEN V1730 waveform digitizer with appropriate settings (fetched from a run-level database). This model digitizes the pulses and passes the set of digitized waveforms to a waveform analysis tool. This tool is identical to the one used to process the data, described in Sec.~\ref{sec:data-processing}. The outputs of the waveform analysis are the extracted hit times and charges, which are passed to the reconstruction algorithms described below. This allows for a straightforward, direct comparison between the simulation and data. 

\section{Data Processing}\label{sec:data-processing}

 A complete processing chain, written as part of \rp{}, reads the \texttt{hdf5} files produced by the data acquisition system (Sec.~\ref{sec:detector}) and extracts PMT times and charges that can then be used as inputs to reconstruction and analysis tools. The data processing tool generates detector data in the same format as the simulated events, allowing them to be analyzed in identical ways.

In order to extract the pulse times and charges from the waveforms, above-threshold waveforms are identified by applying a negative 5-mV threshold. If a channel has a downward-going pulse that crosses below threshold, the time associated with the peak is identified and the PMT gets counted toward the global `hit' sum (\nhit). In this manuscript, we refer to the total number of waveforms with a peak that crosses threshold as the \nhit.

\begin{figure}[ht!]
    \centering
    \includegraphics[width=0.45\textwidth]{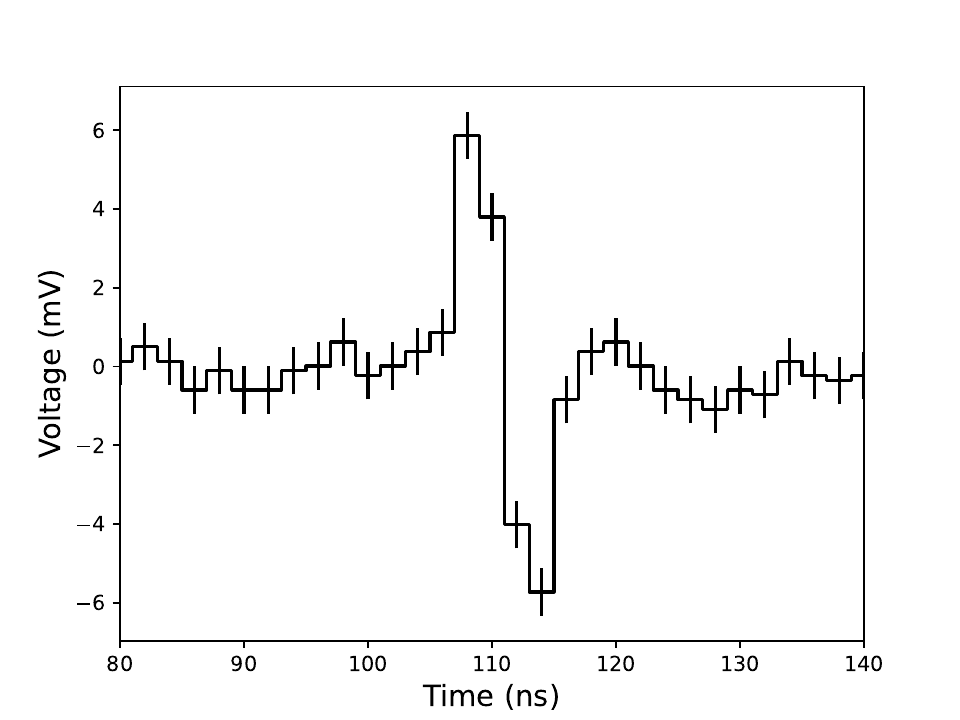}
    \includegraphics[width=0.45\textwidth]{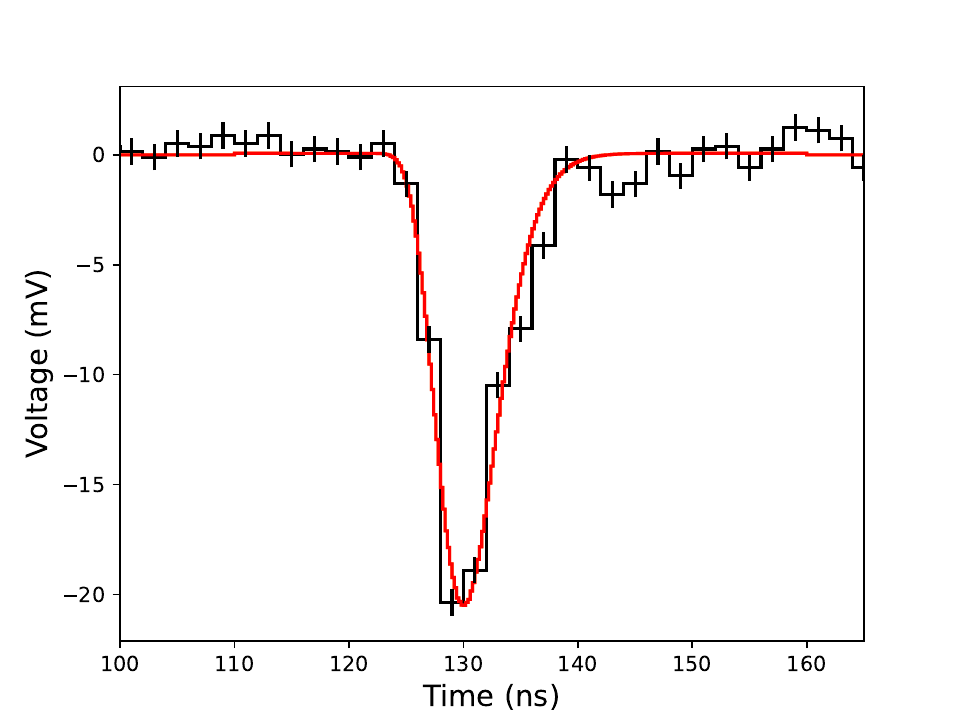}
    \caption{(Left) An example digitized waveform from pickup that passes the 5-mV hit threshold, collected during a cosmic run. This hit is removed by the cross-talk cut described in Sec.~\ref{sec:data-processing}. (Right) An example digitized waveform for an R14688-100 PMT from \eos{} laserball data. The red lognormal fit is performed according to Eq.~(\ref{eq:lognormal}) and is used to extract the arrival time of the pulse. }
    \label{fig:waveform}
\end{figure}

After the initial \nhit{} calculation, a cross-talk cut is applied to remove unwanted, bipolar signals from inductive pickup. An example of such a signal is shown in Fig.~\ref{fig:waveform} (left). This cut is particularly necessary for high-energy cosmic events, where large currents in a single channel can induce threshold crossing waveforms on many neighboring channels. The cross-talk cut is defined by the ratio of the negative to positive amplitude of each pulse. For induced signals this ratio is close to unity, while for pulses from photoelectrons the ratio is significantly larger. Single photoelectron laserball data (Sec.~\ref{sec:laserballsource}) was used to determine the optimal cut value, which provided a clear separation between photoelectron signals and cross-talk. Waveforms that do not pass this cut are removed from the \nhit{} of the event.
 
A lognormal function of the following form is fit in a range of $[-10, +15]$~ns around the peak of the above-threshold pulse: 
\begin{equation}\label{eq:lognormal}
    f\pp{t} = B + \frac{Q}{(t - \tz)\sqrt{2\pi}\sigma} e^{-\frac{1}{2}\pp{\log\pp{\frac{t - \tz}{m}}/\sigma}^{2}},
\end{equation}
where $B$ is the baseline of the waveform, $Q$ is the charge contained in the pulse, $\tz$ is the arrival time of the pulse, $\sigma$ is the shape parameter, and $m$ is a scale parameter that sets the median of the distribution. In the fit, the values of $m$, $B$, $t_{0}$, and $Q$ are floated and the value of $\sigma$ is fixed to 0.15. Fig.~\ref{fig:waveform} (right) shows an example waveform from an R14688-100 PMT, with the associated lognormal fit. The fit captures the shape of the PMT pulses in the data and the \chindf values are typically around one. The PMT time is calculated from the fit using 
\begin{equation}\label{eq:tpmt}
t_{\rm{PMT}} = t_{0} + m.
\end{equation}

Other methods to extract the pulse timing have been explored, including applying a constant-fraction discriminator, fitting a Gaussian function, and applying intersample sinc interpolation (see the methods described in Ref.~\cite{Warburton:2017}). The lognormal fit achieves better or equivalent timing when studying calibration data, and is the method used to extract the PMT times in this manuscript. 

In the data collected for this paper, the PMTs primarily detect single photoelectrons (SPEs). Therefore, the lognormal fits extract only a single hit-time for each waveform. Other methods using machine learning (ML) ~\cite{Jiang:2024wph} and other sophisticated approaches~\cite{Akashi-Ronquest:2014jga,Tang:2024jfs,Xu:2021whl} are currently under development. These tools will be appropriate for assigning hit-times and counting photoelectrons (PEs) for waveforms with multiple photoelectrons (MPEs).

\section{Data Collection}\label{sec:data-collection}

The data collected during the \eos{} water phase consist of deployed optical and radioactive sources. These sources are lowered into the detector via a calibration control system capable of deploying sources at fixed locations along the central vertical axis ($z$). In addition to deployed sources, \eos{} collected cosmic data to tag Michel electrons, which provides a well-understood, uniformly distributed source of events. \eos{} has also deployed several additional sources not described in this manuscript, including an internal acrylic Cherenkov source and an external $\gamma$ source. The sources each have associated detector settings that get loaded at the start of each data collection `run'. These settings include the trigger threshold and trigger scheme (e.g., if a coincidence trigger is required). Table~\ref{tab:calibration-details} summarizes the run plan for each of the calibration sources described in this manuscript. The total number of events collected and position of the sources were determined on a source-by-source basis based on calibration and reconstruction testing needs. Typically, several runs per calibration source position are collected, and repeated runs are used to constrain systematics. Significantly more data (mostly with the laserball) than presented in this manuscript has been collected with \eos{}, but it has been primarily used for ensuring the stability of the detector performance with time. The activity of all radioactive sources is low enough such that we expect negligible pile-up (more than one decay in a 280~ns trigger window).

\begin{table}[ht!]
    \centering
    \begin{tabular}{l|c|c|c|c}
         \hline \hline
         Source & $z$ range (cm) & $z$ step (cm) & Events per run ($10^{5}$) & Trigger scheme \\ \hline
         Laserball  & [-60, 60] & 10 & 2 & Pulsed \\ 
         Thorium & [-50, 50] & 10 & 5 & Single threshold \\
         AmBe & [-30, 30] & 30 & 10 & Coincidence \\ 
         Directional & [-50, 50] & various & 0.5 & Self-triggered \\ \hline \hline
    \end{tabular}
    \caption{Details of calibration source deployments in \eos{}. The $z$ range and step size indicate where data were collected along the central axis of the detector. The number of events collected is per calibration source position and was selected based on the trigger scheme and the relevant background rates. The laserball scan was collected with four different wavelengths, described in Sec.~\ref{sec:laserballsource}. The directional source data was collected with two different radioactive sources, $^{90}$Sr and $^{106}$Ru, as described in Sec.~\ref{sec:directional-source}.}
    \label{tab:calibration-details}
\end{table}

\subsection{Laserball source}\label{sec:laserballsource}

The laserball source provides isotropic light at a fixed wavelength and tunable intensity. This is achieved by pulsing one of four lasers (374~nm, 408~nm, 442~nm, and 515~nm) into a single-mode fiber that is connected directly into a 9.5~mm diameter diffusing ball. The laserball is made of Teflon which scatters the light sufficiently to create an isotropic output. While none of the measurements in this manuscript rely on the laserball being perfectly isotropic, the deviations from a perfectly isotropic source are tested on the benchtop and show less than 5\% variations as a function of output angle.

The intensity of the laser is tuned by measuring the coincidence rate (between the laser pulse and a PMT pulse) for any given PMT and requiring that it is below 5\%. This ensures that the light detected at the PMTs is highly SPE. The pulsed trigger for the laser is used to trigger the digitizers, ensuring that the PMT light arrives back to the digitizers at a fixed time after the trigger signal. The laserball is used for most of the detector calibrations described in Sec.~\ref{sec:detector-calibration}.

\subsection{Thorium source}\label{sec:thorium-source}

The thorium source consists of ten cylindrical tungsten rods, each 17.5~cm long with a radius of 2.4~mm, arranged in a ring inside a cylindrical container of black Delrin. The tungsten rods contain 4\% $^{232}$Th by mass. The $^{232}$Th decay chain produces several $\beta$, $\alpha$, and $\gamma$ emitters, the former two of which are absorbed in the opaque Delrin. The $\gamma$-rays exit the source and Compton scatter in the detector, producing Cherenkov light. The primary $\gamma$-rays come from $^{208}$Tl (2.6~MeV). Lower energy $\gamma$'s are produced by $^{228}$Ac, $^{212}$Bi, and other isotopes, altogether resulting in more than 10 times fewer observable interactions in the detector than the $^{208}$Tl $\gamma$-ray, based on \rp{} simulations. 
The thorium source provides a sample of nearly monoenergetic $\gamma$'s with a known original position distribution (from within the source), allowing for the study of position and direction reconstruction performance. 

To trigger on the thorium source data, a trigger threshold on the CASB is set. This threshold is selected to study the spectrum of events emitted by the thorium source, without being so low that the dark-noise and other PMT related backgrounds start to dominate. A threshold of 100~mV was selected based on studies of the trigger efficiency as a function of \nhit, described in Sec.~\ref{sec:trigger-eff}. This threshold roughly corresponds to five in-time (within the 16~ns trigger pulse width) \nhit. To sustain the trigger rate at this threshold the CTB prescales the data by a factor of 32 (i.e., only 1 in 32 triggered events are collected), which yields a trigger rate around 500~Hz while the thorium source is deployed.



\subsection{Directional $\beta$ Sources}\label{sec:directional-source}

\begin{figure}[ht!]
    \centering
    \begin{minipage}{0.45\textwidth} 
    \includegraphics[width=\linewidth]{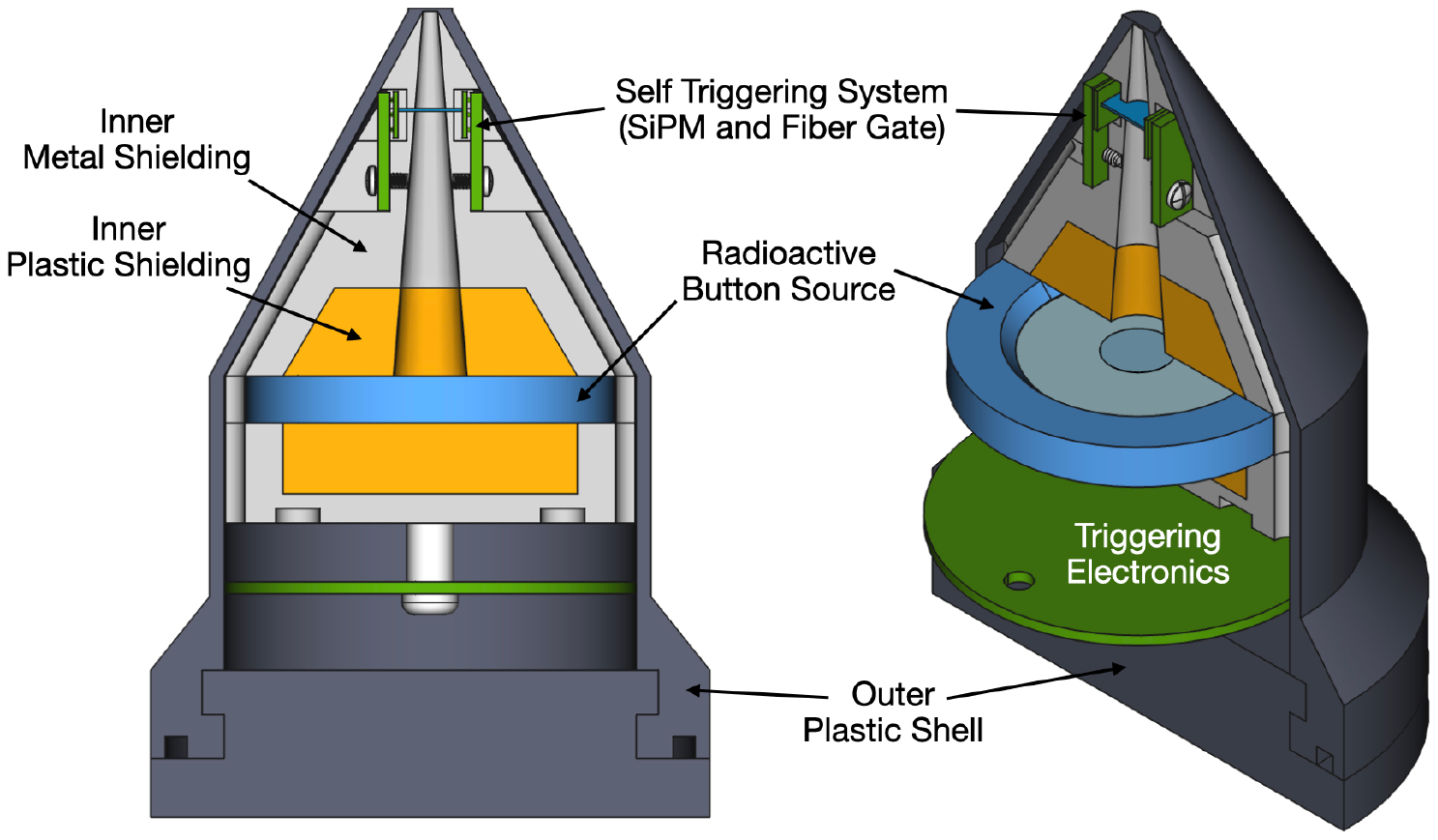}
    \end{minipage} 
    \begin{minipage}{0.45\textwidth} 
    \includegraphics[width=0.50\linewidth]{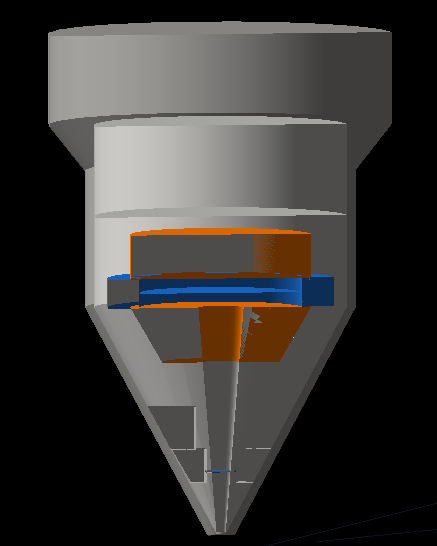}
    \end{minipage} 
    \caption{(Left) A schematic of the 30~mm directional source that shows the internal components necessary for shielding and self-triggering. (Right) Design of the 30~mm directional source in \rp{}.}
    \label{fig:dir-source}
\end{figure}

The directional $\beta$ sources, shown schematically for one design in Fig.~\ref{fig:dir-source} (left), consists of $\beta$-emitting isotopes surrounded by metal and delrin shielding with boreholes that allow $\beta$'s to exit the sources as a collimated stream in a fixed direction. Two different isotopes are used, $^{90}$Sr and $^{106}$Ru. The two isotopes emit $\beta$'s with Q-values of 2.28 MeV and 3.5 MeV respectively, with $^{106}$Ru emitting deexcitation $\gamma$s for 20\% of the decays. The sources can be rotated to desired zenith angle ($\theta$) and azimuthal angle ($\phi$). The \rp{} model of the source is shown in Fig.~\ref{fig:dir-source} (right).

A self-triggering system, consisting of a scintillating fiber ribbon and two silicon photomultipliers (SiPMs), tags exiting $\beta$'s with an approximately 50\% efficiency. Electronics inside the source discriminate the SiPM pulses and emit trigger pulses that are plugged directly into the CTB. The CTB is configured to emit global triggers when the directional source self-trigger occurs in coincidence with a CASB-generated trigger. A 128~ns window is used for this coincidence, which was tuned to provide a pure sample of tagged directional source events. The CASB threshold for the directional source data is typically set to 50~mV. The rate of exiting $\beta$'s out of the source into the detector is on the order of a few Hz and the coincidence between the self-trigger and the detector trigger provides a clean dataset with less than 1\% background.

The source is constructed in three different sizes to study the effect of the shadowing caused by the source body in future scintillator phases. The results described in this paper use both 30~mm sources and the 20~mm $^{106}$Ru source. These directional sources provide a sample of $\beta$ events with a known position, energy distribution, and direction, making them useful for studying reconstruction performance, particularly direction reconstruction.

\subsection{AmBe Source}\label{sec:ambe-source}
The $^{241}$Am$^9$Be (referred to as AmBe) source produces around 900~Hz of neutrons, more than half of which are accompanied by a 4.4-MeV $\gamma$.  An $\alpha$ from the decay of $^{241}$Am is absorbed by $^9$Be to produce $^{12}$C* and a neutron, where the $^{12}$C* de-excites with the 4.4-MeV $\gamma$.  The neutron will capture on a H atom in the water and produce a 2.2-MeV $\gamma$ with a mean capture time of around 210~$\mu$s.  The delayed coincidence of these two $\gamma$'s allows for the pure selection of AmBe signals.  

$^{241}$Am also produces a large number of X-rays, most abundantly at 59~keV.  To absorb most of these, the AmBe source is enclosed in 1.5~mm of lead.  Furthermore, the neutrons produced can have energies up to around 10~MeV, which can induce scintillation light (when the target is a scintillator) via elastic scatters with protons.  To prevent contamination of the prompt 4.4-MeV $\gamma$ signal by these scatters, the AmBe source is further enclosed in a large cylinder of opaque black acrylic with a height of 13~cm and a diameter of 14~cm.  
Thus, the AmBe source has the potential to provide two mono-energetic $\gamma$'s with a known origin and an understood distribution, allowing the study of position, direction, and energy reconstruction performance.

The trigger scheme for the AmBe relies on the coincidence between the multiple gain paths from the CASB, as performed by the CTB. A threshold of 800 mV is configured for the lowest gain path, which is used primarily to tag muons passing through the detector. Any event that causes this gain path on the CASB to trigger is excluded and no global trigger is emitted. Additionally, all triggers are locked out for approximately 1~$\mu$s after a high threshold trigger on the low gain path, to reduce contamination produced by ringing, after-pulsing, and short-lived followers after a cosmic muon interaction.

A threshold of 120 mV is set on the medium gain path (used for the prompt events) and a threshold of 70 mV is set on the high gain path (used for the delayed events). The logic on the CTB waits for a trigger on the medium gain path (with no accompanying trigger on the low gain path) and then masks in the high gain path for 400~$\mu$s. More precisely, the high gain (low threshold) path is enabled in a window of [1, 401]~$\mu$s after a medium gain trigger. Excluding the first 1~$\mu$s rejects any reflections or other noise on the trigger without significantly impacting the efficiency of the selection. This trigger scheme is specifically designed to select the prompt $\gamma$-rays from the AmBe (4.4~MeV signal) succeeded by the neutron capture on a H atom (2.2~MeV signal), while attempting to reduce backgrounds from cosmic muon followers. The selection of the thresholds for the prompt and delayed event are configured as low as possible such that triggering without a prescale is possible. 

\subsection{Cosmic ray muons}\label{sec:cosmics}

Due to minimal overburden, \eos{} is bombarded by cosmic ray muons at a rate of several kHz. These muons and the spallation products created by them can cause significant backgrounds across a wide range of energies. However, they also provide an opportunity to act as an additional calibration source. In particular, both Michel electrons and neutrons are easy to tag by looking for high energy events followed closely in time by a second event.

To trigger on cosmic muons and their followers, a primary threshold of 800~mV, equivalent to roughly 40-50 \nhit, is used to tag high energy events. Similar to the trigger scheme for the AmBe, a low energy follower window is opened for 400~$\mu$s with a threshold of 140~mV, excluding the first 1~$\mu$s after the primary trigger. This high energy follower trigger provides a sample of easily tagged Michel electrons.

\subsection{Backgrounds}\label{sec:backgrounds}

\begin{figure}[ht!]
    \centering
    \includegraphics[width=0.45\linewidth]{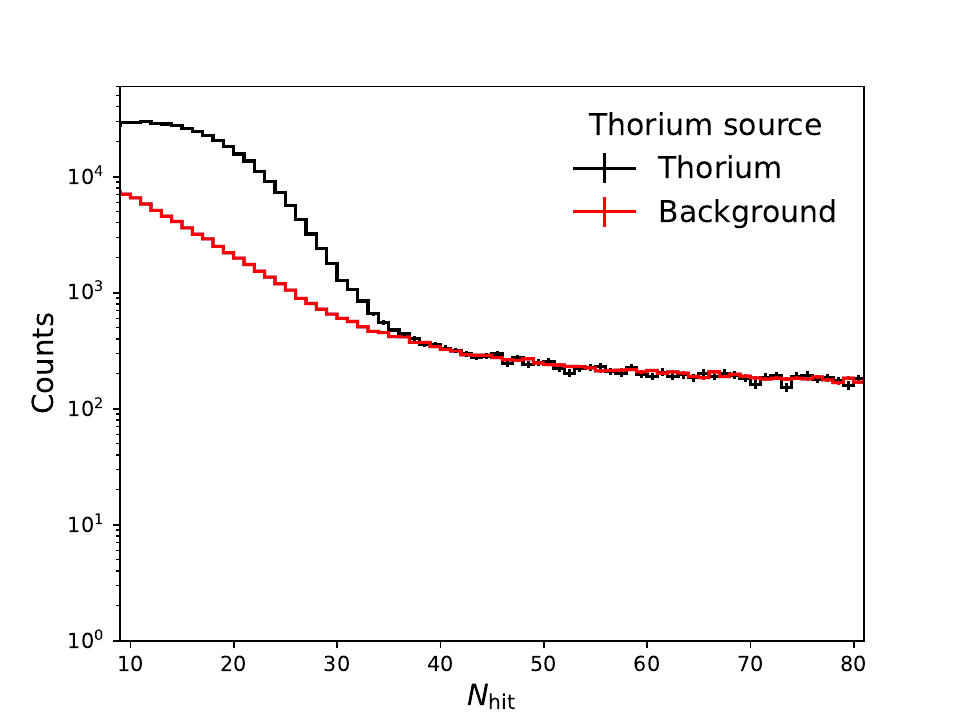}
    \includegraphics[width=0.45\linewidth]{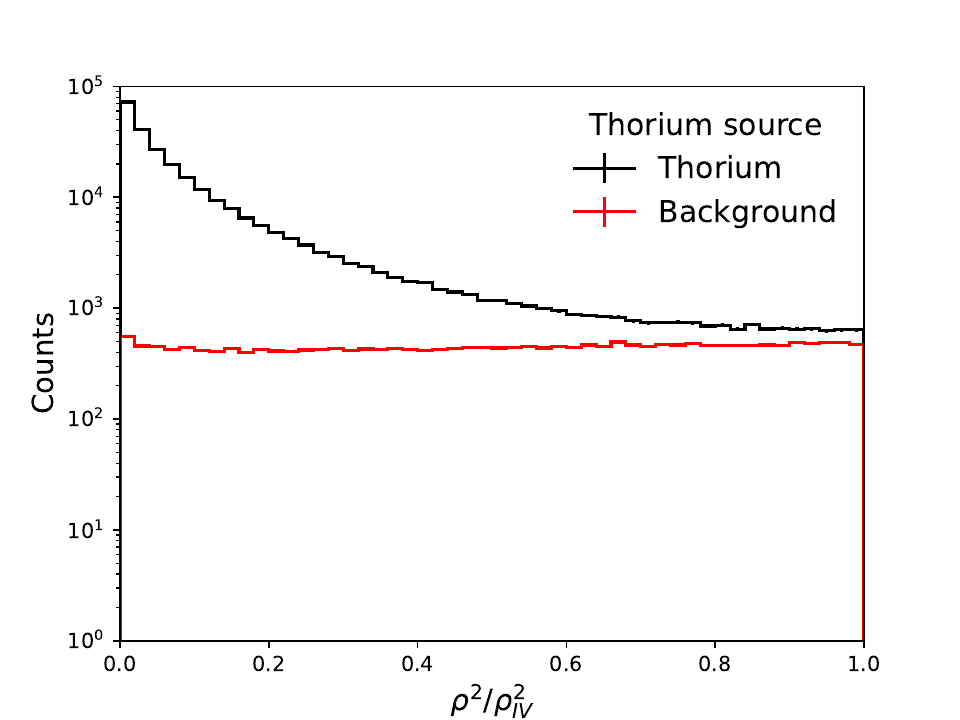}
    \caption{(Left) The \nhit{} of a central thorium source run compared to the associated background run, between 10 and 80 \nhit. The data for the background run is scaled by the relative live time. (Right) The reconstructed $\rho^{2}/\rho^{2}_{IV}$ distribution is from the \snd{} vertex fitter (see Sec.~\ref{sec:vertex-recon}), for central thorium and background data, selecting events in an \nhit{} range between 15 and 40.}
    \label{fig:thorium-background}
\end{figure}

Other than the directional source, the radioactive sources described in this paper are not tagged sources (e.g., by a sensor inside the source). Therefore, \eos{} relies on the relatively high decay rates of the calibration sources to overcome the background rates, for example, from radioactive decay products of uranium and thorium in the water or PMT glass or from spallation products from muons. To account for the effect of the backgrounds on the measured quantities, dedicated runs are collected with no source in the detector and with the detector settings configured identically to the associated calibration source run. For example, background runs are collected directly before and after the thorium source data collection campaign with detector settings identical to the thorium source runs. The \nhit{} and $\rho^{2}/\rho^{2}_{\rm{IV}}$ (where $\rho = \sqrt{x^{2} + y^{2}}$ and $\rho_{\rm{IV}} = \sqrt{91.2^{2} + 91.2^{2}}$ uses the dimensions of the IV, in centimeters) of a central thorium run ($z$ = 0) are compared to the associated background run in Fig.~\ref{fig:thorium-background}. The background distributions are scaled using the relative data collection live times. In the radioactive source analyses, the background data is statistically subtracted from the calibration source data. In several of the analyses, a fiducial volume cut is used to reject events at high radii, which are predominantly background.

\section{Event Reconstruction}\label{sec:analysis}

As part of the primary detector goals, \eos{} is being used to develop and test both traditional and novel reconstruction methods utilizing both the Cherenkov and scintillation light in a tonne-scale hybrid detector. Towards this goal, the reconstruction methods have been developed to run in a pure Cherenkov detector and were characterized during the water phase. In this section we present the details of each of the algorithms. In Sec.~\ref{sec:results} we detail the reconstruction performance for each of the various reconstruction algorithms, and show data and simulation comparisons.

Reconstructing events in \eos{} poses unique opportunities and challenges that will help develop methods for future detectors, such as \theia{}. Considering only Cherenkov light for this manuscript, the reconstruction challenges include an asymmetric detector orientation (e.g., different PMTs at the top and bottom), large calibration source sizes (relative to the size of the detector), and the relatively large sensor to detector size ratio.

The reconstruction methods described in this section all share several features. First, all take as input the PMT timing information produced by the lognormal fitting procedure performed over the waveforms, as described in Sec.~\ref{sec:data-processing}. All methods use the same set of hit PMTs, including the removal of cross-talk channels, and assume all hits are SPEs. Because the water data is dominated by SPE hits, no PMT charge information is currently used. The likelihood algorithms and the ML-based algorithm are flexible to enable the use of PMT charge information in the future. Currently, only the ML-based algorithm, described in Sec.~\ref{sec:hitman}, simultaneously fits for the vertex and direction, and upgrades to the likelihood-based approaches could enable them to fit for both event position and direction at the same time.

The likelihood-based fitting methods use simulated events to produce probability density functions (PDFs). In all cases, a set of PDFs is generated using simulated electrons, distributed uniformly in the detector, with isotropically selected directions, and ranging in energies from 0.5 to 15.0 MeV. A total of twenty million events are simulated for PDF construction and for training the ML algorithm described in Sec.~\ref{sec:hitman}. This method for PDF generation was selected because MeV-scale neutrino interactions are typically detected in Cherenkov detectors through neutrino-electron scattering, and thus are reconstructed under the hypothesis that the event is an electron. By simulating uniformly in position and direction and across a wide energy range, we avoid biasing the fit methods. Other methods for generating the PDFs were investigated, such as using simulations of the calibration sources themselves, and have only a very small impact on the reconstruction results. 

\subsection{Vertex reconstruction}\label{sec:vertex-recon}

There are several algorithms, described below, that run as part of the \eos{} analysis chain to produce reconstructed event vertices. In all cases, the event position is defined by three Cartesian coordinates, with $z$ as the vertical axis of the detector, an event time ($t_{\rm{recon}}$ in Equation~\ref{eq:timeRes}), and the origin (0, 0, 0) of the detector at the center of the IV. An important quantity that is calculated and used by reconstruction methods is the PMT time-residual. These residuals for a hit PMT are calculated as 
\begin{equation}\label{eq:timeRes}
t_{\rm{res}} = t_{\rm{PMT}} - t_{\rm{tof}} - t_{\rm{recon}},
\end{equation}
where $t_{\rm{PMT}}$ is extracted using the lognormal fitting procedure and is defined in Equation~(\ref{eq:tpmt}), $t_{\rm{tof}}$ is the photon time-of-flight from the reconstructed position to the PMT, and $t_{\rm{recon}}$ is the reconstructed time of the event. In general, $t_{\rm{recon}}$ captures a DAQ- and trigger-dependent arbitrary offset of the time-residuals from zero. To calculate the time-of-flight, the speed of 400-nm photons in water is set to 222.69~mm/ns (using $n$ = 1.344)~\cite{Daimon:07} and straight paths are assumed. The value at 400~nm is used because it roughly corresponds to where the PMT quantum efficiency distribution peaks, and selecting other reasonable fixed values of $n$ has negligible impacts on the results.  

In certain scenarios (e.g., for some calibration methods), the known calibration source position, rather than the reconstructed position, is used to generate the time-residuals, in which case the value of $t_{\rm{recon}}$ is set to zero and the time-residuals are indicated as $t^{\rm{source}}_{\rm{res}}$.

\subsubsection{Quad fitter}

The ``quad fitter'' runs on events with four or more PMT hits. The method selects four PMTs from the set of hit PMTs and calculates the event vertex based on the hit-times of those four PMTs, assuming the light traveled in a straight line from a single interaction point in the detector. This method is then repeated for a new set of four PMTs to produce a second event vertex position. A ``cloud'' of estimated vertex points is generated and the median of that cloud is selected as the event vertex. The quad fitter has the advantages of being fast and very robust, and is thus used to seed the position of some of the slower, more precise fitting algorithms. This method has been used by large Cherenkov experiments, such as SNO~\cite{coulter-thesis}. 

\subsubsection{Likelihood-Based Approaches}\label{sec:likelihood-vertex-fitters}

\eos{} utilizes two independently developed likelihood-based vertex fitters called \snd{} and \mimir{}. These frameworks perform likelihood-based event reconstruction strategies in water Cherenkov and scintillation detectors. The methods both reconstruct the position of the event by minimizing a PDF in time residual. Formally, the vertex $\boldsymbol{\theta} = \left(\vec{x}, {t}\right)$ is estimated by maximizing the following likelihood:
\begin{equation}
    \mathcal{L}\left( \boldsymbol{\theta} | \left\{ t_{res}^{i}\right\} \right) = \prod_i \mathcal{P}\left(t_{res}^{i}\right)
\end{equation}
where $t_{res}^{i}$ is the time residual of the \emph{i}th hit of the event. In this manuscript, \snd{} uses a single PDF in time residual and \mimir{} uses distinct PDFs for each PMT type (R14688-100, R11780, and R7081). Both likelihood fitters are seeded with the result from the quad fitter. The optimization used by \snd{} is the \texttt{NLopt} \cite{NLopt} implementation of the Subplex algorithm, \texttt{Sbplx} \cite{SUBPLEX}. \mimir{} uses the \texttt{ROOT} optimizer \texttt{MINUIT2} \cite{minuit}. Because \mimir{} and \snd{} perform similar calculations they are expected to yield similar results in \eos{}. Having two similar algorithms is helpful for providing robustness checks and identifying bugs and outliers in individual algorithms. Additionally, \mimir{} internally continues to fit for the event direction (Sec.~\ref{sec:dir-recon}) using the vertex fitting result as the hypothesis.

\subsection{Direction reconstruction}\label{sec:dir-recon}

As with the algorithms that are used to fit event vertex, there are several algorithms, described below, that run as part of the \eos{} analysis chain to produce a reconstructed event direction. The performance of these algorithms is primarily characterized on the directional sources described in Sec.~\ref{sec:directional-source}. For these sources, the reconstruction of event direction is studied by calculating: 
\begin{equation}
\alpha = \hat{d}_{\text{source}} \cdot \hat{d}_{\text{recon}}
\end{equation}
where $\hat{d}_{\text{source}}$ is the known direction in which the source is pointing and $\hat{d}_{\text{recon}}$ is the reconstructed direction.  This metric is appropriate for the directional source, where the average emission angle of the electrons from the source is known. We expect to find a peak in $\alpha$ at a value of one. This metric does not provide an absolute measure of reconstruction resolution in \eos.  Instead, a comparison of data to simulation allows us to validate the model, which can then be used to evaluate direction resolution for a defined event type.

The distribution of $\alpha$ over many reconstructed events tends to be exponential and can be quantified more easily by fitting an empirical function to the $\alpha$ distributions. The function form, $f$, is an exponential convolved with a Gaussian where the exponential time-constant and the Gaussian width and mean are floated in the fit. After performing this fit, the 1$\sigma$ (68\%) directional resolution ($\alpha_{1\sigma}$) is calculated by finding the value when $\int_{\alpha_{1\sigma}}^{1} f d\alpha = 0.68$. Conceptually $\alpha_{1\sigma}$ quantifies the fraction of events that reconstruct with directions similar to the true direction. The fit also provides the uncertainties on $\alpha_{1\sigma}$. 

For each algorithm the direction reconstruction is calculated using only hits in a prompt time-residual window (between -3 and 5~ns), where the direct Cherenkov photons are expected to arrive. The hits outside this window are likely from PMT dark current or scattered/reflected photons, the inclusion of which would smear out the direction resolution.

\subsubsection{Average direction}

The average direction fitter takes the average direction of the vectors that point from an input event position to the hit PMT positions. The average direction algorithm has the advantages of being fast and robust, and thus can be used to seed the direction of the likelihood-based fitting algorithms. The  event position input for the average direction fitter is taken from the \snd{} result. 

\subsubsection{\AERO{}}

The likelihood-based fitter \AERO{} (Angular Event Reconstruction via likelihood Optimization) utilizes HEALPix \cite{healpix}, an algorithm designed to discretize a spherical surface into equal-area pixels, allowing uniform sampling of directions in three-dimensional space. To create the PDFs for the likelihood-based optimization, events are binned into appropriate HEALPix pixels based on the true initial direction of the electron. For each of the 192 HEALPix pixels a 2D PDF of time-residuals vs PMT ID is generated. Conceptually, these PDFs represent the PMT timing and occupancy information as a function of the event direction (binned into 192 directions on a sphere).

When evaluating the likelihood for a hypothesized direction, \AERO{} averages the PDFs of the 20 nearest pixels, weighted by their angular distances to the evaluation direction, to generate a new, interpolated PDF. This interpolation method allows the reconstruction of events across the full sphere, rather than only at 192 fixed directions. The likelihood of an event in the evaluation direction producing PMT and time-residual patterns seen in the PDF is calculated. The direction that minimizes the negative log-likelihood is selected as the best-fit hypothesis. \AERO{} uses the \texttt{ROOT} optimizer \texttt{MINUIT2} and an input event position from the quad fitter result.

\subsubsection{\mimir{}}

Following the general framework and vertex fitting routine described in Sec.~\ref{sec:likelihood-vertex-fitters}, a direction fitting strategy is performed. In this step, the direction of the particle $\boldsymbol{\theta}_{\textrm{dir}} = \hat{d}_{\mimir{}}$ is estimated by maximizing the likelihood:
\begin{equation}
    \mathcal{L}\left(\left\{ \hat{d^i} \right\}|\boldsymbol{\theta}_{\textrm{dir}}\right) = \prod_i \mathcal{P}_{\textrm{type}}\left(\hat{d}_{\mimir{}} \cdot \hat{d^i} \right)
\end{equation}
where $\hat{d^i}$ is the unit vector pointing from the event vertex to the position of the $i$th PMT hit in the event. 




\subsection{Deep-Learning-based Reconstruction}\label{sec:hitman}

\hitman{} (Highly-parallelizable Inference Technique for Monitoring AntiNeutrinos) is a simulation-based inference tool that uses deep learning techniques to generate likelihood contours for particle kinematics $\boldsymbol{\theta}$ given photosensor data $\mathbf{x}$~\cite{Eller:2022xvi}. It provides a surrogate model for the detector likelihood using Neural Ratio Estimation (NRE). 

\subsubsection{Neural Ratio Estimation}

Unlike regression models that typically provide point estimates, the NRE learns the entire likelihood function. This is accomplished by training a classifier to distinguish correlated simulation pairs $\{\mathbf{x}, \boldsymbol{\theta}\}$ from uncorrelated pairs, approximating the likelihood-to-evidence ratio~\cite{hermans2020likelihood}. To efficiently handle sparse, variable-sized PMT data, we decompose the likelihood using an extended maximum likelihood formulation~\cite{Eller:2022xvi}: 
\begin{equation}
\mathcal{L}(\boldsymbol{\theta}|\mathbf{x}) = \left[\prod_{i=1}^{N_{\mathrm{tot}}} q(x_i | \boldsymbol{\theta}) \right] \mathcal{P}(N_{\mathrm{tot}} | \Lambda_\mathrm{tot}(\boldsymbol{\theta})).
\label{eq:hitman_eml}
\end{equation}
Here, $\mathcal{L}(\boldsymbol{\theta}|\mathbf{x})$ is the likelihood function with real data $\mathbf{x}$ under some hypothesis $\mathbf{\theta}$,  $N_{\mathrm{tot}}$ is the total observed photons in the entire detector with expected value $\Lambda_\mathrm{tot}(\boldsymbol{\theta})$, and $q(x_i | \boldsymbol{\theta})$ is the probability density for a single photon measurement $i$ with features $x_i$ (detection position and time).

Two NRE networks are trained to approximate $q(x_i| \boldsymbol{\theta})$ and $\mathcal{P}(N_{\mathrm{tot}} | \Lambda_\mathrm{tot}(\boldsymbol{\theta}))$:
\begin{itemize}
    \item \textbf{HitNet}: Approximates the per-photon term $q(x_i| \boldsymbol{\theta})$, processing individual photon detection features relative to $\boldsymbol{\theta}$.
    \item \textbf{ChargeNet}: Approximates the Poisson-like term $\mathcal{P}(N_{\mathrm{tot}} | \Lambda_\mathrm{tot}(\boldsymbol{\theta}))$, modeling the total number of detected photons (does not use charge in this manuscript, but is capable of doing so in the future).
\end{itemize}
The total log-likelihood is the sum of the ChargeNet output and the HitNet outputs summed over all photons. 

\subsubsection{Implementation and Parameter Estimation}

For the water phase analysis, the hypothesis $\boldsymbol{\theta}$ includes seven parameters: initial vertex ($x, y, z$), time ($t_{\rm{recon}}$), direction ($\phi, \vartheta$), and energy ($E$). The networks are comprised of dense layers with Mish activations \cite{misra2019mish}, trained on 20 million simulated electron events (0.5--15.0 MeV) described in Sec.~\ref{sec:analysis}.

Network inputs are feature-engineered; for example, HitNet uses the time residual between the photon detection and hypothesis so the network obeys event time translation symmetries. Currently, HitNet and ChargeNet rely on the single photoelectron assumption dictated by the current waveform analysis, meaning each hit PMT corresponds to a single photon detection ($N_{\mathrm{tot}} = N_\mathrm{hit}$). This approximation is sufficient for energy depositions up to a few MeV in water.  Once MPE extraction is implemented (Sec.~\ref{sec:data-processing}) these results will be naturally included in the current framework. 

For parameter estimation, we find the Maximum Likelihood Estimate (MLE) $\hat{\boldsymbol{\theta}}$ by minimizing the negative log-likelihood generated by HitNet and ChargeNet. A seed from the Quad fitter (Sec.~\ref{sec:vertex-recon}) is refined using the gradient-free Sbplx algorithm.

\section {Event Selection}\label{sec:event-selection}

Several techniques are used in \eos{} to reject ambient and instrumental backgrounds. For example, fiducial volume cuts, to select events near the center of the detector, avoiding the regions of the detector that are less well-understood and contain higher rates of backgrounds. The procedure to statistically subtract background distributions, described in Sec.~\ref{sec:backgrounds} is also applied for several sources. This section defines additional event selection criteria, which are specific to each source. The cuts are described in the following sections and summarized for each source in Table~\ref{tab:cuts}.

\subsection{Prompt \nhit{} Selection}\label{sec:prompt-nhit}

It is useful in \eos{} to define a prompt \nhit{} (\np). The prompt \nhit{} criterion selects hits in a narrow time window around the peak of the time-residual distribution, rather than counting the total number of hits over the full 220~ns trigger window. Using \np{} for calibrations greatly reduces the impact of mis-modelings of late light from the PMT transit response or from reflections in the detector. In this manuscript, \np{} is determined by selecting hits $-2~\mathrm{ns} < t^\mathrm{peak}_{\mathrm{res}} < 4$~ns, where $t^\mathrm{peak}_{\mathrm{res}}$ is the peak of the time-residual distribution (which is typically centered at 0~ns). Different \np{} cuts are applied for each dataset, depending on the energy region of interest. These cuts are necessary for several of the radioactive sources in order to avoid the trigger inefficiencies described later in Sec.~\ref{sec:trigger-eff}, which are not modeled in detail. 

\subsection{Directional Source Selection}\label{sec:dir-source-selection}

\begin{figure}[ht!]
    \centering
    \begin{minipage}{0.45\textwidth}
    
    \includegraphics[width=\linewidth]{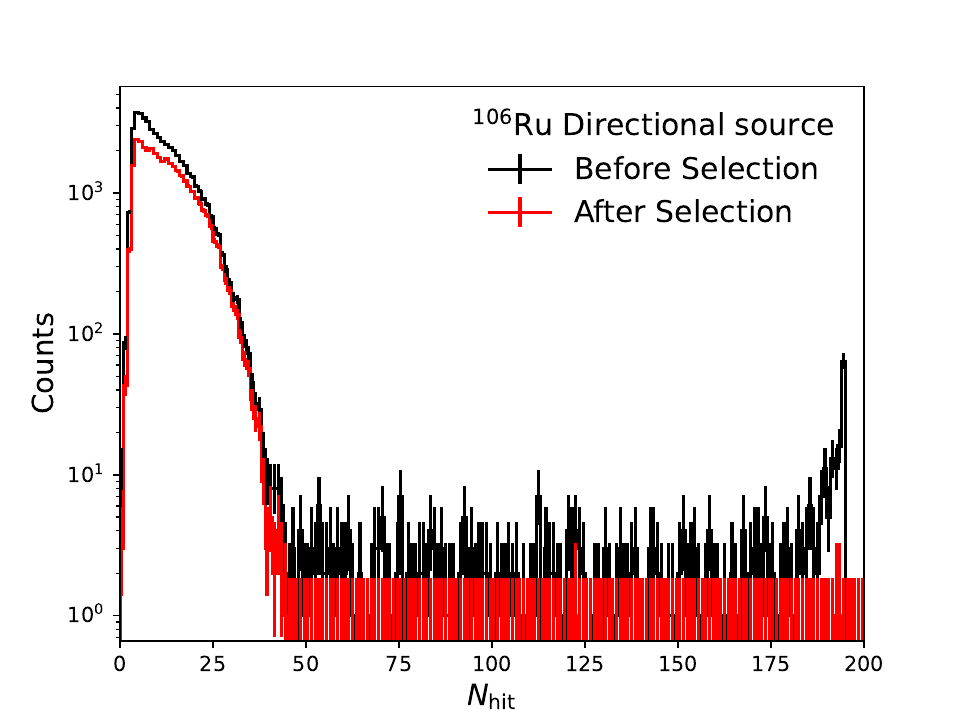}
    \end{minipage}
    \begin{minipage}{0.45\textwidth}
    
    \includegraphics[width=0.66\linewidth]{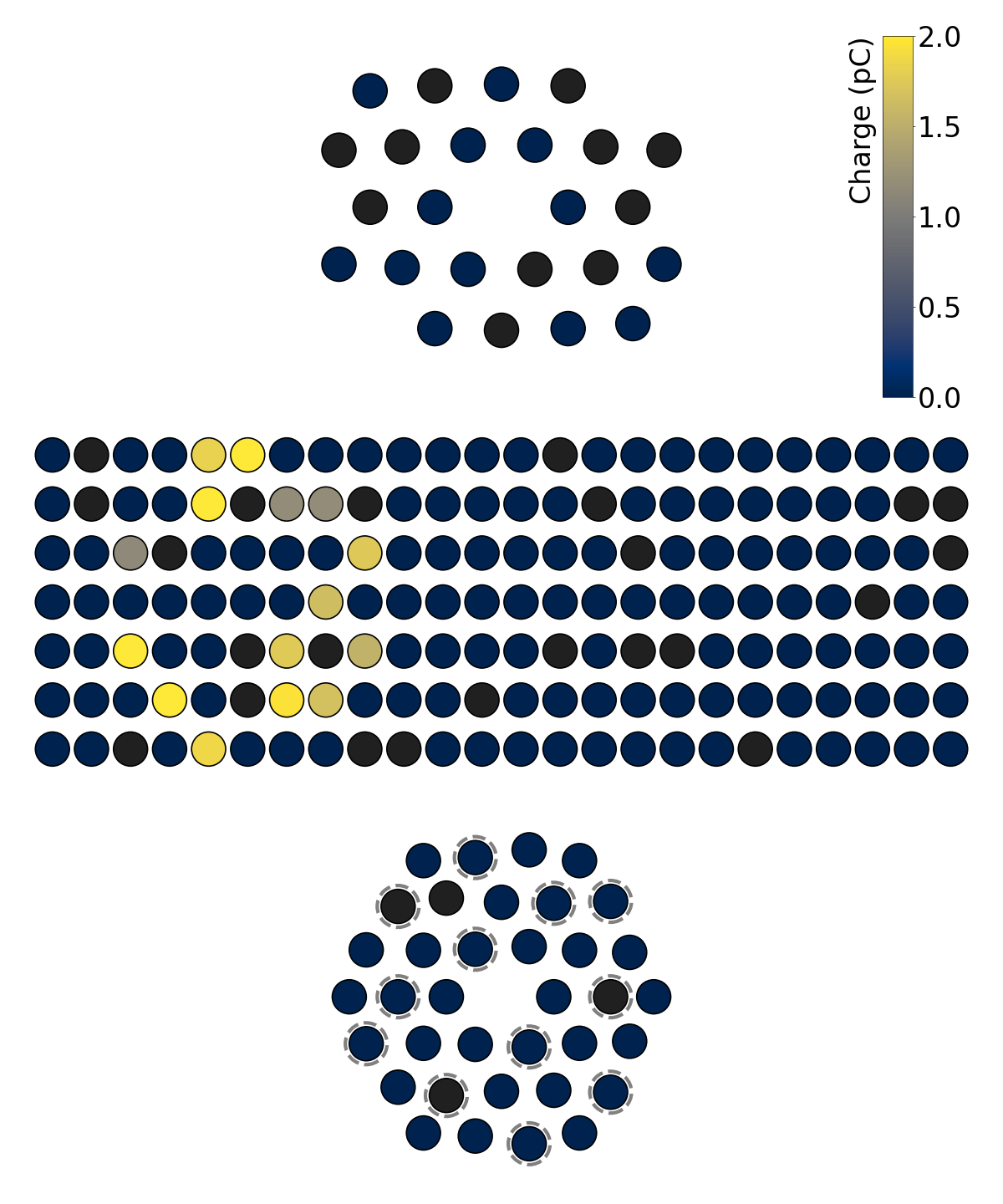}
    
    \end{minipage}
    \caption{(Left) The \nhit{} distribution for the 20~mm $^{106}$Ru directional source data before and after applying the directional source selection cuts described in Sec.~\ref{sec:dir-recon}. The peak around 200 \nhit{} is from accidental coincidences from cosmic muons, which are mostly removed through the directional source event selection. The uncertainties are statistical only. (Right) Event display of a tagged $\beta$ event from the $^{90}$Sr directional beta source, located at the center of the detector and pointed sideways, towards the barrel PMTs. The color scale indicates the charge of prompt hits, capped at 2~pC.}
    \label{fig:dir-source-event-display}
\end{figure}

As described in Sec.~\ref{sec:directional-source}, the directional sources use a self-triggering system consisting of a SiPM and scintillating fiber ribbon at the tip of the source. The time difference between the self-triggering time and the average PMT hit time for each event is calculated. A Gaussian fit to a histogram of this time-difference, across the full datasets, yields an average expected time-difference and events within $\pm2~\sigma$ are accepted. This cut removes the majority of the accidental coincidences between the detector trigger and the self-trigger from the directional sources. Fig.~\ref{fig:dir-source-event-display} (left) shows the \nhit{} distribution for the 20~mm $^{106}$Ru directional source before and after applying the cut on the time difference. Fig.~\ref{fig:dir-source-event-display} (right) shows the \eos{} event display for the sideways pointed directional source and a clear Cherenkov ring is visible.

\subsection{AmBe Source Selection}

A special coincidence tag is developed for the AmBe source, described in Sec.~\ref{sec:ambe-source}. The prompt-event candidates, from the 4.4~MeV $^{12}$C$^{*}$ de-excitation $\gamma$-rays, are selected using an \np{} range of 22 to 40 and with a reconstructed radial position of $r$ < 75~cm (using \snd{}). For each prompt event candidate, delayed events are identified in a subsequent 400~$\mu$s window with $8<$ \np{} $<22$, $r$ < 75~cm, and radial distrance from the prompt event of $\Delta r$ < 1~m. The cut values were informed using the simulation to maintain a high efficiency for selecting coincidences, while removing the vast majority of accidental backgrounds. The low value of the cut on \np{} for the delayed event is acceptable because of the trigger scheme described in \ref{sec:ambe-source}, which considerably lowers the trigger threshold on delayed events. 

If either a prompt-like event or a high-\nhit{} (above 70~\nhit) muon-like event is identified in the delayed event window, that prompt-event candidate is rejected. In situations with multiple delayed-like events in the follower window, all delayed events are accepted. The high rate of low-energy events in \eos{}, due to its location on the surface and the relatively high levels of radioactivity, make it unfeasible to apply a multiplicity cut to remove events with multiple delayed-like events. In a deep-underground, clean detector the event selection would be simplified due to significantly lower backgrounds. 

\begin{figure}[ht!]
    \centering
    \begin{minipage}{0.45\textwidth}
    \includegraphics[width=\linewidth]{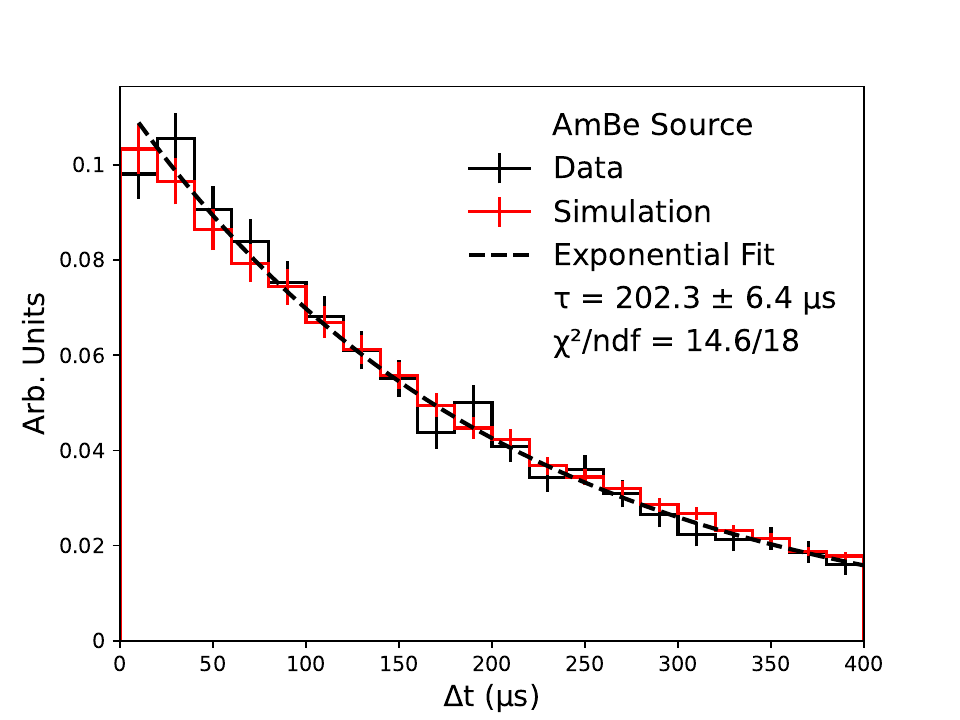}
    \end{minipage}
    \begin{minipage}{0.45\textwidth}
    \includegraphics[width=0.66\linewidth]{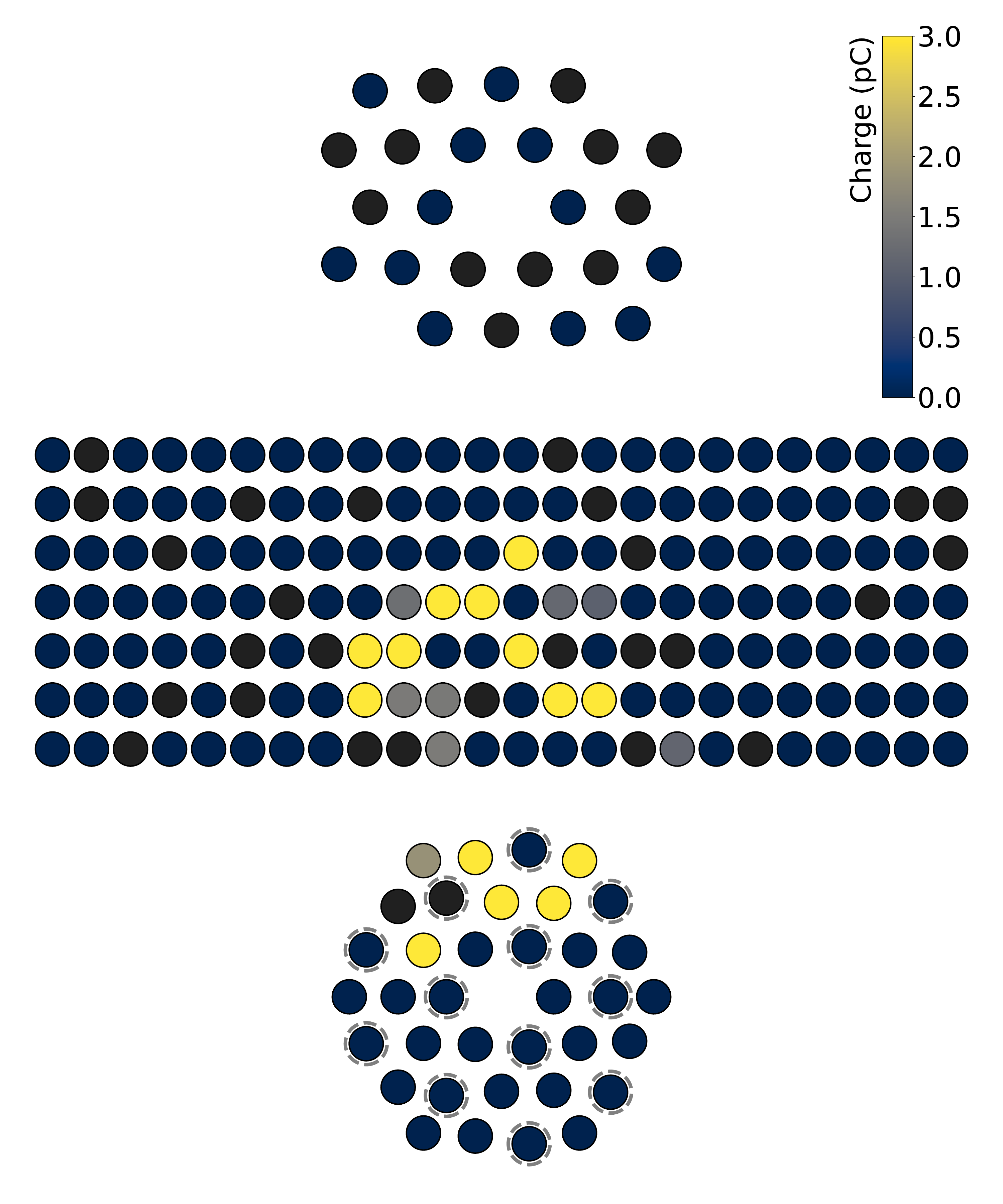}
     \end{minipage}
    \caption{(Left) The time-difference between the prompt and delayed events for a central AmBe run, shown for both data and MC. An exponential fit to the data yields the expected neutron capture time. The uncertainties are statistical only. (Right) Event display for a tagged prompt event from the central AmBe run. The color scale indicates the charge (in pC) of prompt hits and the dichroicon PMTs at the bottom are indicated by the dashed circles.}
    \label{fig:ambe-deltat}
\end{figure}

Fig. \ref{fig:ambe-deltat} (left) shows the time difference ($\Delta t$) between the selected prompt and delayed events for a central AmBe run. Good agreement between data and simulation is observed and a fitted neutron capture time yields $\tau$ = 202.3 $\pm$ 6.4 $\mu$s, consistent with expectations. The 400~$\mu$s cut-off is caused by the size of the coincidence window implemented in the triggering logic (for details, see Sec.~\ref{sec:ambe-source}). This indicates that the coincidence between the prompt 4.4 MeV $\gamma$ and delayed neutron capture, producing a 2.2 MeV $\gamma$ can be selected. Fig.~\ref{fig:ambe-deltat} (right) shows the \eos{} event display for a tagged prompt event from the centrally deployed AmBe source.

\subsection{Michel Electron Selection}

Michel electrons can be efficiently tagged in \textsc{Eos} using the high-energy follower trigger described in Sec.~\ref{sec:cosmics}. Events above 100 \np{} are classified as muon candidates and Michel electron followers are searched for in a subsequent 10~$\mu$s window. The Michel electron \np{} window is selected between 50 and 100 based on MC studies. A multiplicity cut ensures exactly one candidate Michel electron occurs  after the high-energy trigger in a 10~$\mu$s window. The $\Delta t$ after the prompt muon is required to be larger than 1~$\mu$s, which removes reflections and other noise in the trigger system, prevalent primarily for high-energy events. Additionally the requirement that the $\Delta t < 5$~$\mu$s selects a higher purity of Michel electrons.  

\begin{figure}[ht!]
    \centering
    \begin{minipage}{0.45\textwidth}
        \includegraphics[width=\linewidth]{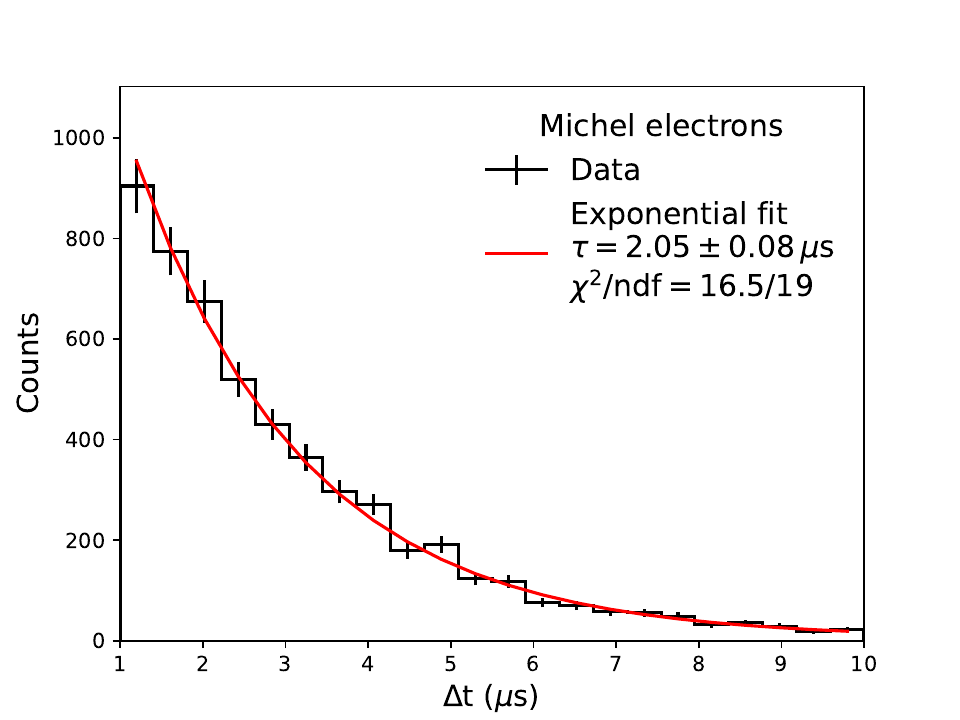}
    \end{minipage}
    \begin{minipage}{0.45\textwidth}
        \includegraphics[width=0.66\linewidth]{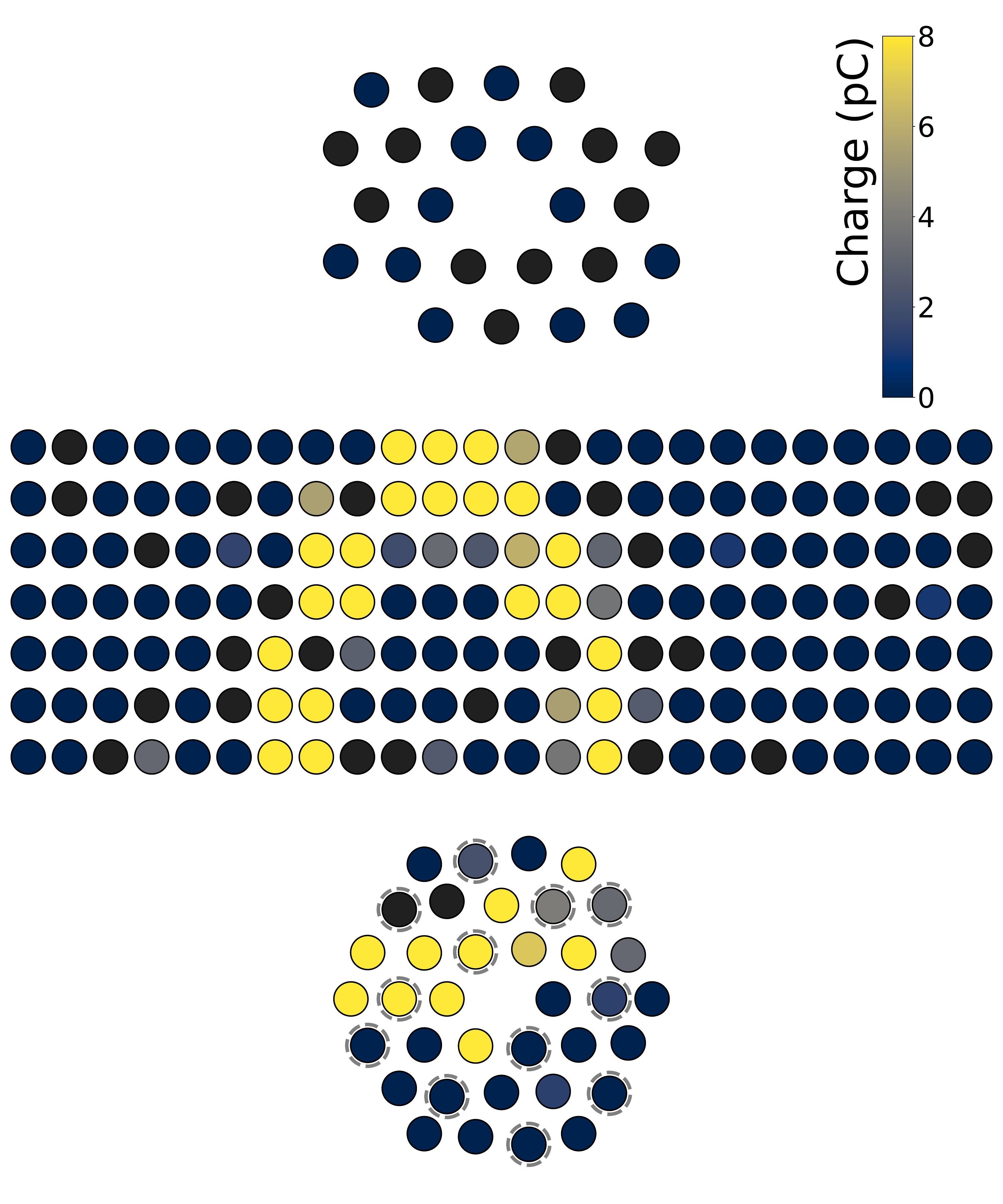}
    \end{minipage}
   \caption{(Left) The time difference between selected cosmic muons and the Michel electron followers, which is fit with an exponential plus a constant background. The uncertainties are statistical only. (Right) Event display of a Michel electron candidate. The color scale indicates the charge (in pC) of prompt hits and the dichroicon PMTs at the bottom are indicated by the dashed circles.} 
   \label{fig:michel-dt}
\end{figure}

Fig.~\ref{fig:michel-dt} (left) shows the $\Delta t$ between the cosmic muons and the Michel electron followers in a timing window of 10~$\mu$s. The timing spectrum is fitted to an exponential decay function plus a constant background. This yields a muon lifetime of $\tau = 2.05 \pm 0.08~\mu$s with $\chi^{2}/\mathrm{dof} = \text{24/19}$. The expected muon lifetime is calculated in Appendix~\ref{sec:appendix-B} to be 2.04 $\pm$ 0.01 $\mu$s, which is in agreement with the fitted result. Fig.~\ref{fig:michel-dt} (right) shows the event display of a Michel electron event detected with the high-energy follower trigger, in which a clear Cherenkov ring is identified.

\subsection{Summary}

Table \ref{tab:cuts} summarizes the analysis selections for each of the calibration sources. Identical selection criteria were applied on both the simulated and measured datasets.

\begin{table}[ht!]
    \centering
    \begin{tabular}{l|c|c|c|c|c|c} \hline \hline 
         Source & \np{} & $|z|$ (cm) & $\rho$ (cm) & $\Delta r$ (m) & $\Delta t$ ($\mu$s) & Bkg. subtract  \\ \hline 
         Laserball & - & - & - & - & - & - \\
         Thorium & $>$ 10 & $<$ 70 & $<$ 70 & - & - & $\checkmark$ \\
         Directional & $>$ 10 & $<$ 70 & $<$ 70 & - & - & - \\
         AmBe Prompt & [23, 39] & $<$ 70 & $<$ 70 & - & - & $\checkmark$  \\
         AmBe Delayed & [9, 21] & $<$ 70 & $<$ 70 & $<$ 1 & $<$ 400 & $\checkmark$ \\
         Michel $e^{-}$ & [51, 99] & $<$ 70 & $<$ 70 & - & $>$ 1 \&  $<$ 5 & - \\ \hline \hline
    \end{tabular}
    \caption{A summary of the various analysis cuts for each of the calibration sources. Note that the directional sources also utilize the coincidence cut described in Sec.~\ref{sec:dir-source-selection}.}
    \label{tab:cuts}
\end{table}

\section{Detector Calibration}\label{sec:detector-calibration}

The calibration of the detector response is critical for tuning the simulation to better understand the \eos{} reconstruction performance in water and to prepare for future scintillator phases. This section focuses on the calibration of the PMT responses using the laserball and thorium sources. The output of these calibration efforts are used as inputs to the \rp{} detector model. In Secs.~\ref{sec:model-validation} and~\ref{sec:results}, comparisons between data and simulation for several important observables are presented after applying these calibrations. 

\subsection{Timing Delays}\label{sec:cable-delays}

The duration between the time when a photon hits the photocathode and the time that is extracted from the digitized waveform varies from PMT to PMT. This variation is caused by different PMT transit times, cable lengths and path lengths along the electronics boards. This variation is calibrated individually for each PMT with laserball data collected using the 515~nm laser. It is necessary to use the 515~nm laser as the PMTs behind the dichroicons only detect enough light to perform this calibration at long-wavelengths (above 450~nm). 

\begin{figure}[ht!]
    \centering
    \includegraphics[width=0.45\textwidth]{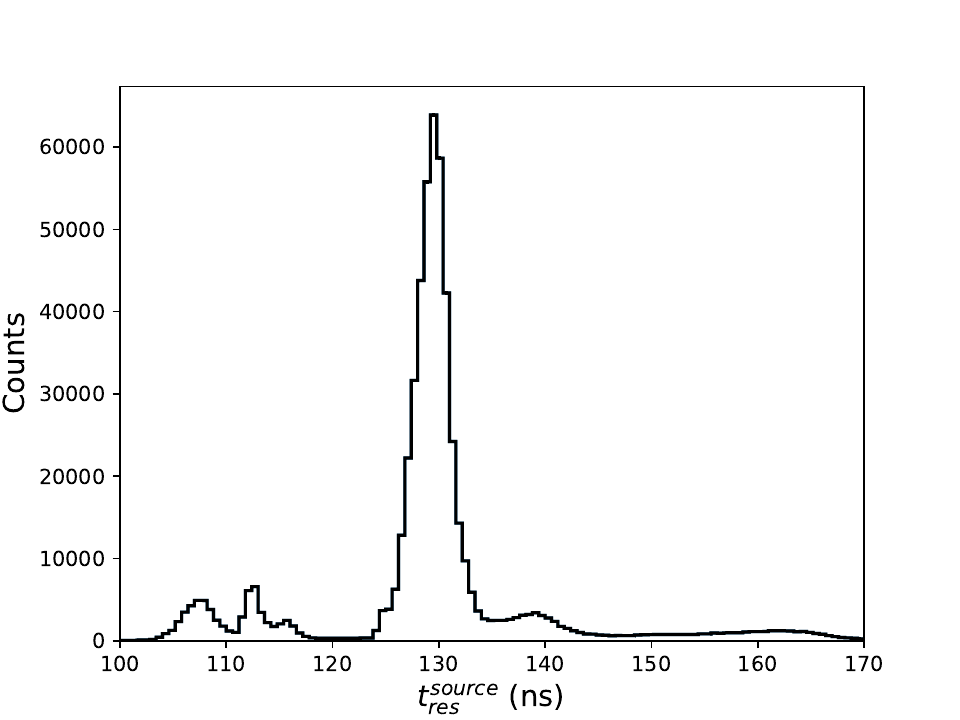}
    \includegraphics[width=0.45\textwidth]{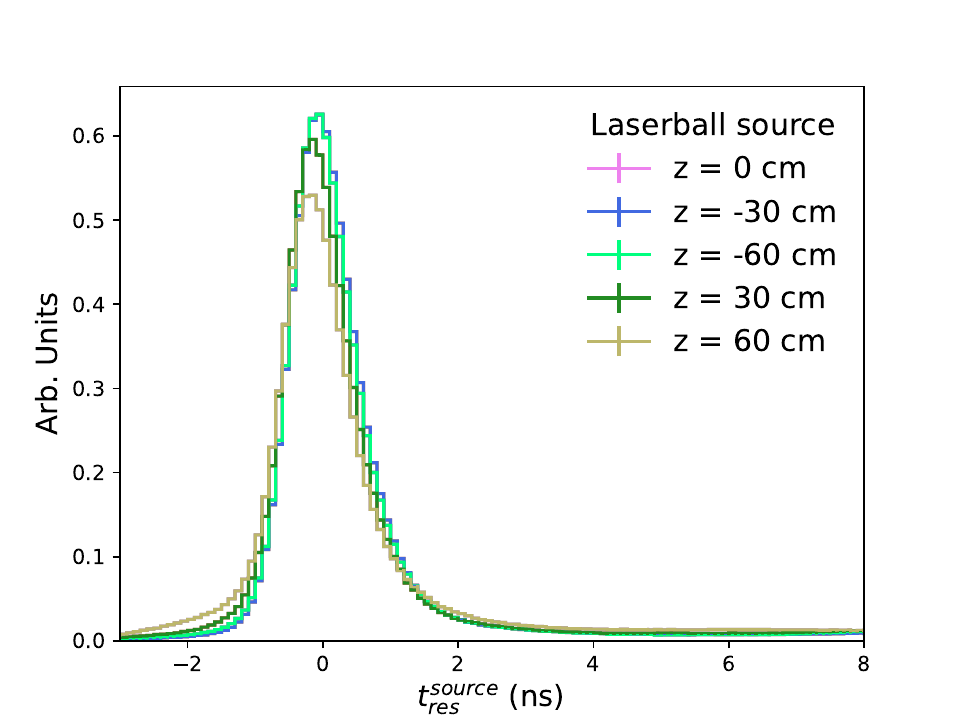}
    \caption{(Left) The laserball time-residuals for all PMTs in \eos{} prior to applying the timing delay calibration. The multiple peaks are in part due to the different cable lengths for the different PMT types. (Right) The laserball time-residuals for all PMTs in \eos{} after applying the timing delay calibration. The broader distribution at larger $z$-positions is due to the proximity to the R11780 PMTs at the top of the detector, which have significantly broader transit-time spreads than the R14688-100 PMTs.}
    \label{fig:laser-timing-delays}
\end{figure}

The time for each PMT signal is extracted using the method described in Sec.~\ref{sec:data-processing}, accounting for the known position of the laserball source to subtract the photon time-of-flight. This time-residual distribution, combined across all PMTs, is shown in Fig.~\ref{fig:laser-timing-delays} (left).

To extract the per-PMT timing delay, the time-residual is calculated for each individual PMT over all the events in the laserball run. A Gaussian is fitted $\pm$0.5~ns around the peak of each PMT's time-residual distribution. The mean of the fit is selected as the timing delay for that PMT. The validity of the timing delay calibration is tested by applying the calibration to runs varying the $z$-position of the laserball between 60~cm and -60~cm in steps of 10~cm. The result of applying this correction is demonstrated by plotting the time-residuals for all PMTs and for several laserball position in Fig.~\ref{fig:laser-timing-delays} (right). After applying the delay calibration for a central laserball run, a Gaussian is fitted $\pm$~0.5~ns around the peak of the time-residuals. The $\sigma$ of the fit to the central data is 450~ps. This resolution includes broadening effects from the laser jitter and scattering inside the laserball. 

These timing offsets are applied directly to correct the data, and the simulation does not include any channel to channel differences in the delays. The time-residuals include the per-PMT cable delay corrections for all analyses in this manuscript. 

\subsection{PMT Pulse Shape}\label{sec:pulse-shapes}

The PMT pulse shape is modeled as a Gaussian function in simulation. Using a Gaussian function (as opposed to a lognormal) allows us to more easily calibrate for the variation in pulse shape, 
and the skewness of observed PMT pulses is small enough that a Gaussian pulse model has minimal impact on fitted values such as charge and time.

A spectrum of fitted pulse widths is extracted from laserball data by fitting Gaussian functions and extracting the standard deviation, $\sigma$, from individual pulses. This $\sigma$ spectrum is used in simulation to choose the shape of the simulated pulse on a hit-by-hit basis. The spectrum is adjusted to account for the broadening that occurs due to the process of re-fitting a function on a sampled pulse. It is then shifted by a multiplicative scaling factor $S$ for each PMT. 

It should be emphasized that this method is different than that used to extract the PMT hit-times (fitting a Gaussian rather than a lognormal distribution to the PMT pulse); however, the Gaussian model was simpler to implement and resulted in a negligible difference in the final pulse-shape calibration results (when comparing to the results from the lognormal pulse shapes).

\subsection{PMT Charge}\label{sec:pmt-charge}

The PMT gain is set to $1 \times 10^{7}$ based on Hamamatsu recommendations for optimal performance. The high voltage delivered to each PMT to achieve this gain is determined from bench-top measurements of each PMT \cite{Kaptanoglu:2023ayz}. The PMT to PMT variation in the gain is calibrated in-situ using laserball data. Two different charge distributions are studied: the charge integrated in a window [-10, 20] samples around the peak of the waveform and the fitted charge, $Q$, from Equation~(\ref{eq:lognormal}). For the purpose of the water phase, we focus on the latter charge and show here the results for the fitted charge. For future phases where MPE contributions are more significant, the integrated charge is expected to be more useful, and is calibrated similarly. As previously described, \nhit{} are counted as pulses that cross a threshold of $-5$~mV, so it is important to accurately model the PMT charge and pulse-heights. 

\begin{figure}[ht!]
    \centering
    \includegraphics[width=0.45\textwidth]{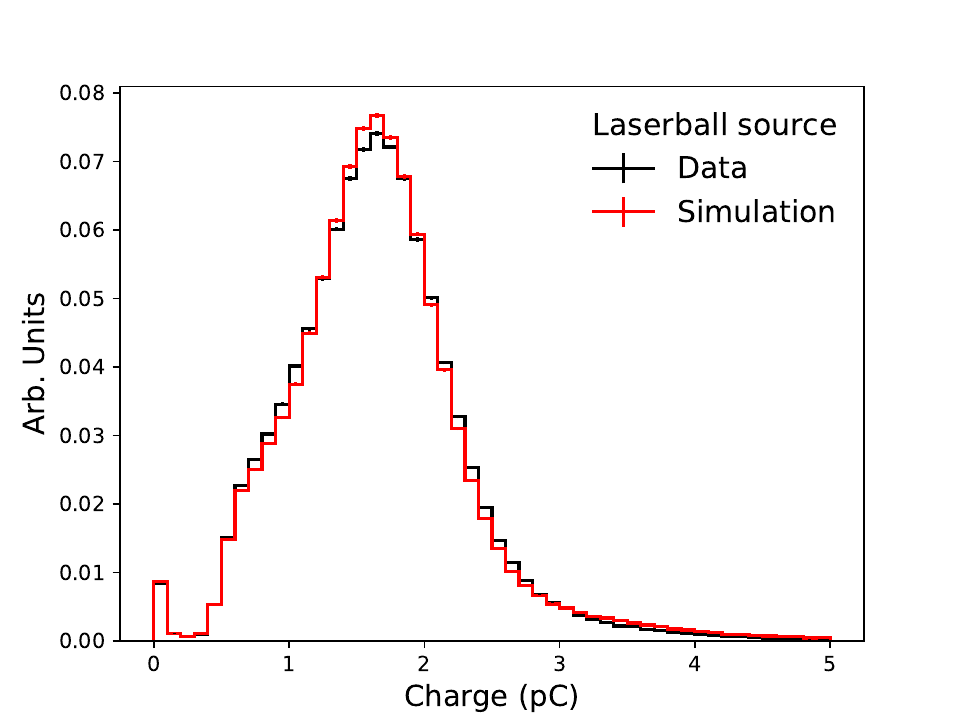}
    \includegraphics[width=0.45\textwidth]{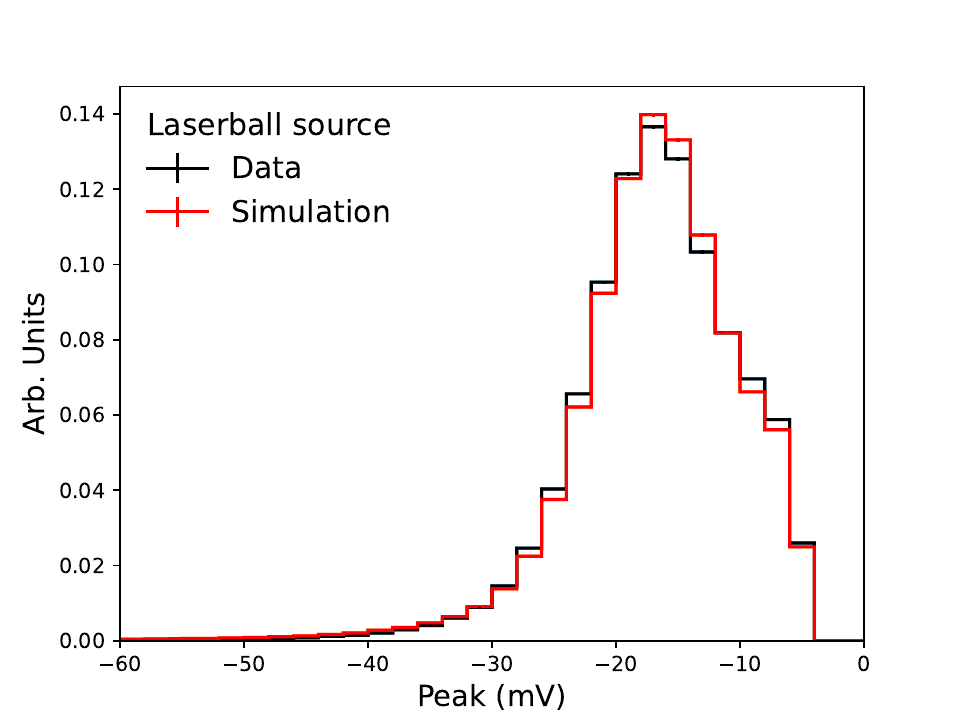}
    \caption{(Left) The data to simulation comparison for a central 408~nm laserball run after applying the charge calibration, for all online PMTs. (Right) The same for the peak of the waveform.}
    \label{fig:charge-spe}
\end{figure}

In the simulation the initial value of $Q$ is selected by drawing from a histogram that is generated from bench-top SPE data collected for each PMT in the \eos{} detector~\cite{Kaptanoglu:2023ayz}. A multiplicative scaling factor is introduced that shifts the value of $Q$ on a hit-by-hit basis. This effectively shifts the $Q$ distribution higher or lower in average charge. The value of this factor is adjusted using central laserball data. A histogram of $Q$ is fitted with a Gaussian around the peak for both the data and simulation. The value of $S$ is selected to shift the fitted mean value in the simulation to match the data. The distribution of $Q$ for all PMTs for a central 408~nm laserball run, compared between data and simulation, is shown in Fig.~\ref{fig:charge-spe} (left). After applying this calibration, the PMT pulse-heights are in good agreement between data and simulation, shown in Fig.~\ref{fig:charge-spe} (right).

\subsection{Dark-rate}

The PMT dark-rates are calibrated by looking in a pre-trigger window in the laserball data. This early window is guaranteed to be prior to the laser firing, so contains only hits from the PMT dark noise. The digitizer windows are intentionally configured such that at least 80~ns of the window is prior to the trigger. The laserball source is selected for the dark-rate calibration as the data is collected every week, allowing a regular calibration of the dark-rates.

A 60~ns window is selected for each PMT type (slightly different cable delays require slight differences in the windows) such that the window contains only the dark-noise. The total number of PMT hits in that window for each PMT over all trigger events is calculated and converted to a rate. The R11780, R14688-100, and R7081 PMTs have a measured average dark-rate of 8~kHz, 14~kHz and 4~kHz respectively. These fixed rates are input into the simulation, where the dark-rate is added as a uniform, random background for each PMT. Laserball runs collected throughout the dataset show that the dark-rate was stable over time. 

\subsection{Trigger Efficiency}\label{sec:trigger-eff}

This efficiency of the trigger system as a function of the number of PMT hits can be studied using the laserball. The laserball operates by using a fixed rate, pulsed trigger that gets propagated to the CAEN digitizers regardless of the number of PMTs that detected light. The analog global trigger sum is digitized on an unused CAEN channel, and the pulse height of the sum is measured in an offline analysis. A map between the number of PMTs that detected light and the trigger sum is generated. This map can be used to measure the efficiency of the trigger, given a specified trigger threshold, as a function of the \nhit{}. It is assumed in the analysis that 100\% of the pulses that cross the trigger threshold would have triggered the detector.
 
\begin{figure}[ht!]
    \centering
    \includegraphics[width=0.6\textwidth]{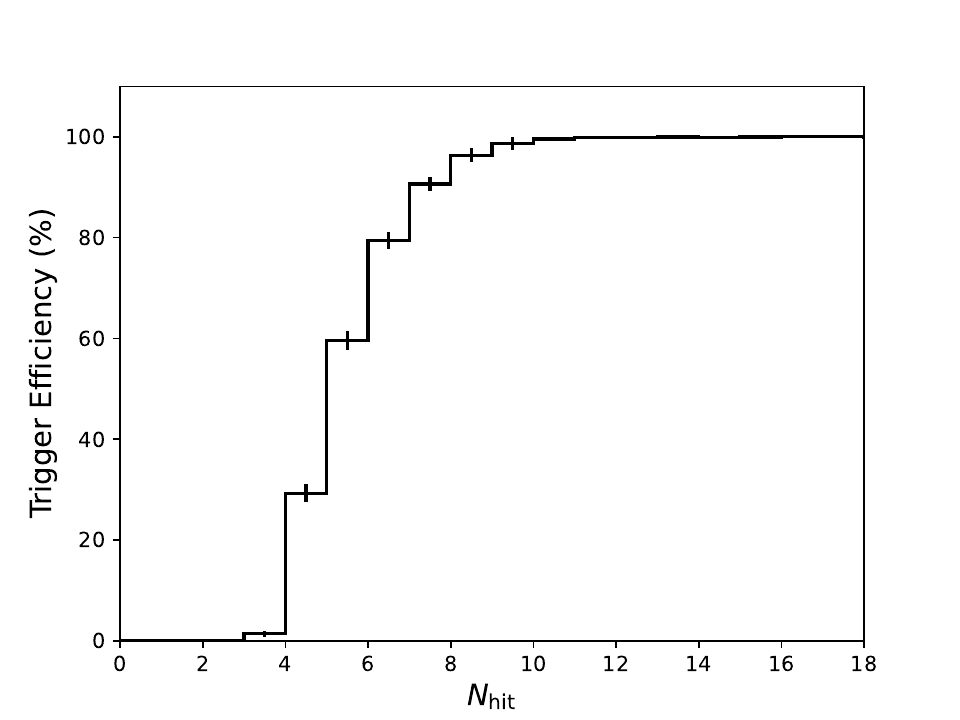}
    \caption{The trigger efficiency as a function of the \nhit{} for an assumed global trigger threshold of 100~mV (typical for a thorium source run).}
    \label{fig:trigger-efficiency}
\end{figure}

 Fig.~\ref{fig:trigger-efficiency} shows the trigger efficiency as a function of \nhit{} for a threshold of 100~mV. The efficiency is calculated by finding the peak of the digitized global trigger sum, which is compared to a 100~mV threshold. If it falls below the threshold, it is assumed that no global trigger would have been generated, and vice versa. This assumption is based on bench-top testing and validation of the CASB trigger performance. Notably, for a 100~mV threshold, typical for thorium source runs, the trigger efficiency reaches 100\% above 10~\nhit.

\subsection{PMT Efficiency}\label{sec:pmt-efficiency}

The wavelength-dependent PMT quantum efficiencies for each PMT type are extracted directly from the Hamamatsu datasheets~\cite{ham_datasheet_r14688}. The collection efficiency -- the efficiency for which a PE created at the photocathode reaches the last dynode and produces a signal -- is not measured ex-situ for any of the PMTs. A scale correction to the overall detection efficiency, applied for each PMT type, is intended to calibrate the unknown collection efficiencies. These corrections can also account for other mis-modeling that may contribute to a disagreement in the total amount of light collected by the PMTs, between the data and the simulation. 

In order to calibrate these efficiency scales, a central deployment of the thorium source is used. The total number of photons emitted by the laserball is not known and is thus not appropriate for this calibration; however, the thorium source produces predominantly 2.6-MeV $\gamma$-rays that Compton scatter in the water and create Cherenkov light with a well-understood intensity. 

\begin{figure}[ht!]
    \centering
    \includegraphics[width=0.45\linewidth]{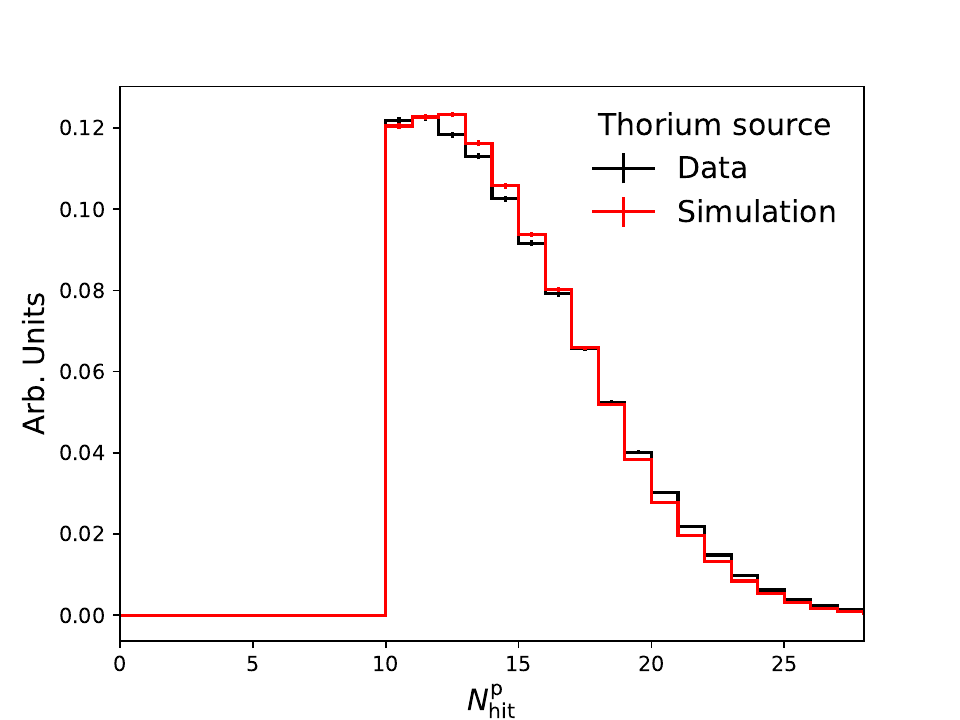}
    \includegraphics[width=0.45\linewidth]{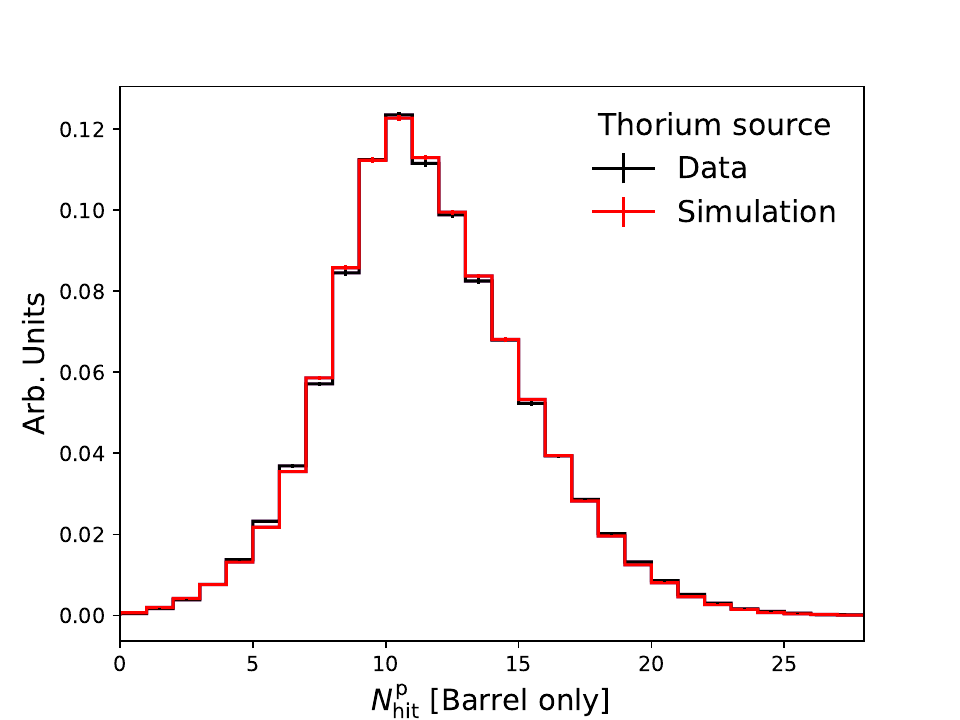}
    \caption{The \np{} distributions for a central thorium run, after applying the PMT efficiency calibrations. The left figure shows the data and simulation comparison for all PMTs and the right figure shows the barrel PMTs only. The uncertainties are statistical only.}
    \label{fig:thorium-nhit}
\end{figure}

For each PMT group, the resulting \np{} histogram from the simulation is linearly interpolated into finer bins, and a least-squares fit is performed with the data to determine the scale factor that brings the simulation in to closest agreement.  The fitted scale factors for each PMT group are then input into the simulation and the process is repeated until sufficient convergence is achieved.  Three efficiency correction factors are obtained: 0.78 for the 168 barrel PMTs (R11468), 0.90 for the 24 top PMTs (R11780), and 0.88 for the 49 bottom PMTs (all bottom PMTs are grouped together due to their low \nhit{} and the more complex detector structure at the bottom). Fig.~\ref{fig:thorium-nhit} shows the resulting agreement between data and MC for all PMTs (left) and for the barrel PMTs (right) after applying the efficiency correction calibration.

\subsection{Deployment System Tilt}\label{sec:tilt}

As described in Sec.~\ref{sec:detector}, a small tilt of the calibration tower relative to the IV is possible, based on the detector construction, and can be measured using the laserball source deployed at varying $z$ positions. By evaluating the difference in the time-of-flights for opposite sides of the detector, it is determined that the laserball is traveling in the ($+x, +y$) direction as it moves up the detector. The size of the tilt is measured to have a zenith angle $\theta = 0.65 \pm 0.35$ degrees. The uncertainty is evaluated by estimating the size of the tilt for several different laserball deployment scans. Due to the large uncertainty and lack of precision as to the orientation of the tilt in the azimuthal angle $\phi$, this measurement is later used to estimate a systematic on the source position, and all simulations are performed assuming zero tilt.  

\section{Systematics}\label{sec:systematics}

To evaluate detector systematics associated with the stability of detector over time, the data is compared across several deployments of the same source, collected months apart. The data for a central thorium source run are compared and a maximum of 5\% bin-to-bin variations in the \np{} and time-residuals are identified. Other sources show similar sized maximum variations. This effect could be due to varying detector properties, such as the PMT gain and dark-rate. To account for this, all figures comparing histograms of data and simulation in Sec.~\ref{sec:model-validation} and Sec.~\ref{sec:results} include a 5\% systematic uncertainty added to each bin. 

Differences in the reconstruction results over time are checked similarly. Variations in detector performance lead to a maximum difference in the \snd{} vertex reconstruction of the thorium source $z$ position scan of 1.4~cm (mean position) and 0.35~cm (width) for the Cartesian coordinates. In presenting the reconstruction bias and width results with the thorium source, described in Sec.~\ref{sec:vertex-recon}, these maximum differences in performance are conservatively included as systematic uncertainties. The systematic is evaluated separately for each algorithm and is applied across all deployed source positions.

In addition to detector changes, there is an uncertainty in the knowledge of the true source position. This is particularly important to evaluate when comparing the reconstructed position distributions between data and simulation. As discussed in Sec.~\ref{sec:detector} the precision of the measurements of the various detector components, such as the height of the IV, leads to an uncertainty of 1.5~cm on $z$. Additionally, the measured tilt of the calibration deployment system, described in Sec.~\ref{sec:tilt}, is measured to have $\theta < 1^{\circ}$. Conservatively, the maximum $\rho$ deviation is 0.9~cm at a tilt of $1^{\circ}$ (neglibile impact on $z$) and this systematic uncertainty is included in the vertex bias results for every deployed position (it has a negligible impact on the resolution). This systematic on the reconstructed position is summed in quadrature with the uncertainty assigned from the detector stability analysis and treated separately for each Cartesian coordinate and for each reconstruction algorithm.

In the direction reconstruction results presented in Sec.~\ref{sec:dir-recon}, evaluation of the reconstruction is performed for when the source pointed directly downward and when $\theta = 90^{\circ}$. In the latter case, only a rough estimate of the $\phi$ direction was possible at the time of deploying the source, accurate to approximately $\pm 5^{\circ}$. To account for this, the simulations are run scanning the value of $\phi$ in that range. The $\phi$ that produces the best data to simulation agreement is selected and a systematic uncertainty is generated based on the simulation with maximum difference in $\alpha_{1\sigma}$ from that value. In all cases, an additional systematic is estimated by adjusting the fit function to a single exponential, removing the Gaussian convolution. The difference between the measured $\alpha_{1\sigma}$ for the two models yields a systematic on the selected model.

In the following sections, systematic uncertainties are included on the simulation model when comparing data to simulation. Calculations of \chindf between the data and simulation include the systematic uncertainties, added in quadrature with the statistical uncertainty.

\section{Model Validation}\label{sec:model-validation}

The \eos{} calibration methods utilized the laserball source, with positions distributed across the central axis, to generate PMT cable delay, pulse-shape and charge calibration constants for the simulation. Afterward, a central thorium run is used to tune PMT efficiency corrections that are used to improve the \np{} data and simulation agreement. With these calibrated parameters, we can perform checks that the \rp{} model is predictive by testing it using the off-center thorium source data as well as data with the other calibration sources. This section focuses on validating the time-residual and \np{} agreement between data and simulation, as these are the primary inputs into the reconstruction frameworks. While not as integral for the \eos{} reconstruction algorithms, the data and simulation agreement for various dichroicon data is also presented. The modeling of the dichroicons is of particular interest for future large-scale hybrid detector that may utilize these instruments for Cherenkov and scintillation separation.   

\subsection{Time-Residuals}

\begin{figure}[ht!]
    \centering
    \includegraphics[width=0.45\linewidth]{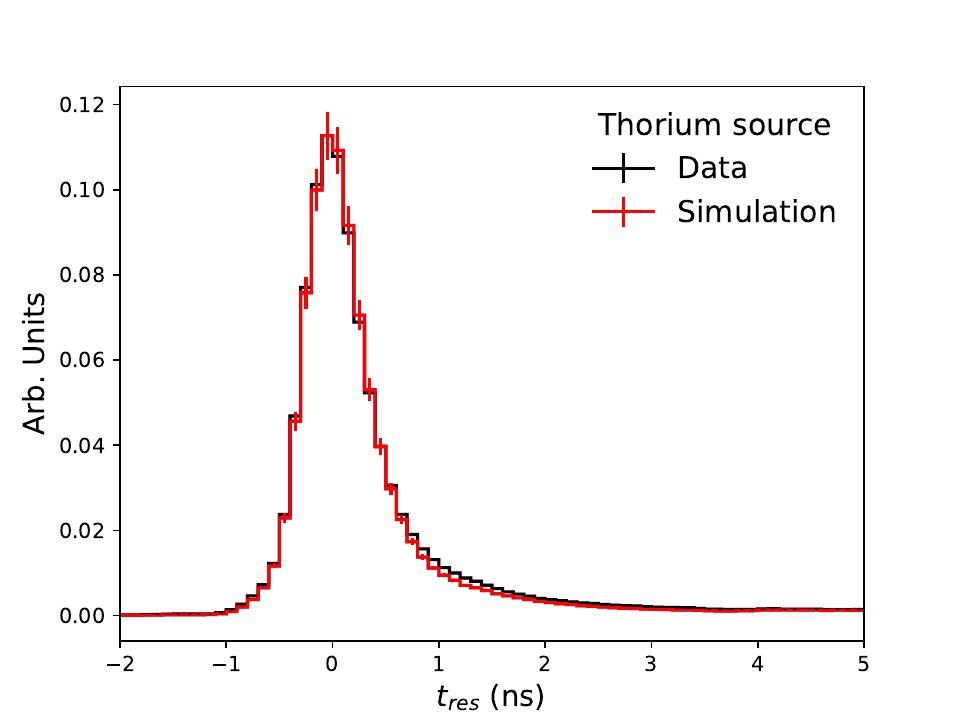}
    \includegraphics[width=0.45\linewidth]{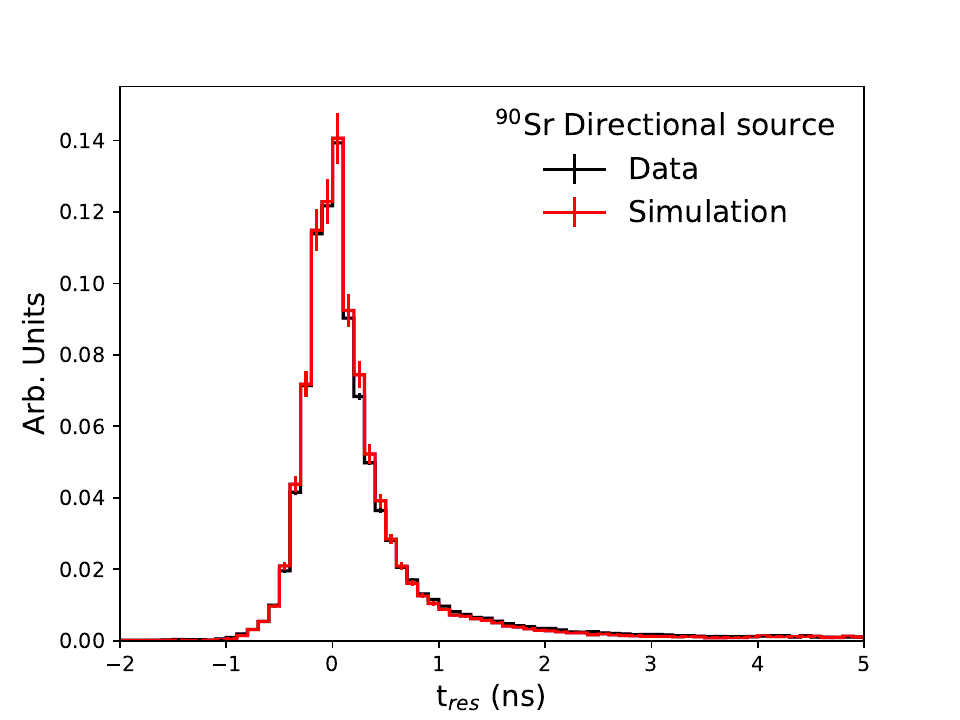}
    \caption{(Left) The time-residuals for the central thorium source data and MC, using \snd{} to reconstruct the event vertex. (Right) The same for the centrally deployed, downward-pointing $^{90}$Sr directional source.}
    \label{fig:time-residual-thorium}
\end{figure}

The time-residuals, described in Sec.~\ref{sec:vertex-recon}, are a critical input into the reconstruction frameworks. Fig.~\ref{fig:time-residual-thorium} shows a histogram of $t_{res}$ for the centrally deployed thorium source (left) and a centrally deployed, downward-pointing $^{90}$Sr directional source (right) for both data and simulation. In this figure, the time-residuals are calculated using the reconstructed vertex from \snd{} (using the output from the other fit methods produces very similar results). The width of the residuals in the thorium and directional source data provide useful benchmarks. From a Gaussian fit to the peak, between -0.5 and 0.5~ns, widths ($\sigma$) of 278 $\pm$ 2~ps and 276 $\pm$ 2~ps are found in the thorium data and simulation, respectively (fit uncertainties only). Similarly, values of 247 $\pm$ 2~ps and 254 $\pm$ 4~ps are found in the directional source data and simulation. As the time-residuals around the prompt peak are the primary input for several of the likelihood-based reconstruction frameworks, the good agreement between data and simulation indicated by these studies is an important check of the modeling.

\subsection{\nhit}

\begin{figure}[ht!]
    \centering
    \includegraphics[width=0.45\textwidth]{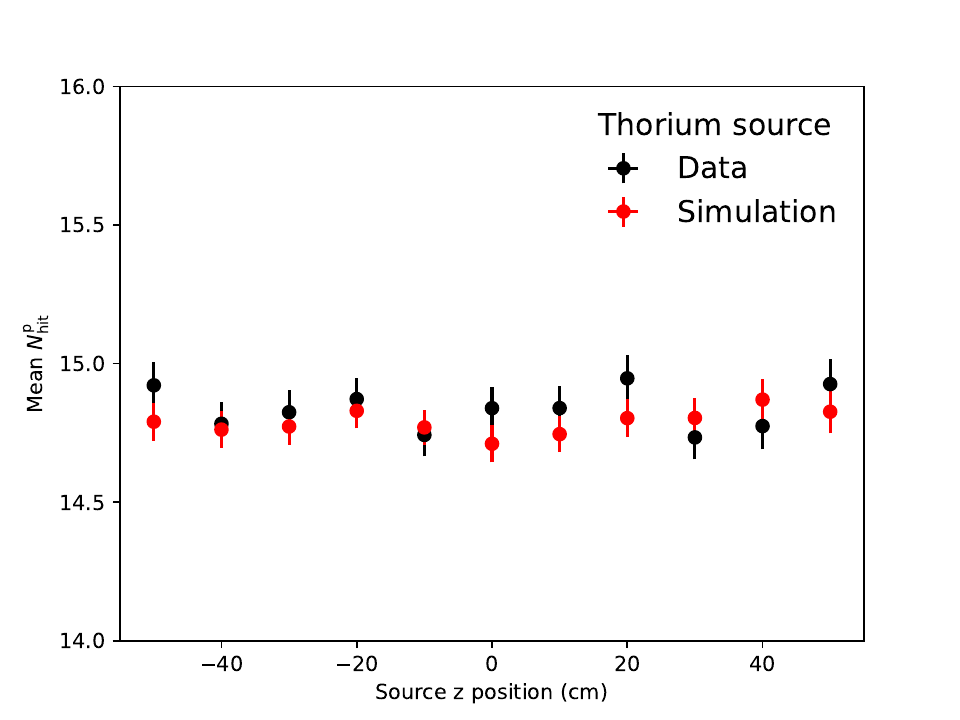}
    \includegraphics[width=0.45\textwidth]{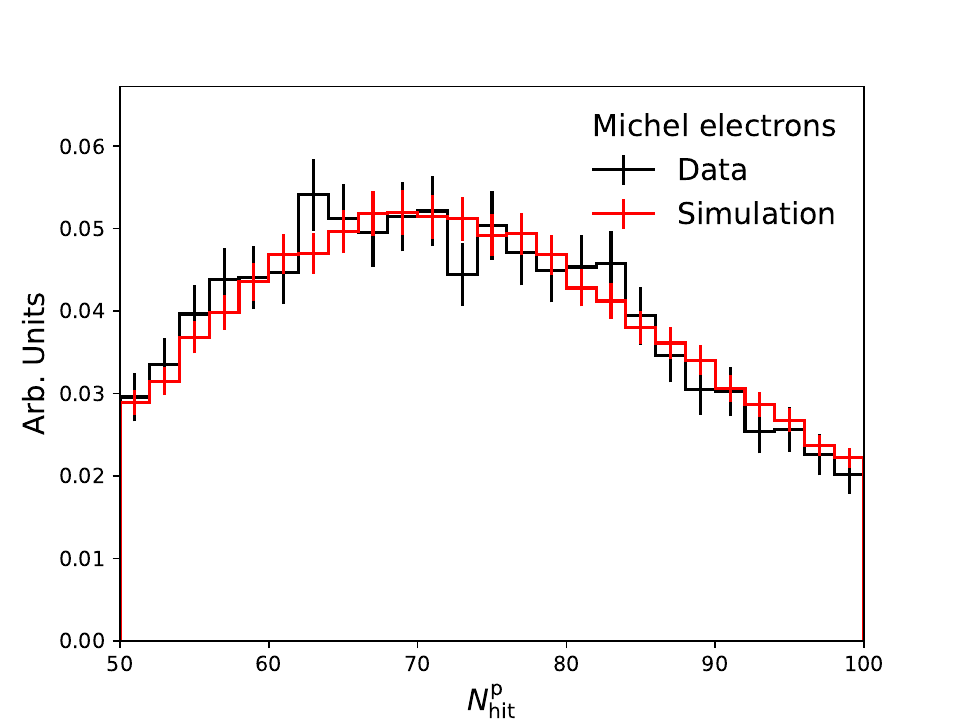}
    \includegraphics[width=0.45\textwidth]{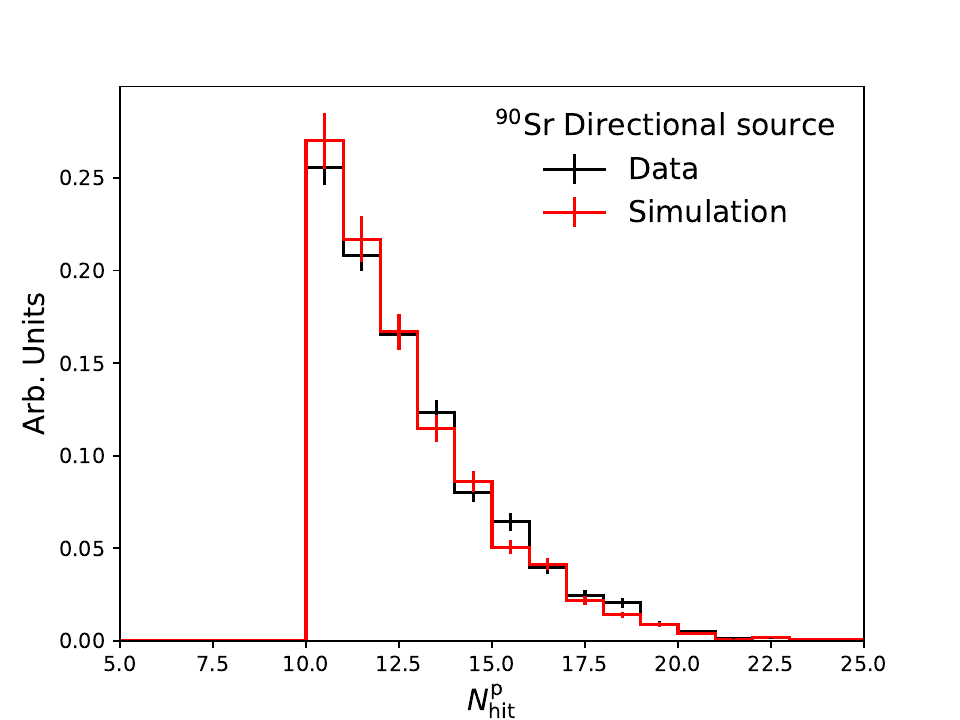}
    \includegraphics[width=0.45\textwidth]{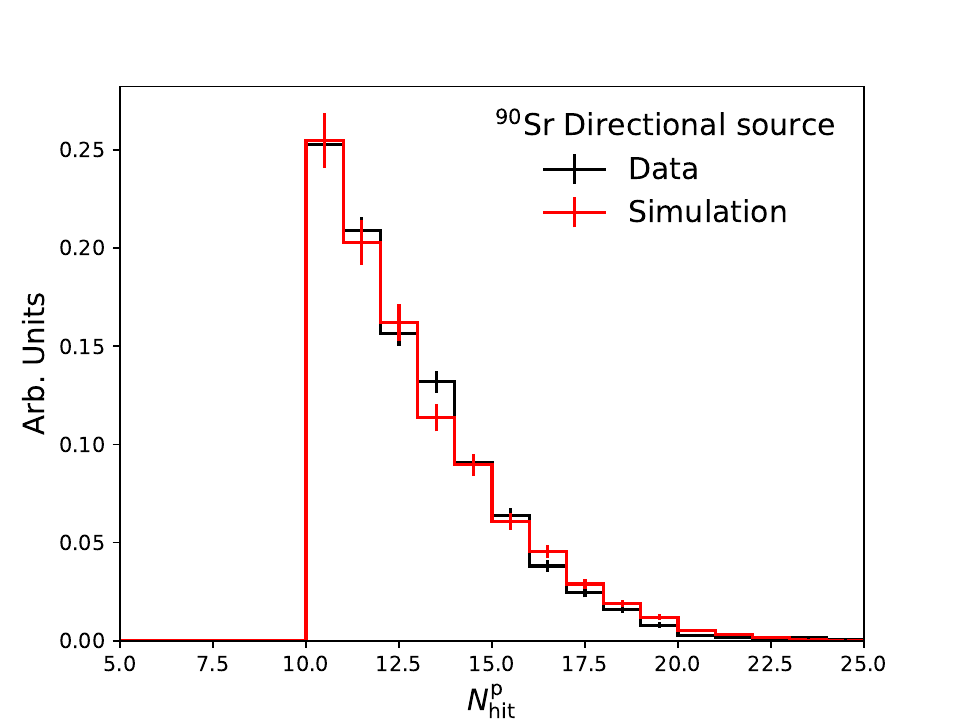}
    \includegraphics[width=0.45\textwidth]{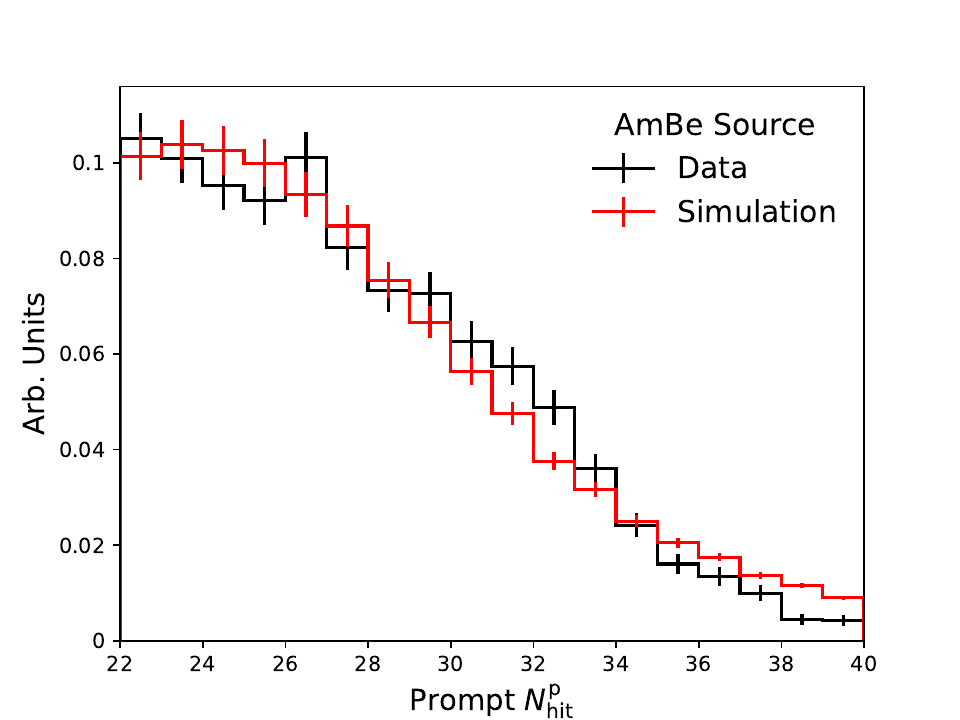}
    \includegraphics[width=0.45\textwidth]{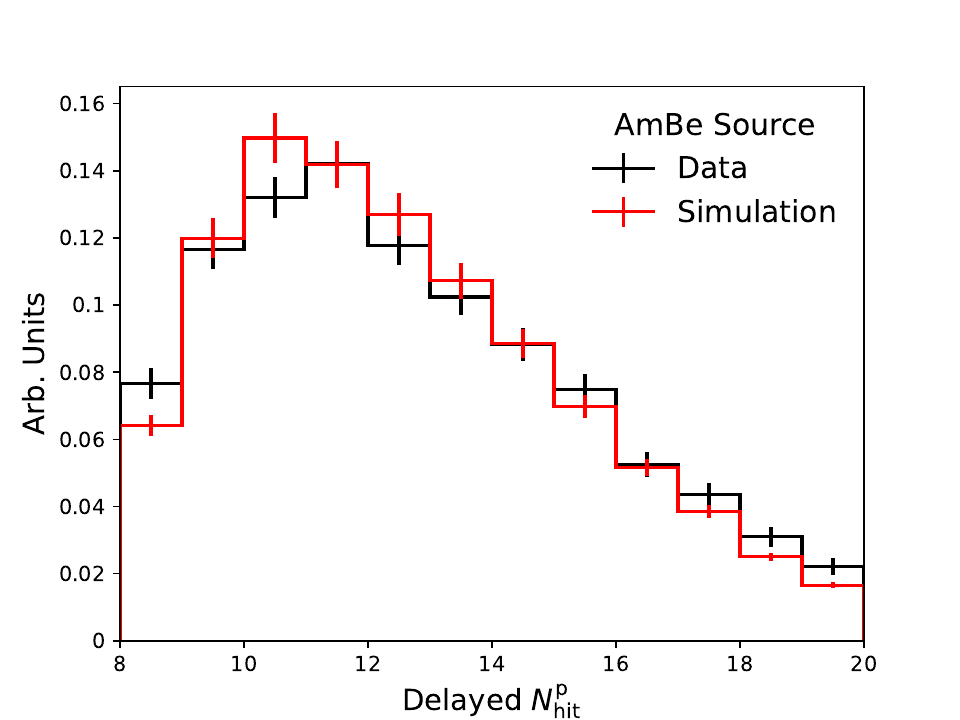}
    \caption{(Top left) The mean \np{} as a function of the source $z$ position for the thorium data and simulation. The uncertainties are taken as the standard error on the mean. (Top right) The \np{} for tagged Michel electrons compared for data and simulation. (Center) The \np{} distributions for a central 30~mm $^{90}$Sr directional run pointed (left) at 90 degrees (toward the barrel PMTs) and (right) directly downward. (Bottom) The prompt (left) and delayed (right) event \np{} distributions from a central AmBe source run.}
    \label{fig:nhit-summary}
\end{figure}

As described in Sec.~\ref{sec:pmt-efficiency}, a central thorium run was used to tune an efficiency correction factor that improves the data and simulation agreement in \np. To validate the data and simulation agreement as a function of event position, we used the data from the thorium scan across the detector, in which data were collected in steps of 10~cm. Fig.~\ref{fig:nhit-summary} shows the mean \np{} value in the data and simulation, demonstrating agreement across a large range of event positions. This further verifies that the data and simulation agreement extends across a wide range of $z$-positions and demonstrates the validity of the optical modeling and efficiency tuning. 

As further verification of the calibration, the data and MC agreement for \np{} is checked for several other sources. This is shown for the selected Michel electrons (top right in Fig. \ref{fig:nhit-summary}), the centrally deployed 30~mm $^{90}$Sr directional source, pointed both downward and sideways (central row of Fig.~\ref{fig:nhit-summary}), and for the tagged prompt and delayed events from a centrally deployed AmBe source (bottom row in \ref{fig:nhit-summary}). For the Michel electron comparison, the \chindf = 11.6/25. For the downward and sideways pointing directional sources, values of \chindf = 16.8/17 and \chindf = 25.5/17 are found respectively. For the AmBe prompt and delayed events, values of \chindf = 81.8/17 and \chindf = 20.7/11 are found, respectively. In general, the data and simulation agreement is very good across all sources, with the poorest agreement identified in the AmBe prompt events, due to a slight difference in the shape of the \np{} distribution.

\subsection{Dichroicon Response}\label{sec:dichroicon-response}

In-situ measurements of the dichroicons were performed using the laserball source and compared to simulation. By varying the laser wavelength and deployment height, the dataset effectively scans over the dichroicons' response both as a function of the photon wavelength and incidence angle. The collected data are compared with laserball simulations performed using \rp{}, as described in Sec.~\ref{sec:detector-modeling}. Spectrophotometry measurements from Ref.~\cite{Bacon:2025efk} are used to model the dichroic filters in the simulation. The Monte Carlo dataset is generated with the laserball positioned at all heights where data have been taken in the detector, and with identical run conditions as data collection. Rather than simulating monoenergetic photons at the specified laser wavelengths, a full wavelength range between 375~nm to 550~nm is simulated (in steps of 1~nm) to account for the finite wavelength spreads of the lasers. To account for laser intensity variations, occupancies of dichroicon PMTs are normalized by the average occupancy of the bottom non-dichroic R14688-100 PMTs in the same run. This is referred to as the ``normalized $N^{\mathrm{p}}_\mathrm{hit}$''.

\begin{figure}[ht!]
    \centering
    \includegraphics[width=0.45\textwidth]{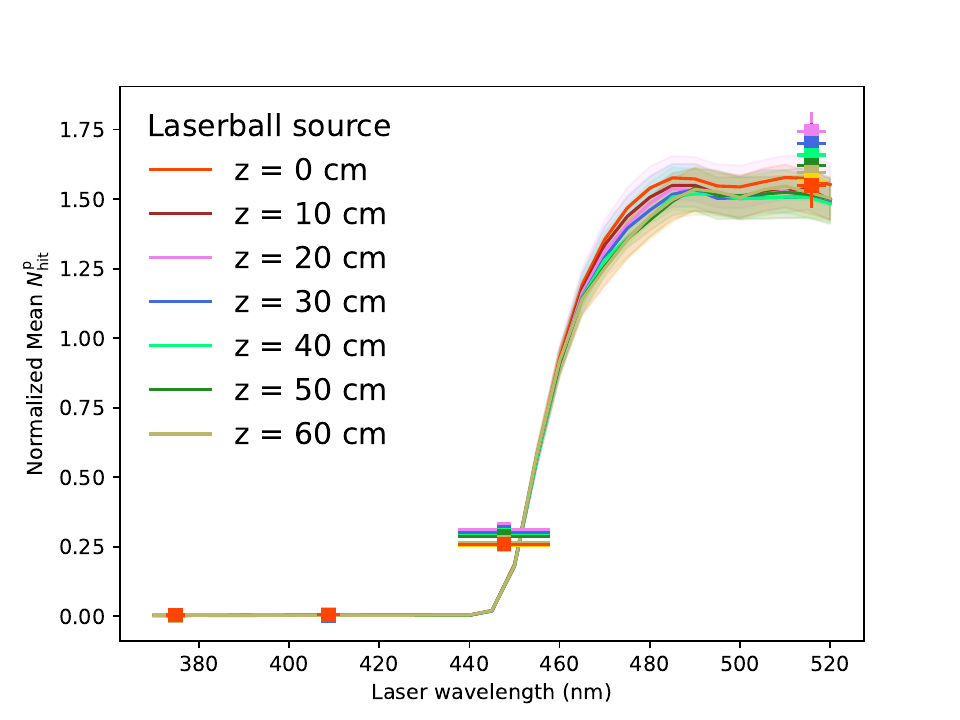}
   \includegraphics[width=0.45\textwidth]{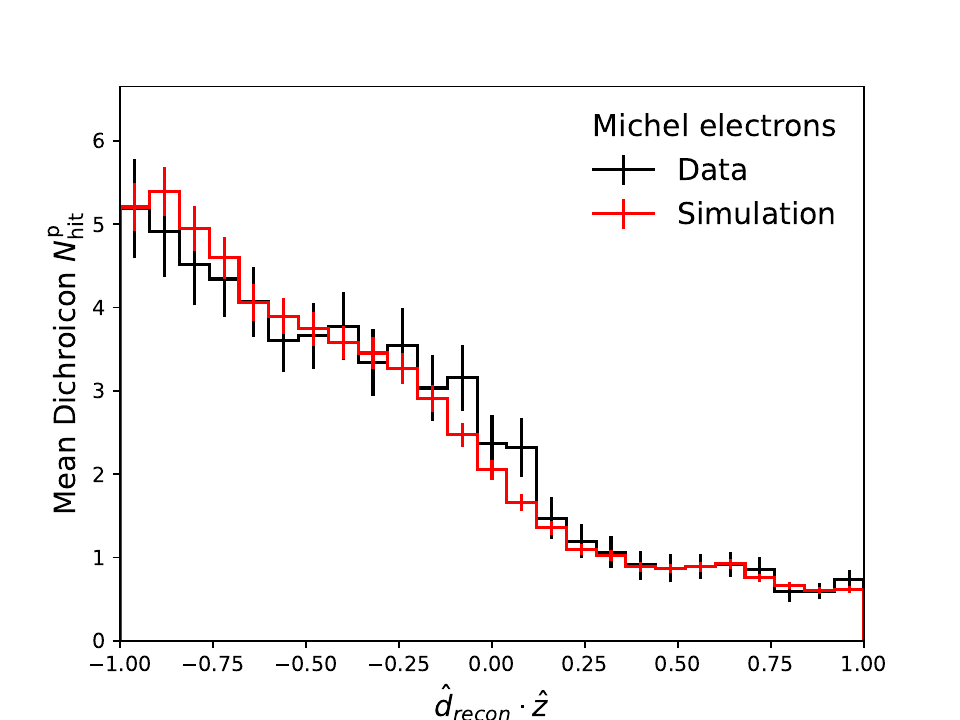}

    \caption{(Left) Data and simulation comparison for the occupancy of dichroicon PMTs at different wavelengths and deployment heights. The occupancy is normalized by the average occupancy of 8 inch PMTs without dichroicons at the bottom PMT array for each run. The horizontal error bars are set based on the width ($\sigma$) of the laser emission distribution (which is described well as a Gaussian with a mean at the indicated wavelength). (Right) The data and simulation comparison of the mean dichroicons \np{} as a function of the reconstructed $z$ direction of the tagged Michel electrons.}
    \label{fig:dichroicon}
\end{figure}

Fig.~\ref{fig:dichroicon} shows the normalized $N^{\mathrm{p}}_\mathrm{hit}$ of the dichroicon PMTs in both data and MC for laserball deployments in the upper half of the detector. Below the cut-on wavelength of 450~nm, the absorbing long-pass filter that covers the PMT absorbs the photons, resulting in effectively zero hits. This is modeled well in the simulation. Around the cut-on wavelength, the optics of the dichroic filters are complicated, with transmission and reflection that depend strongly on the precise wavelength and incidence angle. The laser at 442~nm probes this transition region and shows good agreement between the data and simulation. This laser has the broadest wavelength distribution, which is indicated by the horizontal uncertainty. Above the cut-on wavelength, the absorbing filter becomes transparent, and the dichroic filters become reflective, allowing the dichroicon to act as a concentrator, increasing the effective coverage area of the PMTs. This results in these PMTs seeing more hits than the PMTs without dichroicons, as indicated by normalized occupancies above unity. At 515~nm the simulation slightly under-predicts the data at the center of the detector, and the agreement between data and simulation improves at higher $z$. A \chindf = 9.8/7 is calculated by summing over the seven deployment positions with the 515~nm laser. The agreement between the data and simulation gets worse at source positions below the center and this difference is understood to be caused by the dependence of the filters on incidence angle. There is ongoing work to better understand and further calibrate the angular response of the dichroicon. It should be noted that in a large-scale detector, the distance between the events of interest and the PMTs is typically several meters, where the agreement between the data and MC for the dichroicons is best.

Additionally, the Michel electrons can be used to understand the dichroicon response to Cherenkov light, which spans the full wavelength regime probed by the lasers. The total occupancy of the PMTs with dichroicons in front of them is plotted as a function of the \mimir{} reconstructed z-direction ($\hat{d}_{\text{recon}} \cdot \hat{z}$) of the Michel electron   in Fig.~\ref{fig:dichroicon} (right). A value of \chindf = 11.4/24 is calculated when comparing the data and simulation. As the Michel electrons point more directly downward, the average \np{} of the dichroicons increases. 

\section{Reconstruction Results}\label{sec:results}

A major goal of the \eos{} water fill reconstruction results is to validate the methodology prior to collecting data with target materials that produce scintillation light. Performance metrics (e.g., bias and width distributions) for the reconstruction methods are shown to compare the performance of different algorithms and estimate the level of the data and simulation agreement. The capacity of the reconstruction is highly specific to the \eos{} detector, which is both small and asymmetric. As such, these results provide validation for future phases of \eos{} and for future simulation studies of large-scale hybrid detectors.

 The reconstruction objectives for \eos{} are based on what has been achieved previously in large water Cherenkov detectors, such as SNO, SNO+, and Super-Kamiokande. These experiments find vertex bias and resolution systematics between 0.5 to 10~cm \cite{SNO:2009uok,SNO:2024vjl,Super-Kamiokande:2016yck}. The systematics on the vertex bias and resolution tended to be subdominant in the total uncertainty for solar neutrino measurements presented in these Refs. Using the thorium source, \eos{} is evaluating vertex reconstruction at energies lower than any of these previous detectors. For this manuscript the goal for the agreement between data and MC in both the vertex bias and resolutions is to be within 3~cm, as characterized using the thorium source. This objective is based on the experience of these previous water Cherenkov detectors, accounting for \eos{}'s unique, asymmetric detector configuration and low energy detection capabilities. Additionally, in principle any data and simulation differences could be either corrected for or taken as a systematic for any future \eos{} analysis that goes beyond evaluating detector performance.

The goals for the direction reconstruction, characterized using the directional sources, are determined similarly. For this manuscript, the goal is to provide data and simulation agreement in $\alpha_{1\sigma}$ to within 0.1 across various source $z$ positions. This is similar to the size of the systematic on the angular resolution in Ref. \cite{SNO:2024vjl}, which resulted in a roughly 2\% systematic on the $^{8}$B flux measurement in a water Cherenkov detector. 

\subsection{ Vertex}\label{sec:vertex-results}

\begin{figure}[ht!]
    \centering
\includegraphics[width=0.45\linewidth]{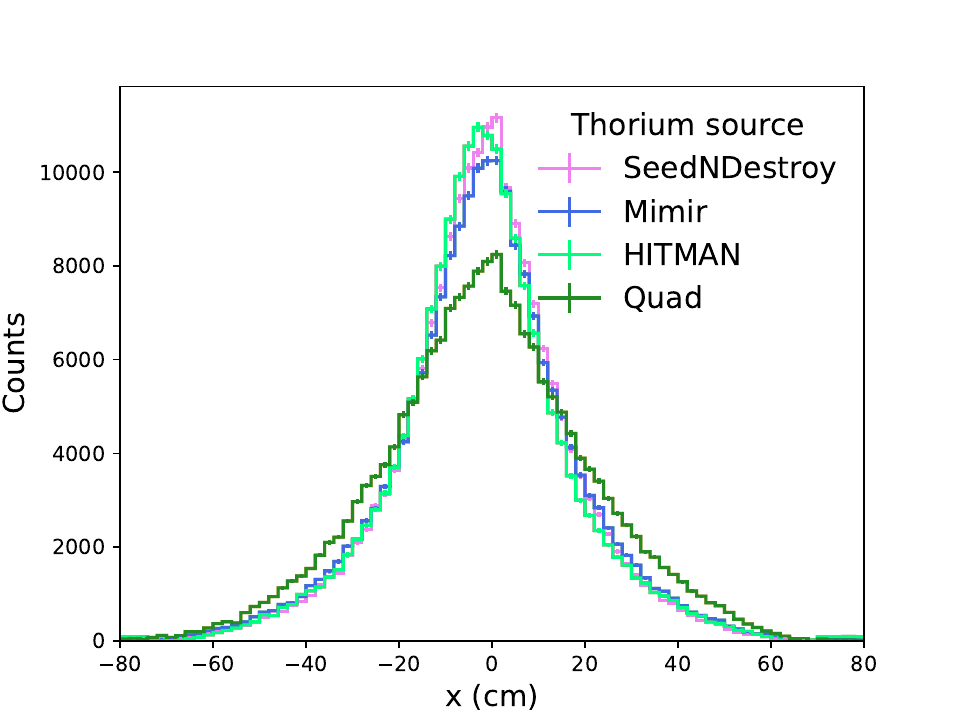}
\includegraphics[width=0.45\linewidth]{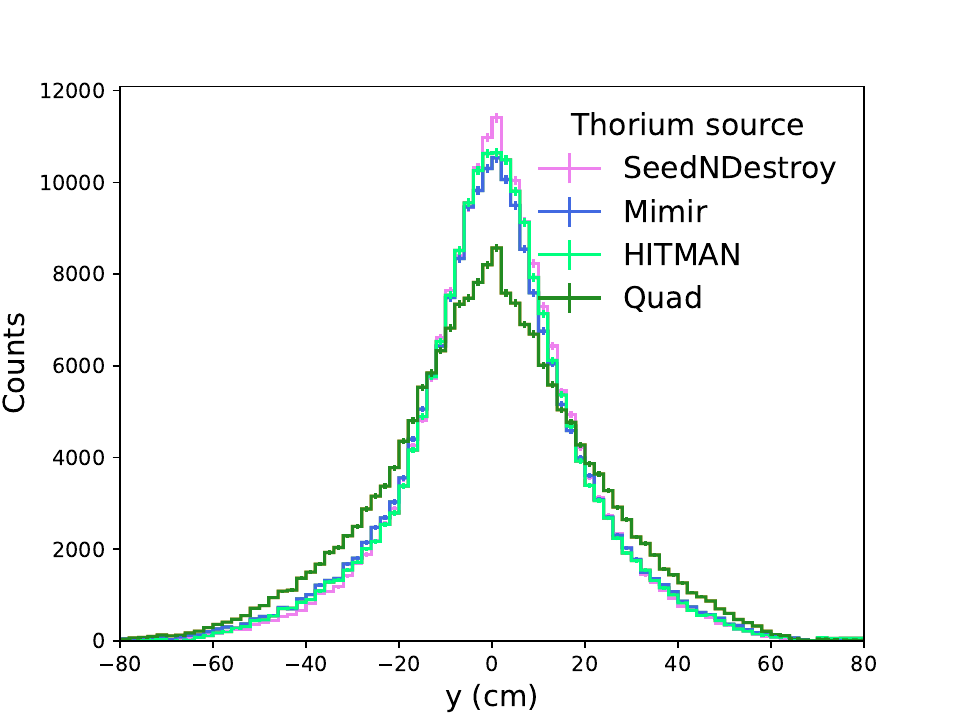}
\includegraphics[width=0.45\linewidth]{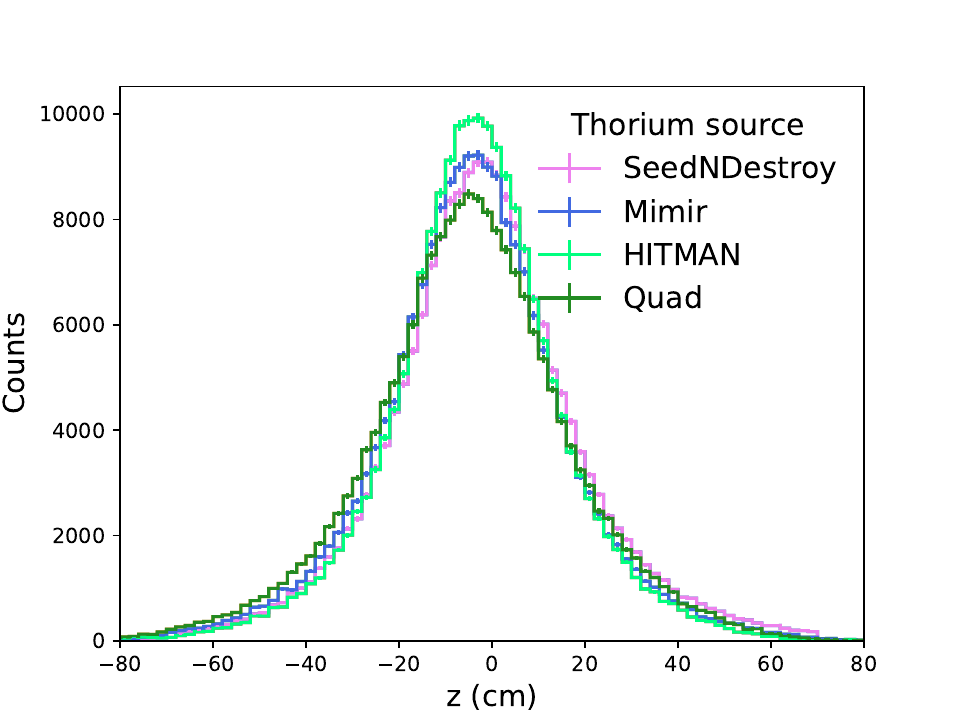}
\includegraphics[width=0.45\linewidth]{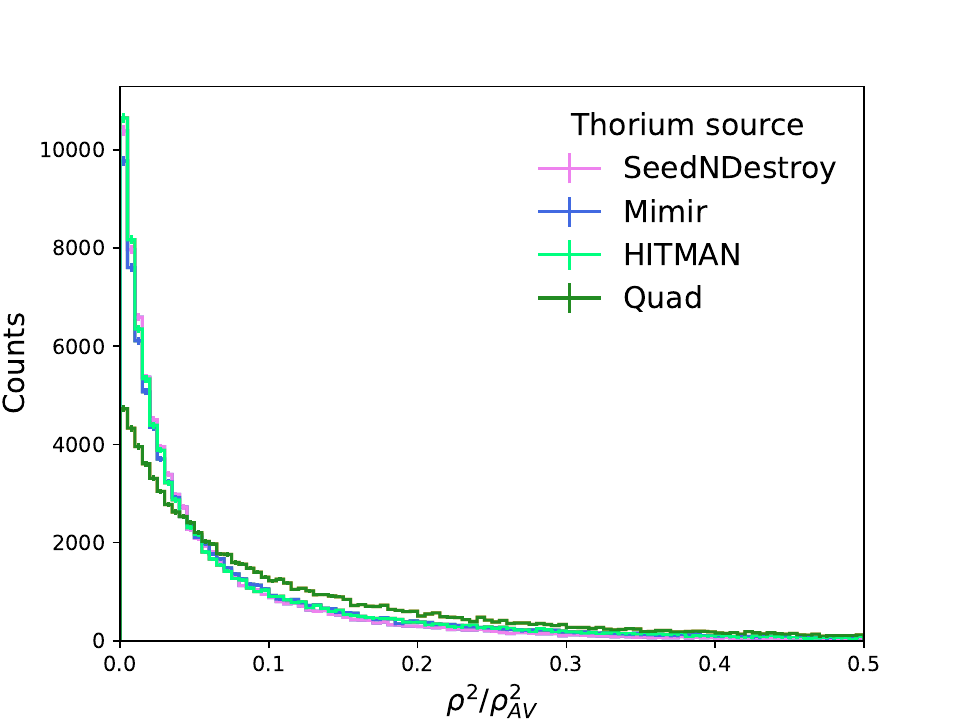}
    \caption{Reconstructed $x$, $y$, $z$, and $\rho^{2}/\rho^{2}_{IV}$ distributions from the four vertex fitters applied to central thorium run data in \eos. The uncertainties are statistical only.}
    \label{fig:vertex-fitter-comp}
\end{figure}

The event vertex is reconstructed using the methods described in Secs.~\ref{sec:vertex-recon} and \ref{sec:hitman}: the quad fitter, \snd{}, \mimir{} and \hitman{}. Fig.~\ref{fig:vertex-fitter-comp} shows the vertex reconstruction results in $x$, $y$, $z$, and $\rho^{2}$ for central thorium source data. In general, the quad fitter has the largest bias and resolution, as expected due to the simplicity of the algorithm. The other three fitters perform similarly, with \hitman{} slightly outperforming the likelihood-based fitters when reconstructing the $z$ position. The reconstruction of the event $z$ position is the most challenging as it is significantly impacted by the \eos{} top-bottom detector asymmetries.

For each of the algorithms, histograms of the reconstructed Cartesian coordinates are fit with a Gaussian around the peak, using a range of $\pm$20~cm. The bias (difference from the true source position) is taken from the mean of the fit and the width ($\sigma$) of the fit provides a metric that estimates the vertex resolution. As is clear in Fig.~\ref{fig:vertex-fitter-comp}, the distributions are not perfectly Gaussian, and the fit does not capture some of the features in the tails. This method is intended to demonstrate the performance for the bulk of the events without focusing on outliers. It should be emphasized that the thorium calibration source produces $\gamma$-rays, and thus, the reconstruction resolution convolves effects from the Compton scattering. As such, the goal here is to show the relative comparison between the algorithms, as well as the overall agreement between data and simulation. 

\begin{figure}[ht!]
    \centering
    \includegraphics[width=0.45\linewidth]{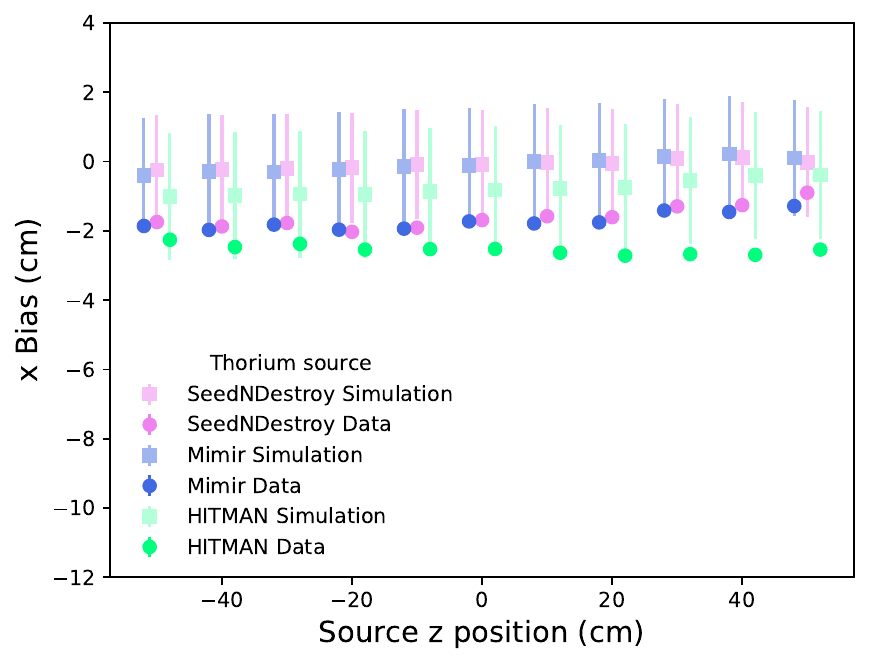}
    \includegraphics[width=0.45\linewidth]{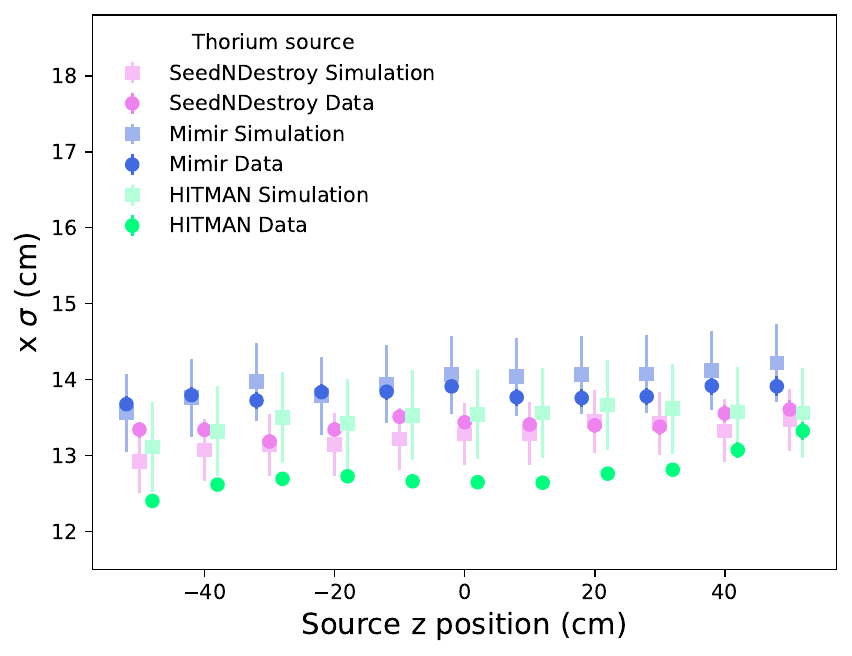}
    \includegraphics[width=0.45\linewidth]{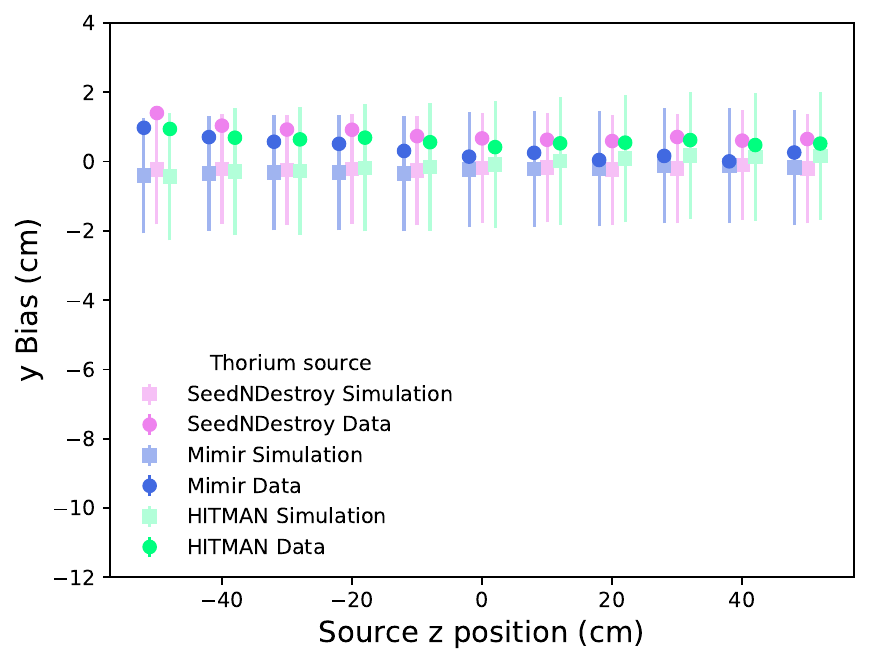}
    \includegraphics[width=0.45\linewidth]{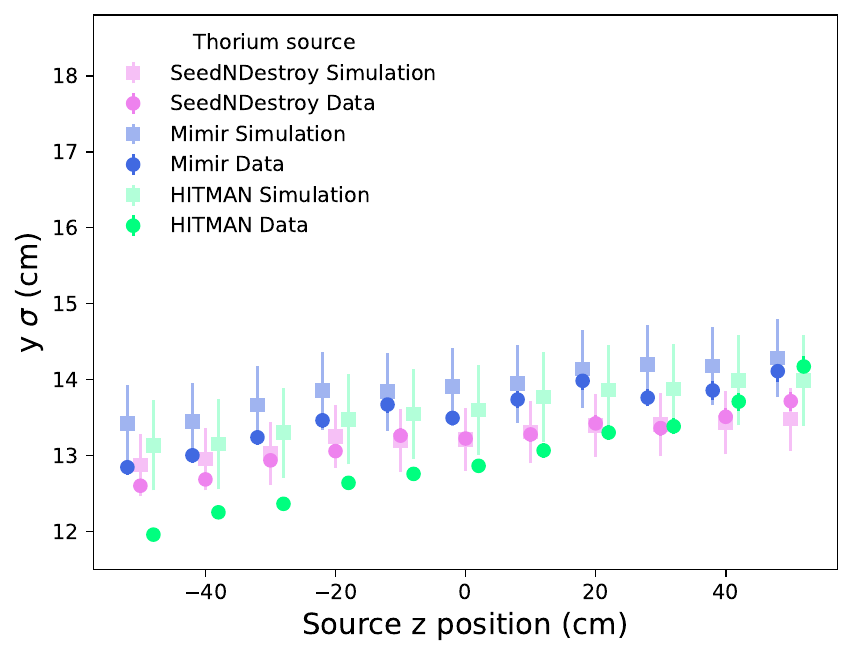}
    \includegraphics[width=0.45\linewidth]{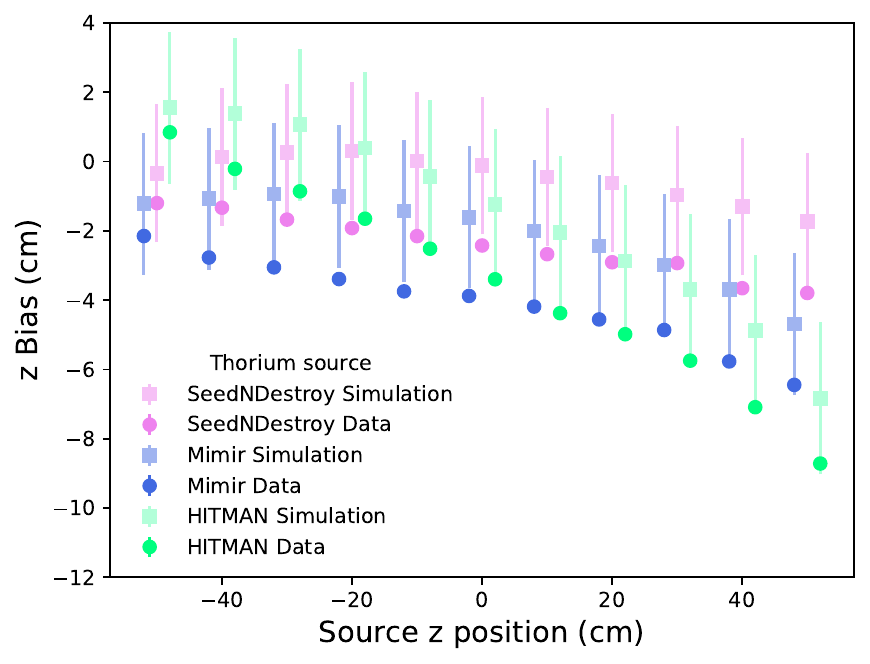}
    \includegraphics[width=0.45\linewidth]{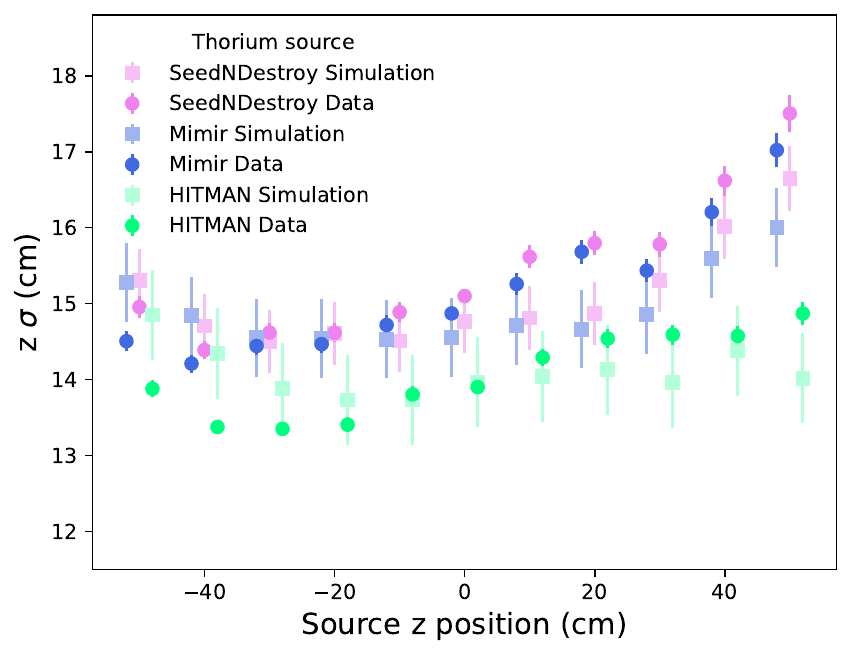}
    \caption{The bias (left) and resolution (right) for the $x$ (top), $y$ (middle), and $z$ (bottom) positions, for \snd{}, \mimir{}, and \hitman{}, as a function of the thorium source $z$-position. The values are taken from Gaussian fits performed over the histogram of the relevant reconstructed quantity. For \mimir{} and \hitman{} the source $z$ position is shifted slightly from the true value on the x-axis to make the figure more legible.}
    \label{fig:vertex-fitter-comparison-all}
\end{figure}

Fig.~\ref{fig:vertex-fitter-comparison-all} shows the bias and resolution results for three fitters as a function of the source $z$-position (the quad fitter has been excluded, as it is used primarily to seed the other algorithms). The $x$ and $y$ bias from the true position for each of the fitters is around 1-2~cm, and does not depend strongly on the source $z$-position. The $z$ bias ranges from 3-10~cm, is strongest in \hitman{}, and increases as the source moves higher in the detector. This is expected, as the PMT coverage towards the bottom of the detector is higher than at the top, and the simulation accurately predicts this effect. The systematic uncertainties on the bias for each coordinate are on the order of 2~cm, as described in Sec.~\ref{sec:systematics}. The \chindf comparisons between the data and simulation are provided for each reconstruction algorithm for both the bias and resolution in Table~\ref{tab:chi2}. In several cases, a very small value of the \chindf is indicative that the systematic uncertainties have been overestimated. This is expected in some cases as the systematics were estimated based on conservative assumptions.

\begin{table}[ht!]
    \centering
\begin{tabular}{c|c|c|c} \hline \hline 
     & \snd{} & \mimir{} & \hitman{} \\ \hline 
    x bias & 10.4/11 & 10.7/11 & 10.6/11 \\
    y bias & 4.8/11 & 2.0/11 & 1.8/11 \\
    z bias & 11.5/11 & 10.6/11 & 8.8/11 \\
    x resolution & 2.7/11 & 1.8/11 & 17.5/11 \\
    y resolution & 1.5/11 & 5.3/11 &  17.0/11 \\
    z resolution & 16.1/11 & 14.3/11 & 10.0/11 \\ \hline
\end{tabular}
\caption{The \chindf values for the data and simulation comparisons presented in Fig. \ref{fig:vertex-fitter-comparison-all}.}
\label{tab:chi2}
\end{table}

\begin{figure}[ht!]
    \centering
    \includegraphics[width=0.45\linewidth]{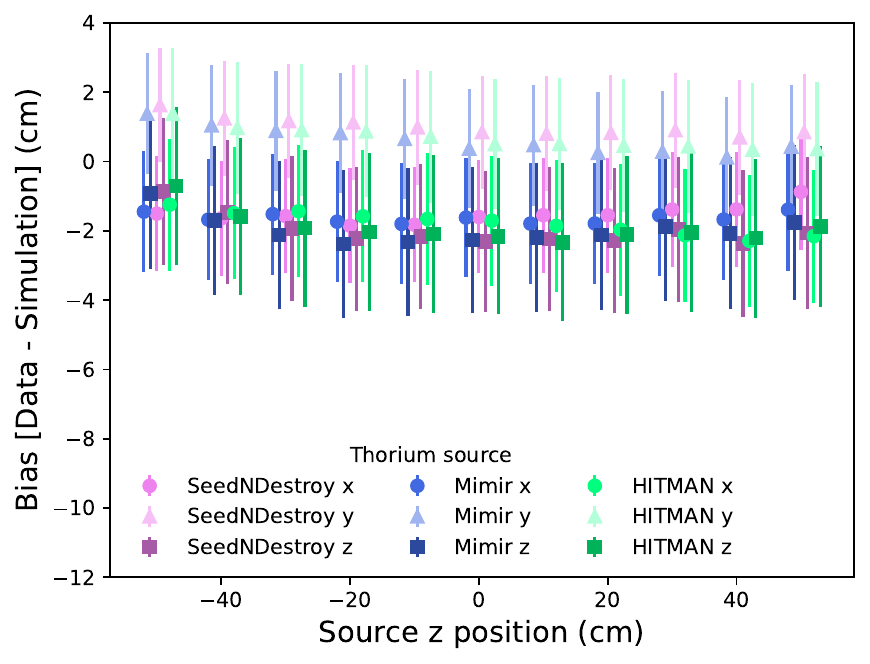}
    \includegraphics[width=0.45\linewidth]{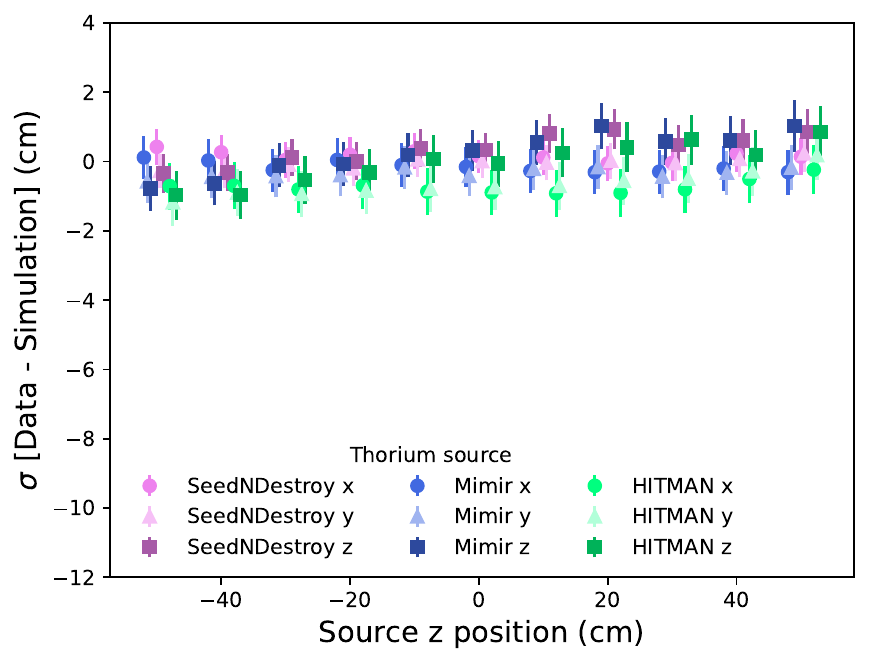}
    \caption{The data minus the simulation for  the $x$, $y$, and $z$ bias and resolutions from the three vertex fitters, for the thorium source. For \mimir{} and \hitman{} the source $z$ positions are shifted slightly from the true value on the x-axis to make the figure more legible.}
    \label{fig:vertex-differences}
\end{figure}

The $x$ and $y$ resolutions ($\sigma$) from the Gaussian fits are around 12-13~cm for all algorithms and the data and simulation agree within 1~cm. The $z$ resolution has an understood dependence on the source $z$-position, and extends from roughly 12~cm near the bottom to 16~cm near the top of the detector. The data and simulation agreement is most clearly delineated in Fig.~\ref{fig:vertex-differences}, which shows, for each algorithm, the difference between the data and simulation for the bias and resolution. In general, deviations between the data and simulation for both the bias and resolution are on the order of 1-2~cm  for each algorithm. Interestingly, the simulation tends to slightly over-predict the resolution when compared to the data.

\begin{figure}[ht!]
    \centering
    \includegraphics[width=0.45\linewidth]{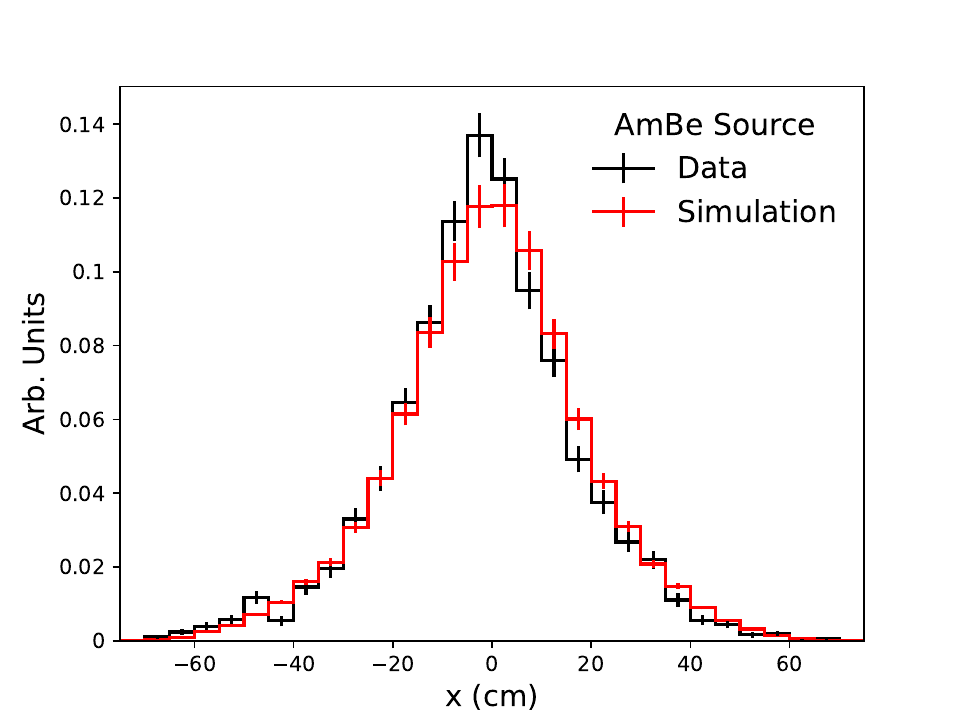}
    \includegraphics[width=0.45\linewidth]{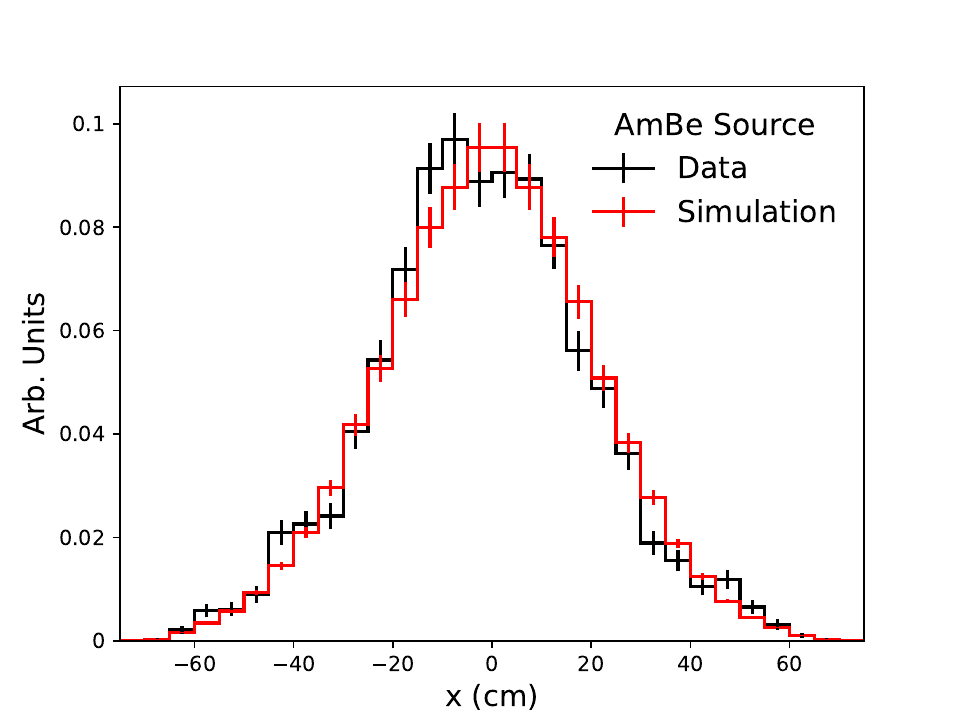}
    \includegraphics[width=0.45\linewidth]{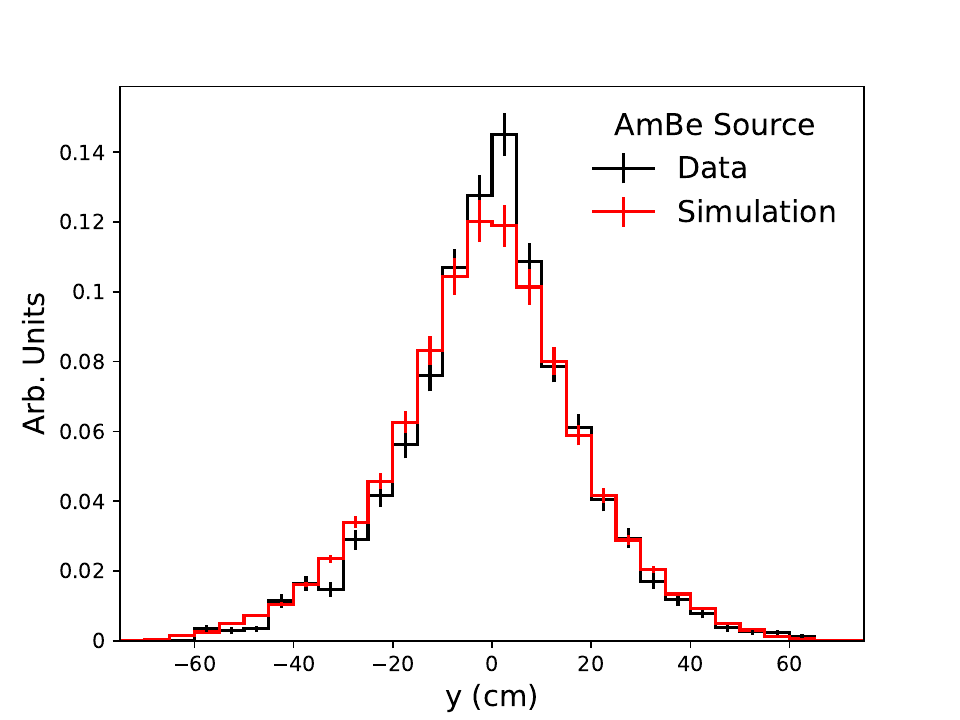}
    \includegraphics[width=0.45\linewidth]{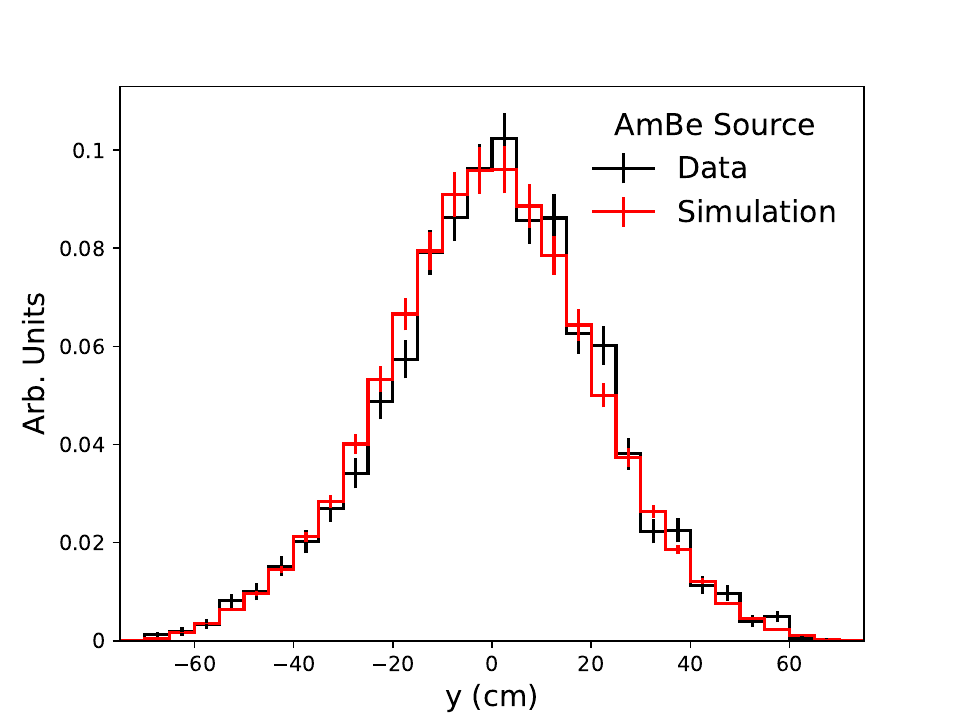}
    \includegraphics[width=0.45\linewidth]{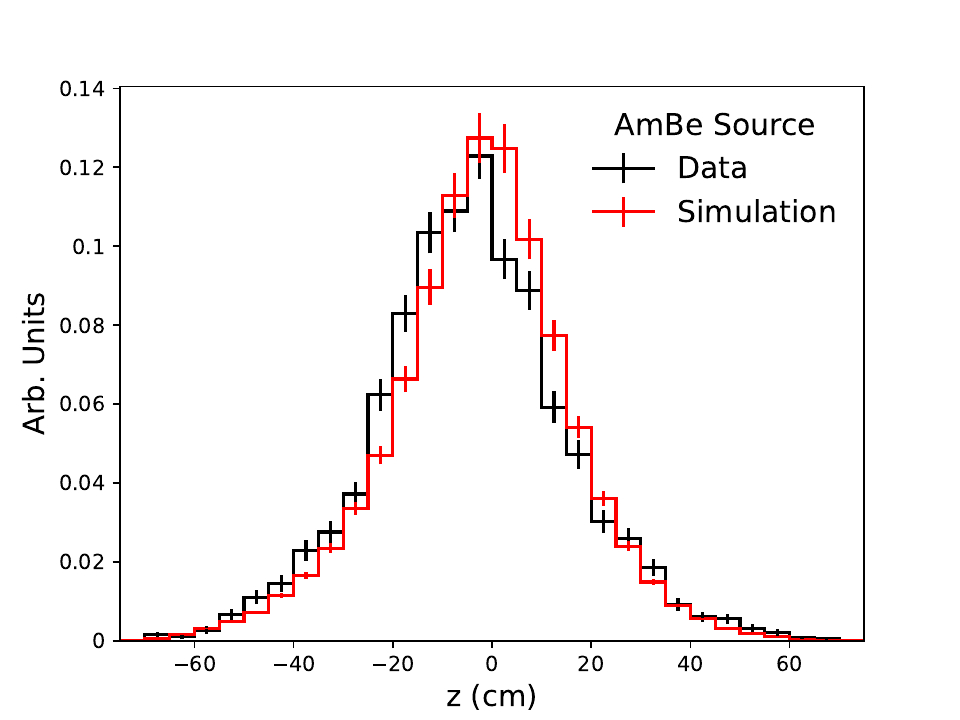}
    \includegraphics[width=0.45\linewidth]{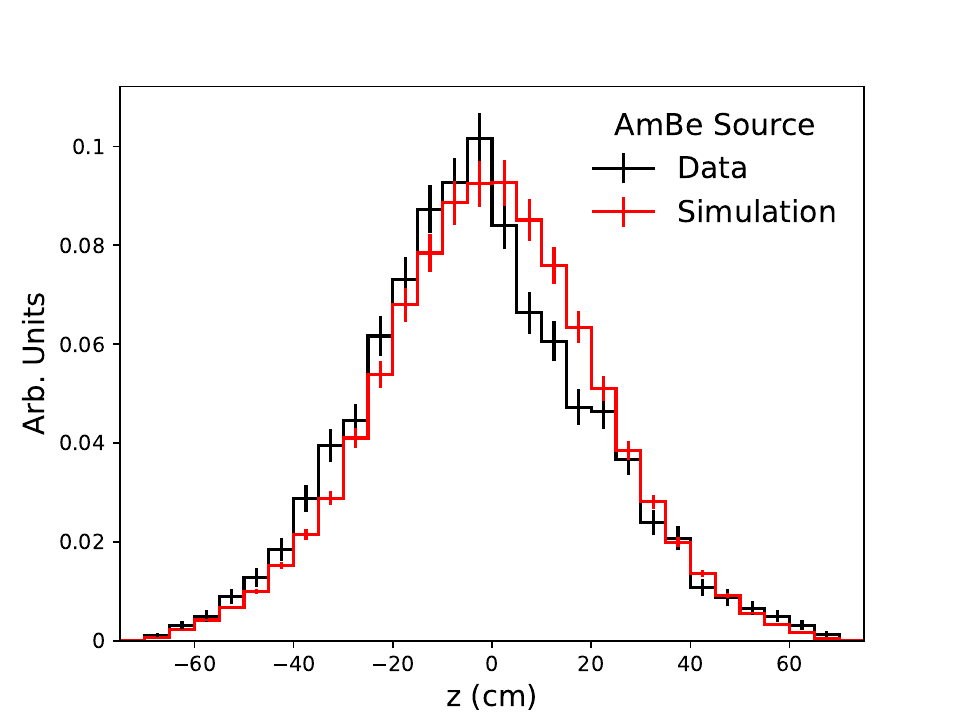}
    \caption{The reconstructed position distributions, using \snd{}, compared between data and simulation for a central AmBe source run. The left column shows the prompt event and the right column shows the delayed event.}
    \label{fig:ambe_position}
\end{figure}

The AmBe source provides an additional method for testing the vertex reconstruction, with a different source geometry (and thus different source shadowing effects) and different energy $\gamma$-rays (both the prompt 4.4~MeV and delayed 2.2~MeV). In Fig.~\ref{fig:ambe_position} the \snd{} reconstructed $x$, $y$, and $z$ position distributions are compared between data and MC for the tagged prompt and delayed events from a centrally deployed AmBe source. The vertex resolution for the prompt/delayed events from the $4.4/2.2$~MeV $\gamma$-rays is around 13/19~cm in the data. The data and simulation for these reconstructed distributions are compared and the \chindf are 64.1/27 (prompt $x$), 66.4/27 (prompt $y$), 85.7/27 (prompt $z$), 47.8/27 (delayed $x$), 31.4/27 (delayed $y$), and 74.6/27 (delayed $z$). Similarly to the \np{} distributions, there remain small mismodelings of the AmBe source in the detector that lead to relatively large \chindf values.

\begin{figure}[ht!]
    \centering
    \includegraphics[width=0.6\linewidth]{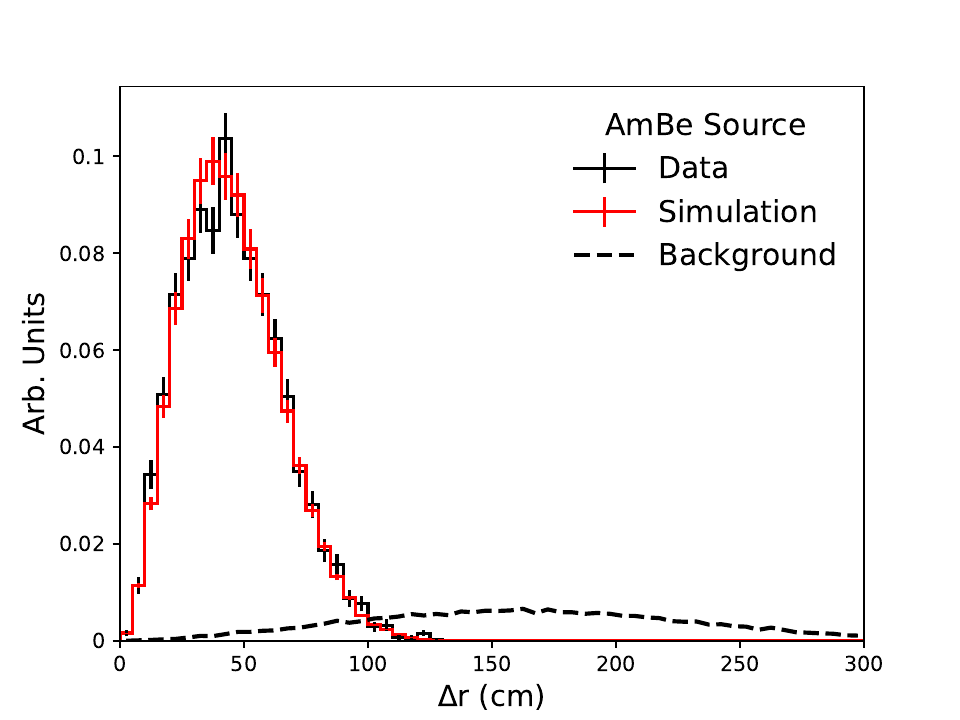}

    \caption{The reconstructed $\Delta r$ distribution using \snd{}, compared between data and simulation for a central AmBe source run. The background distribution is taken from a dedicated run without the source in the detector and is scaled to the same livetime as the AmBe data.}
    \label{fig:ambe_dr}
\end{figure}

The $\Delta r$ distribution between the prompt and delayed events for the AmBe is shown in Fig.~\ref{fig:ambe_dr}. Included is the background distribution from a dedicated run without the source in the detector, scaled to have the same livetime as the AmBe run. For this figure, the fiducial volume is relaxed to show the full background distribution. The $\Delta r$ distribution exemplifies that, despite the surface deployment and water target, \eos{} can measure low-energy coincidences from neutrons with little background. The good data and simulation agreement indicates that while we observe small differences between the data and simulation in the individual Cartesian coordinates, overall the $\Delta r$ distribution is modeled quite well (\chindf = 21.5/26).

In \eos{}, the directional source data provide a test of the vertex reconstruction using electrons. Due to source shadowing, the $\beta$ energy spectrum, and other effects, we must still rely on our simulation model to understand the resolutions. The figures for the directional source results are detailed in Appendix~\ref{sec:appendix-A}. In these figures, we combine across several runs in order to generate enough statistics for the reconstruction tests.  These include data and simulation comparisons for both the $^{90}$Sr and $^{106}$Ru source, the latter of which includes two source sizes (20~mm and 30~mm). For the $^{90}$Sr source, the bias and resolution for both downward pointing and sideways pointing data are displayed. Overall, the $x$ and $y$ resolutions are less than 10~cm for the $^{106}$Ru data across all source $z$-positions. As in all cases, the resolution in $z$ is more complicated and has a stronger dependence on the deployed source $z$-position. 

As a useful metric, we provide the resolutions for simulated central, 2.0~MeV electrons in \eos{} for each of the algorithms in Table~\ref{tab:resolutions-electrons}. For all algorithms other than the quad fitter, the expected resolution is less than 10~cm for central 2.0 MeV electrons. 

The agreement achieved in the bias and resolutions across varying source type, particle type, source size, source position, and source direction (with energies spanning roughly 1 to 4.5~MeV) is typically around 1~cm and is less than 3~cm for all algorithms. These efforts provide additional verification of the \eos{} modeling, reconstruction software, and the model of the sources themselves. This is critical for future phases of \eos{} using various liquid scintillators as the target material. The evaluation of the difference between data and simulation in \eos{} allows us to understand the limitations of our predictions for these future phases and for future detectors, such as \theia{}.

\begin{table}[ht!]
    \centering
    \begin{tabular}{l|c|c|c} \hline \hline 
         Algorithm & $x$ (cm) & $y$ (cm) & $z$ (cm) \\ \hline 
         Quad & 10.9 $\pm$ 0.2 & 10.8 $\pm$ 0.2 & 10.7 $\pm$ 0.2 \\
         \snd{} & 8.9 $\pm$ 0.1 & 8.8 $\pm$ 0.1 & 9.7 $\pm$ 0.1 \\ 
         \mimir &  8.8 $\pm$ 0.1 & 8.8 $\pm$ 0.1 & 9.4 $\pm$ 0.1 \\ 
         \hitman & 7.8 $\pm$ 0.1  & 7.9 $\pm$ 0.1 & 8.3 $\pm$ 0.1 \\ \hline \hline
    \end{tabular}
    \caption{The $x$, $y$, and $z$ resolution from a Gaussian fit to the associated histogram for simulated central, isotropically pointed, 2.0~MeV electrons. The uncertainties are from the fit and are statistical only.}
    \label{tab:resolutions-electrons}
\end{table}

\subsection{Direction}\label{sec:direction-results}

\begin{figure}[ht!]
    \centering
    \includegraphics[width=0.45\linewidth]{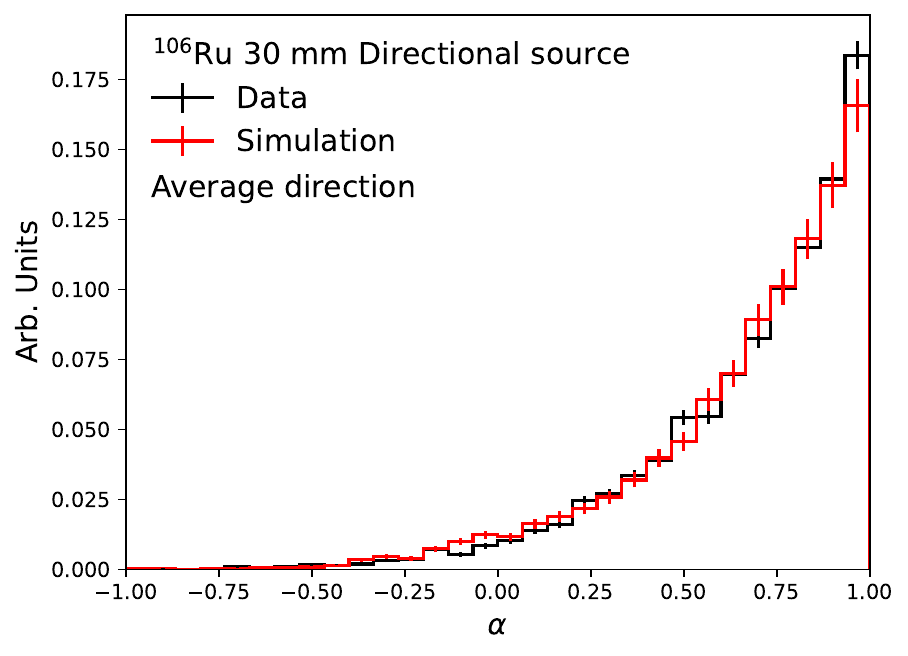}
    \includegraphics[width=0.45\linewidth]{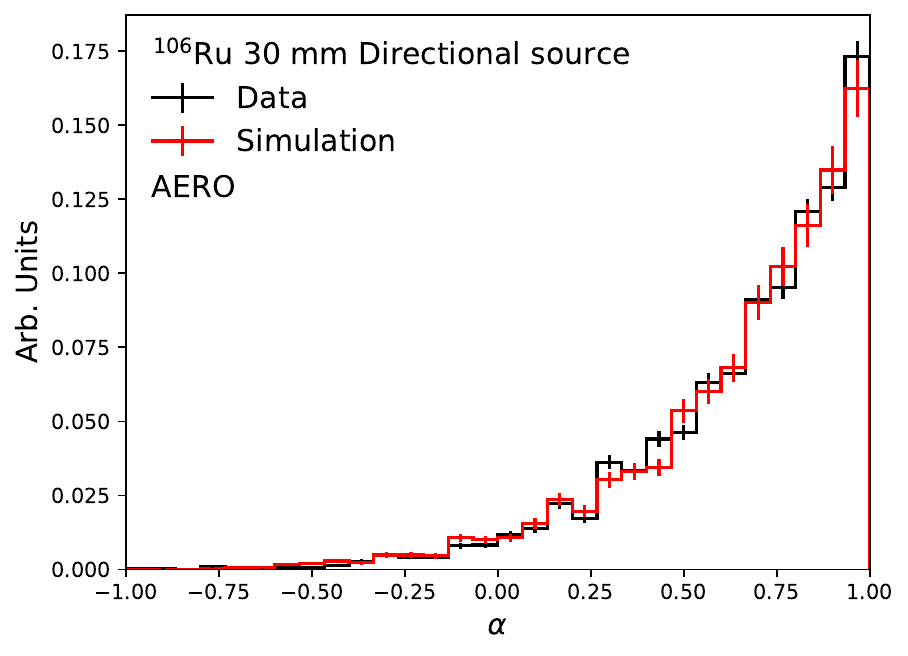}
    \includegraphics[width=0.45\linewidth]{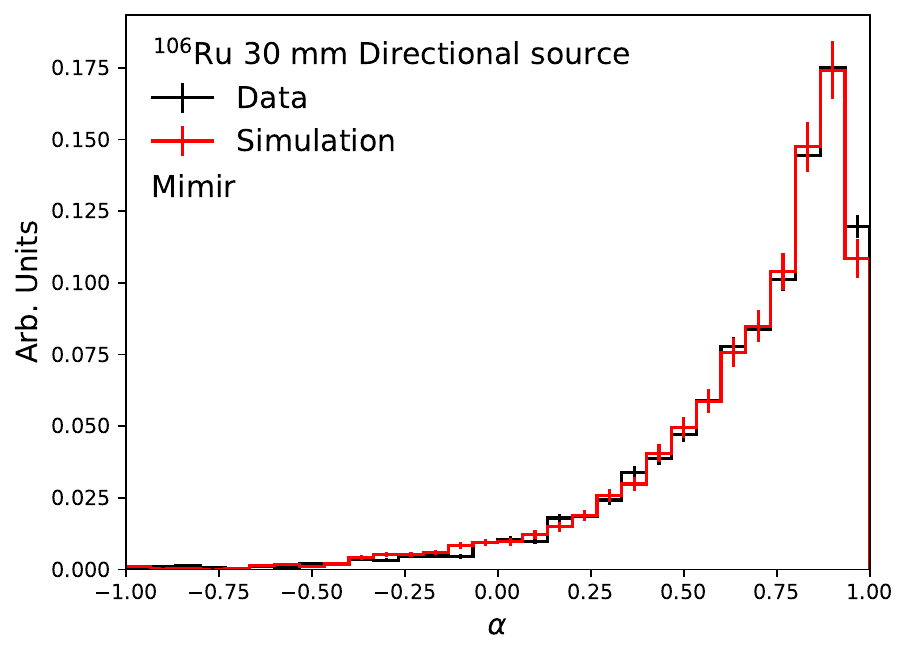}
    \includegraphics[width=0.45\linewidth]{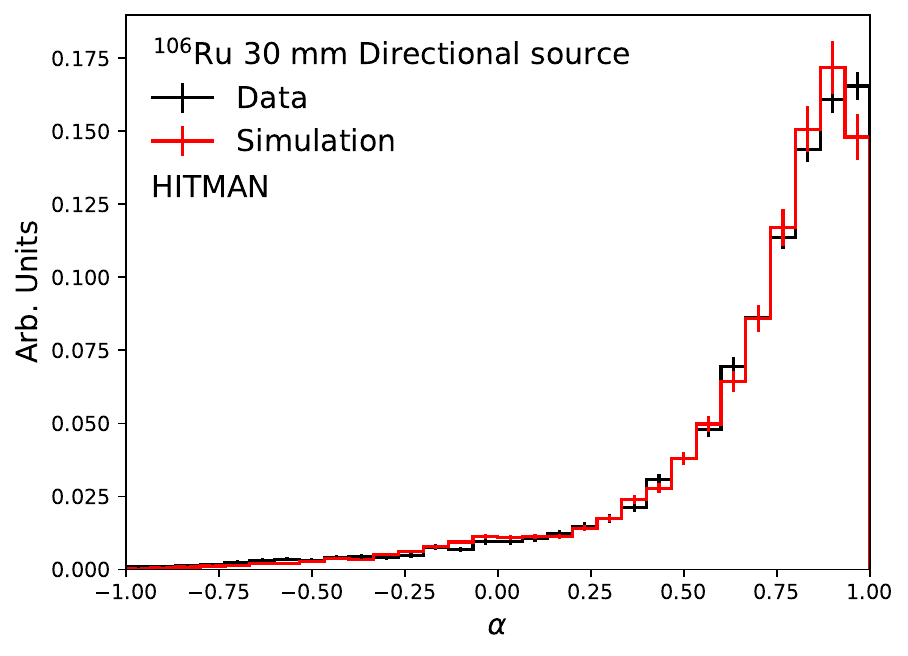}
    \caption{The $\alpha$ reconstruction results for data and simulation, from a central deployment of the 30~mm $^{106}$Ru directional source, pointed downward. All four direction reconstruction algorithms are shown.}
    \label{fig:DS_dir_reco}
\end{figure}

The event direction is reconstructed using the methods described in Sec.~\ref{sec:dir-recon}. Fig.~\ref{fig:DS_dir_reco} compares the direction reconstruction results of $\alpha$ for the data and simulation, applying the four different direction reconstruction algorithms to the central 30~mm $^{106}$Ru directional source pointed downward. Due to the self-triggering mechanism of the directional source, negligible background is identified, which would manifest as a flat component in $\hat{d}_{\text{source}} \cdot \hat{d}_{\text{recon}}$. All four algorithms perform similarly and show comparable levels of agreement between data and simulation. The calculated \chindf values for these data and simulation comparisons are 36.7/28 (average direction), 33.9/30 (\mimir{}), 34.3/29 (\AERO), and 38.6/30 (\hitman{}). Based on these findings, the following directional source results focus on the agreement for a single reconstruction framework (\AERO).

These histograms are fit with an exponential convolved with a Gaussian, described in Sec.~\ref{sec:dir-recon}, to extract $\alpha_{1\sigma}$. In this manuscript, we present selected results for the directional sources: the downward pointing 30~mm $^{90}$Sr and $^{106}$Ru sources, the downward pointing 20~mm $^{106}$Ru source, and the 90$^{\circ}$ sideways pointing 20~mm $^{106}$Ru source. In general, consistent performance is identified with the 20~mm and 30~mm sources, so not all of the collected data is presented here. Additionally, the source $z$ position is limited to [-30, +30]~cm in these studies. Data collected closer to the bottom of the detector shows poorer performance that is not yet well modeled in the simulation, similar to the discussion about the dichroicon performance with the laserball data in Sec.~\ref{sec:dichroicon-response}. There is active, ongoing work to better understand the detector response for sources near the bottom of the detector. For the sideways pointing runs, an additional systematic is included (described in Sec.~\ref{sec:systematics}) based on the fact that no precise measurement of the $\phi$ direction of the source was possible at the time. 

\begin{figure}[ht!]
    \centering
    \includegraphics[width=0.45\linewidth]{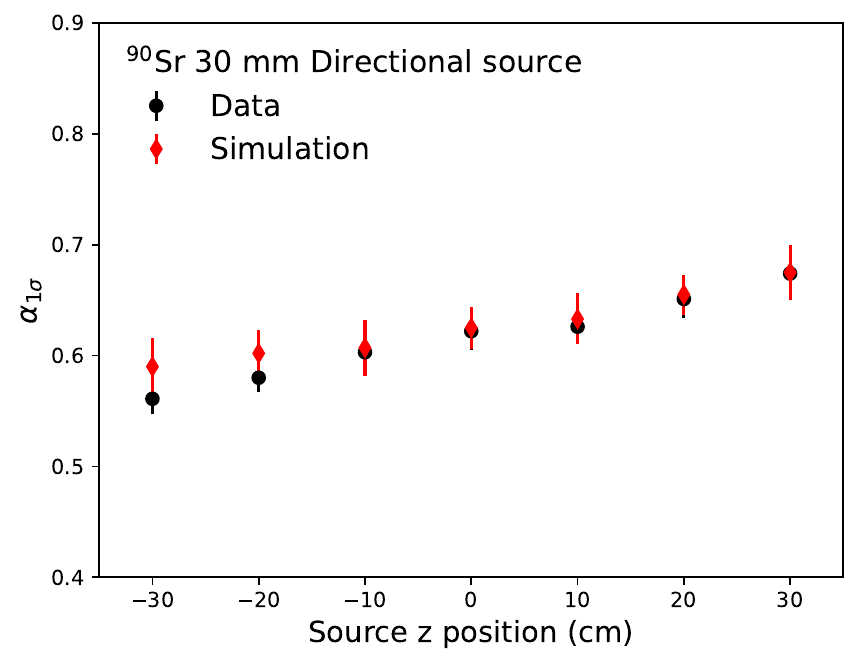}
    \includegraphics[width=0.45\linewidth]{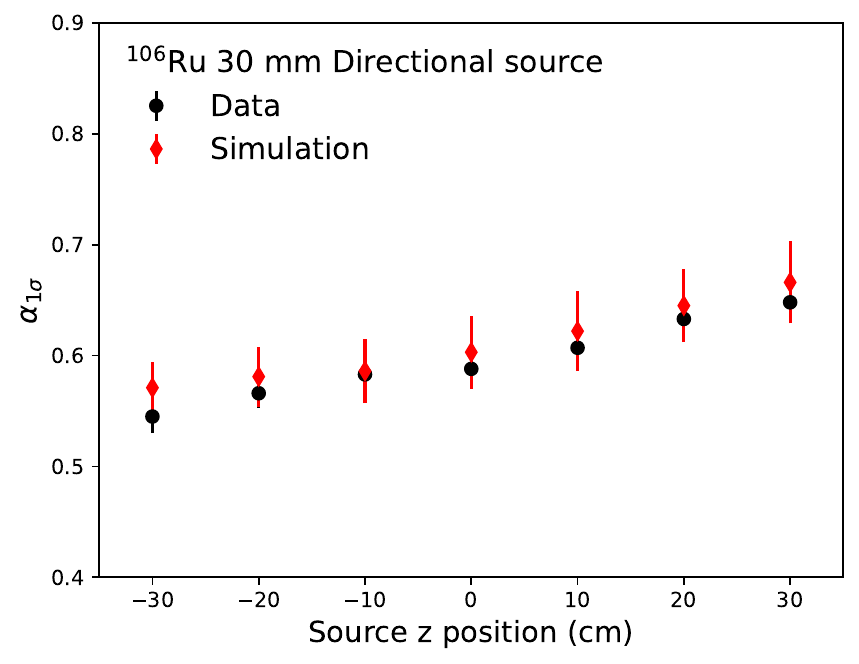}
    \includegraphics[width=0.45\linewidth]{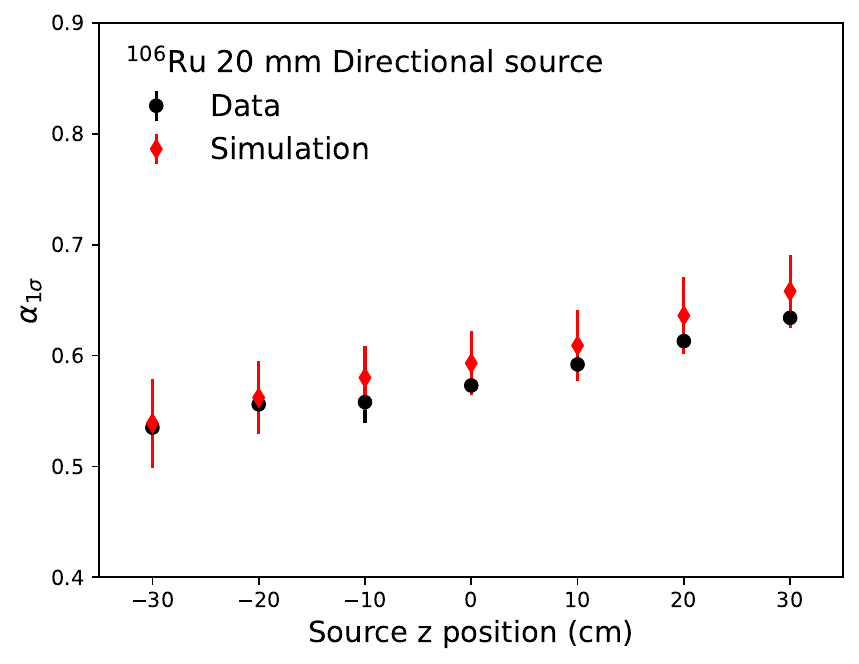}
     \includegraphics[width=0.45\linewidth]{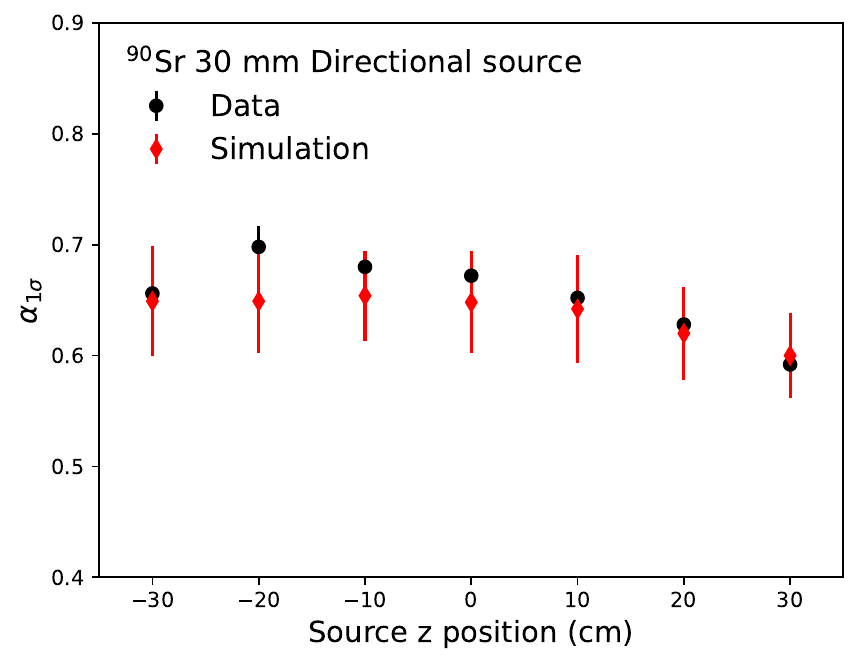}
    
    \caption{The $\alpha_{1\sigma}$  reconstruction results using \AERO{} for data and simulation for the downward pointing $^{90}$Sr 30~mm directional source (top left), the downward pointing $^{106}$Ru 30~mm directional source (top right), the downward pointing $^{106}$Ru 20~mm directional source (bottom left), and the $\theta = 90^{\circ}$ sideways pointing $^{90}$Sr 30~mm directional source (bottom right).}
    \label{fig:dir-source-summary-results}
\end{figure}

The results for $\alpha_{1\sigma}$ for these deployments as a function of the source $z$ position are shown in Fig.~\ref{fig:dir-source-summary-results}. In general, the data and simulation agreement is good and the \chindf{} values for the $^{90}$Sr 30~mm source (downward), $^{106}$Ru 30~mm source (downward), $^{106}$Ru 20~mm source (downward), and $^{106}$Ru 30~mm source (sideways) are 1.8/7, 1.9/7, 2.1/7, and 1.7/7, respectively. Across these results we find that the agreement between the data and simulation is well within the goal of 0.1 on $\alpha_{1\sigma}$, regardless of the source type and direction. 

\begin{figure}[ht!]
   \centering
 \includegraphics[width=0.45\textwidth]{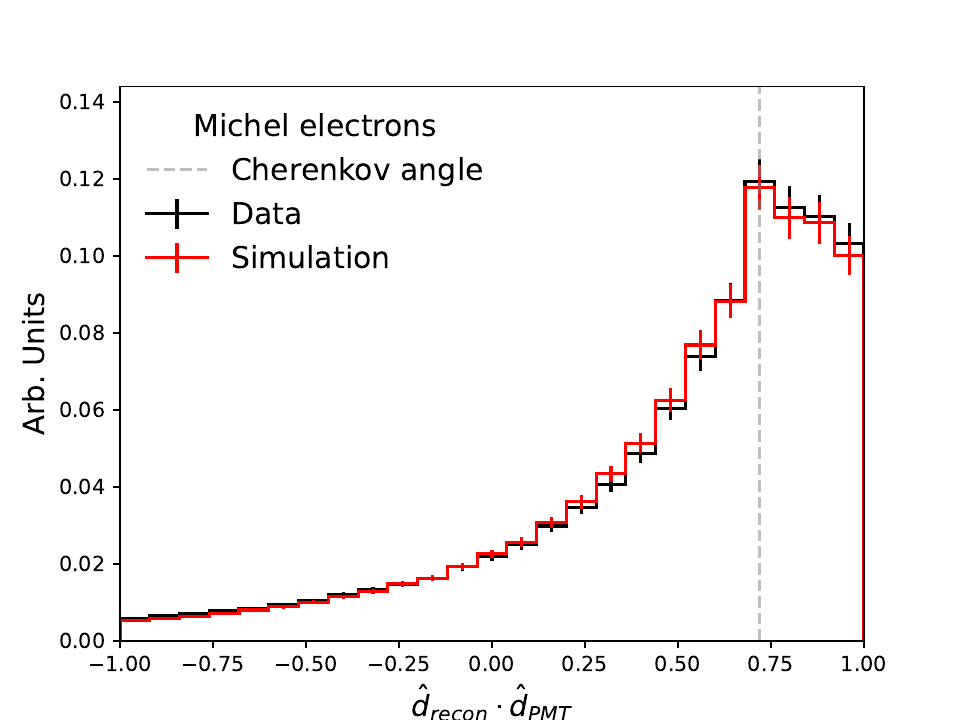}
 \includegraphics[width=0.45\textwidth]{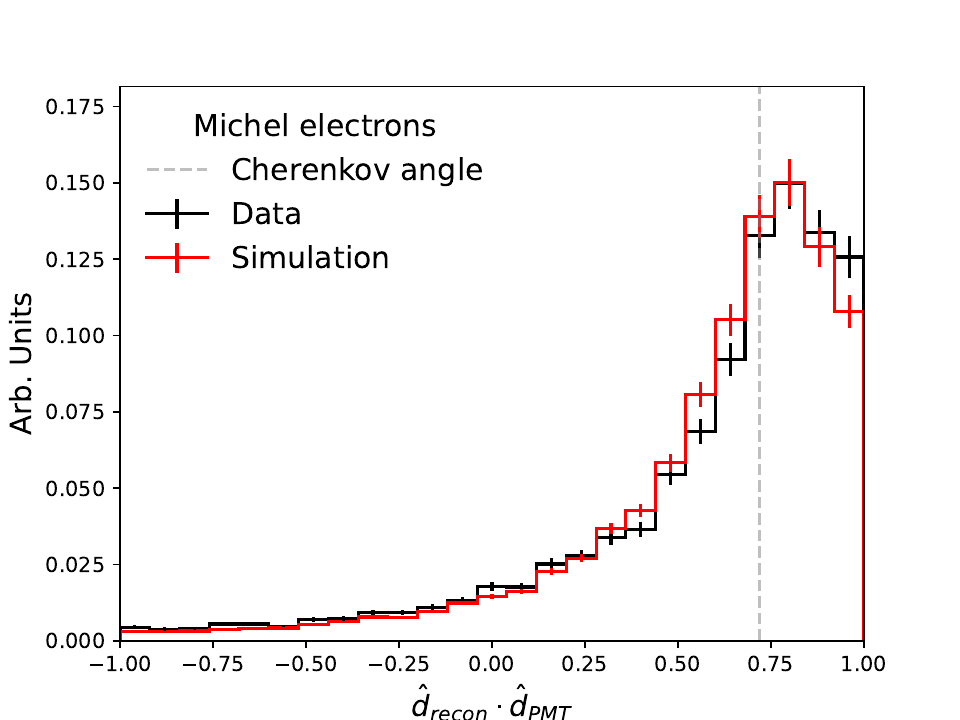}
   \caption{The $\hat{d}_{\text{recon}} \cdot \hat{d}_{\rm{PMT}}$ compared between data and simulation using all PMTs (left) and using only the PMTs behind the dichroicons (right). The dashed gray line denotes the Cherenkov emission angle.} 
   \label{fig:michels-directionality}
\end{figure}

The direction reconstruction can also be tested using Michel electrons, which provide a higher energy, distributed source in which to study the algorithms. Fig.~\ref{fig:michels-directionality} shows $\hat{d}_{\text{recon}} \cdot \hat{d}_{\rm{PMT}}$, where $\hat{d}_{\rm{PMT}}$ is the vector pointing from the hit PMT to the reconstructed event position for the tagged Michel electron. This distribution is shown using both all PMTs (left) and only the PMTs behind the dichroicons (right). These results use \mimir{} for the reconstructed direction. The peak in both distributions around $\hat{d}_{\text{recon}} \cdot \hat{d}_{\rm{PMT}} \sim 0.75$ is from the Cherenkov angle at $\sim$ 41$^{\circ}$. The slight offset in the peak from the Cherenkov angle, in both the data and simulation, in the dichroicon-only result is caused by geometric effects. Values of \chindf = 13.6/24 and \chindf = 53.4/24 are calculated for the $\hat{d}_{\text{recon}} \cdot \hat{d}_{\rm{PMT}}$ results for the all PMTs and dichroicon PMTs respectively. The agreement between data and simulation is further verification of the detector modeling and direction reconstruction, which work well even in the 10's of MeV range for tagged Michel electrons.

\section{Discussion}

\eos{} plays a key role as the first large-scale demonstrator of many of the key technologies that enable future hybrid detectors. This manuscript marks the first use of \eos{} data and demonstrates the development and calibration of simulation models that can be used to make predictions and generate physics sensitivity estimates for \theia{}. These calibrated models will also be invaluable in the subsequent phases of \eos{}, which include the deployment of both water-based liquid scintillator (WbLS) \cite{YehWbLS} and pure liquid scintillator (such as LAB+PPO)~\cite{Anderson:2020xxb}, where the optical properties of the medium in the IV is less well-understood.  

Furthermore, the AmBe results show that \eos{} is able to use the deployed advanced instrumentation and sophisticated triggering to perform clear neutron detection, despite the near-surface deployment, the small detector size, and the water target, which produces only Cherenkov light.  Neutron detection is a key component for the detection of reactor antineutrinos, and thus these results aid in understanding the potential to use this technology for nuclear reactor monitoring and other important physics topics that require the detection of antineutrinos or neutrons.

The \eos{} collaboration is continuing to develop detector modeling and data analysis techniques. This includes actively developing multi-PE counting and timing estimators, utilizing charge in the reconstruction algorithms, improving the modeling of the AmBe source, analyzing additional calibration source data from sources not used for this paper (e.g., an external $\gamma$ source), high energy ($>10$~MeV) reconstruction methods, angular response calibrations of the dichroicons, and improvements to the incorporation of scintillation light into the reconstruction algorithms. This work is paired with ongoing efforts to understand physics sensitivities for a potential future deployment of \eos{} at the Spallation Neutron Source at Oak Ridge National Lab. 

\section{Conclusion}

In this manuscript we describe the data collected with the \eos{} detector while it was filled with water, using a variety of deployed radioactive and optical sources. Laserball data are used to calibrate the PMT timing and charge, tune the model of the dichroicons, and investigate the trigger efficiency. Central thorium data are used to calibrate the PMT efficiencies. The results from these calibrations are input into detailed simulation models, built with the \rp{} software. Using this framework, a variety of checks are performed against the data using the laserball, thorium, AmBe, and several directional sources. These analyses indicate good data and simulation agreement for the PMT time-residuals and \np, across a wide range of source positions. Both likelihood- and ML-based reconstruction methods that utilize the PMT time-residuals, are used to reconstruct the event vertex and direction. The outputs of these reconstruction methods are compared between data and simulation for the thorium, directional, and AmBe sources. Agreement between data and simulation to within the stated goals for \eos{}  is achieved for the reconstructed parameters. This is a powerful statement as these tools are being used to extrapolate to much larger detectors and to evaluate the physics sensitivities of future detectors. Additionally, the detailed calibrations, detector models, and calibration source models developed for the water phase are critical to the success of future phases of \eos{} in which less well-understood target materials will be evaluated. The results from this manuscript will aid in the future demonstration of Cherenkov and scintillation separation in a tonne-scale hybrid detector and are a critical step towards evaluating the performance of many key elements of hybrid detector technology. 

\section{Data Availability Statement}

The \eos{} raw data was collected as \texttt{hdf5} files and access can be provided upon reasonable request. The \eos{} processed data and simulated data are stored in \texttt{ROOT} format, and access can be provided upon reasonable request. All data and simulation is stored on the Savio computing cluster at the University of California, Berkeley. All data processing and simulation generation was performed using the open source \rp{} framework \cite{ratpac-two-github}. 

\section{Acknowledgements}

Work conducted at Lawrence Berkeley National Laboratory was performed under the auspices of the U.S. Department of Energy under Contract DE-AC02-05CH11231. The work conducted at Lawrence Livermore National Laboratory was supported by the U.S. Department of Energy under contract DE-AC52-07NA27344, release number LLNL-JRNL-2020089. The work conducted at Brookhaven National Laboratory was supported by the U.S. Department of Energy under contract DE-AC02-98CH10886. The work conducted at Oak Ridge National Laboratory was supported by the U.S. Department of Energy under contract number DE-AC05-00OR22725. The work conducted at Oxford was supported by the Royal Society. The project was funded by the U.S. Department of Energy, National Nuclear Security Administration, Office of Defense Nuclear Nonproliferation  Research and Development (DNN R\&D). This material is based upon work supported by the U.S. Department of Energy, Office of Science, Office of High Energy Physics, under Award Number DE-SC0018974.  This work is supported as part of the Growth in Research Opportunities With Traineeship in High energy physics at Minority Serving Institutions (GROWTH-MSI) program funded by the DOE Office of Sciences High Energy and Nuclear Physics, Grant number DE-SC0023725. This research used the Savio computational cluster resource provided by the Berkeley Research Computing program at the University of California, Berkeley (supported by the UC Berkeley Chancellor, Vice Chancellor for Research, and Chief Information Officer).

\clearpage

\appendix

\section{Directional Source Vertex Reconstruction}\label{sec:appendix-A}

\begin{figure}[ht!]
    \centering
    \includegraphics[width=0.45\linewidth]{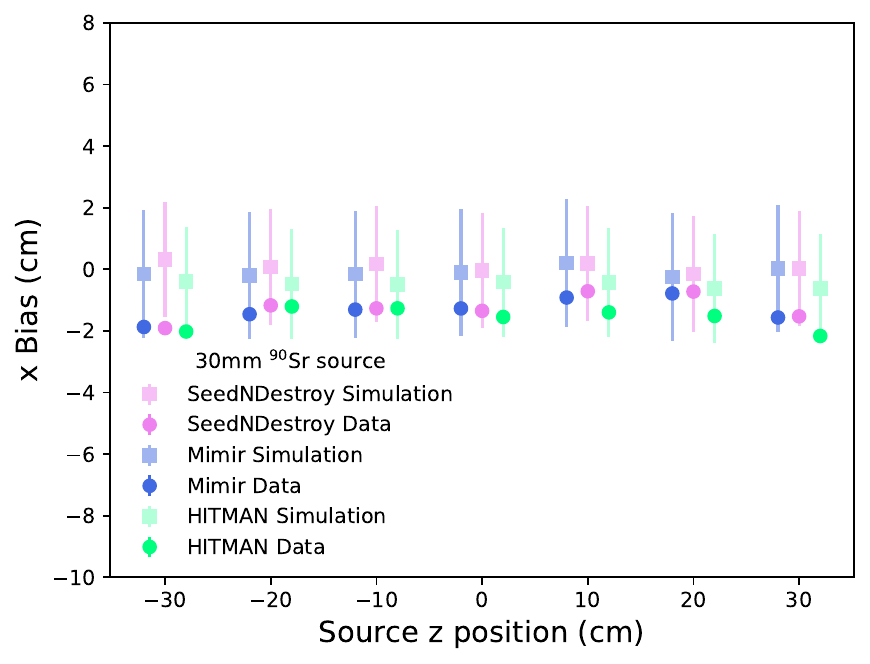}
    \includegraphics[width=0.45\linewidth]{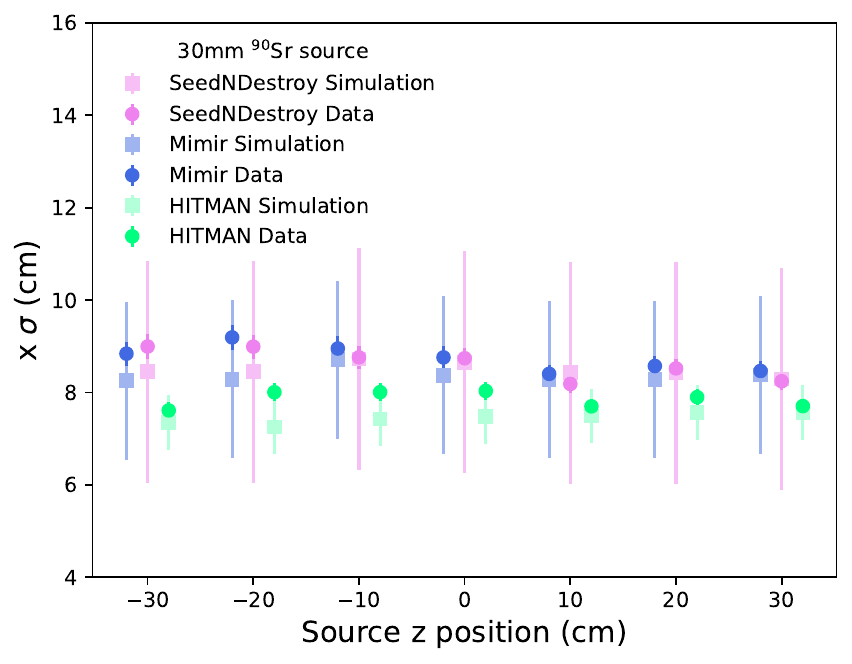}
    \includegraphics[width=0.45\linewidth]{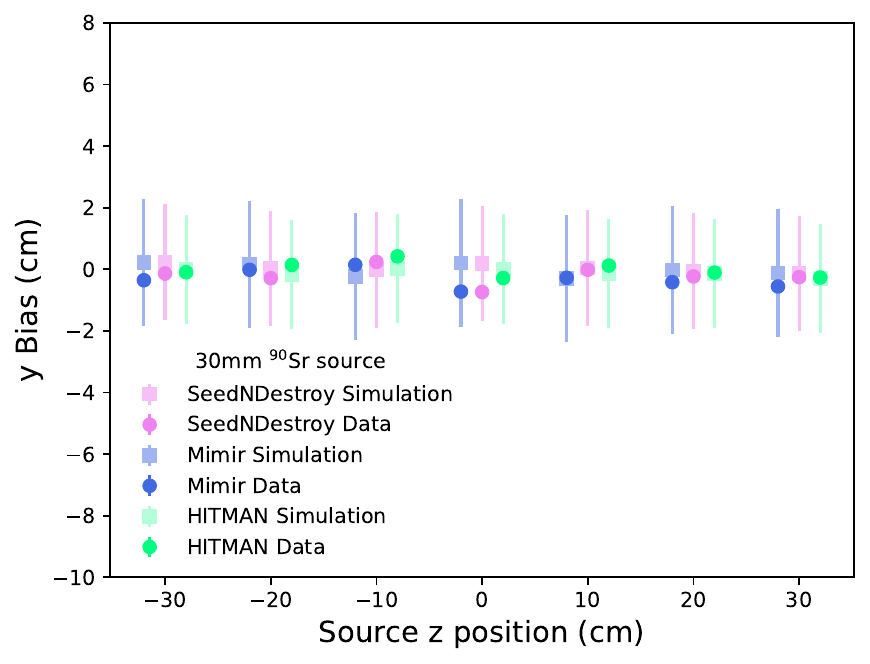}
    \includegraphics[width=0.45\linewidth]{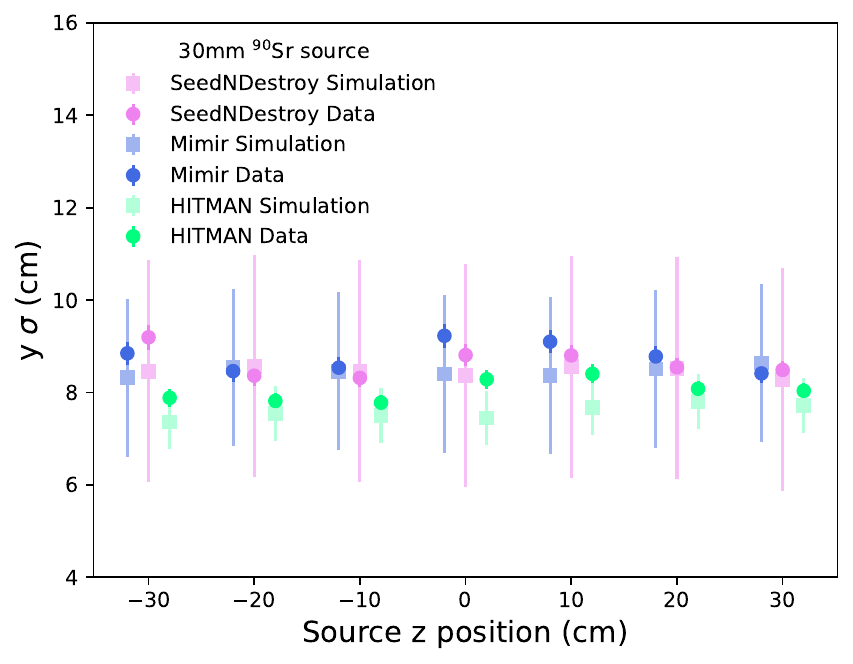}
    \includegraphics[width=0.45\linewidth]{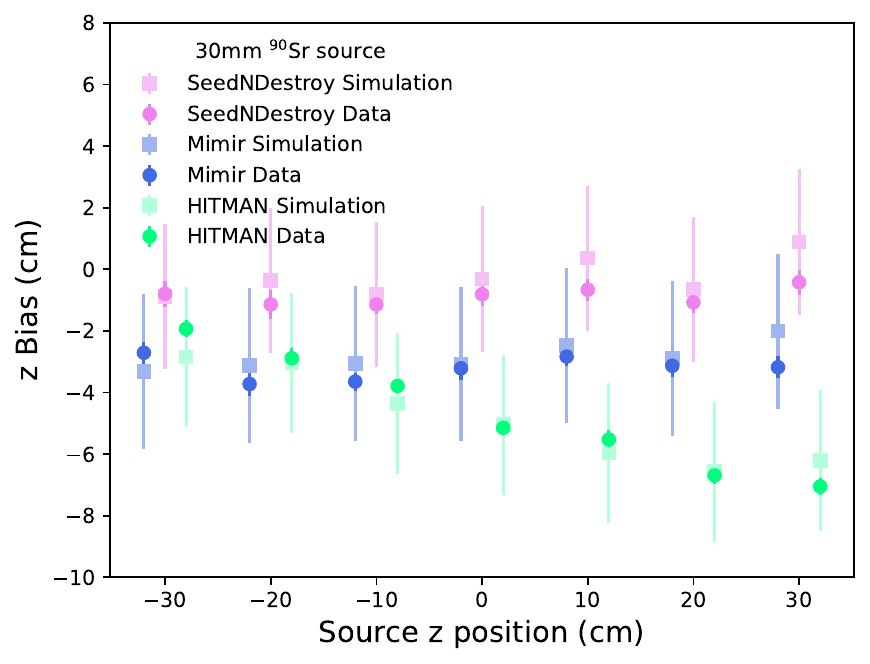}
    \includegraphics[width=0.45\linewidth]{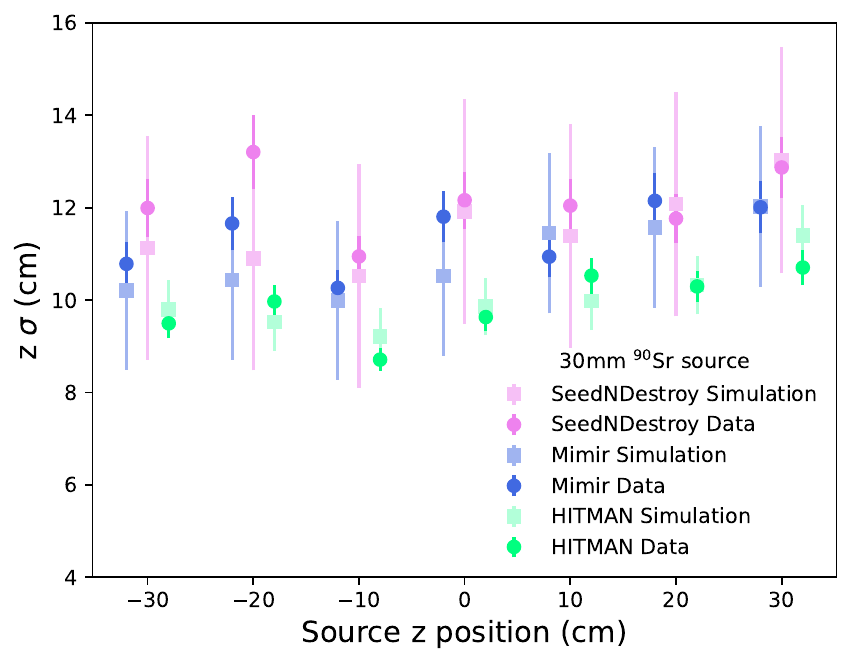}
    \caption{The bias (left) and resolution (right) for the $x$ (top), $y$ (middle), and $z$ (bottom) positions, for \snd{}, \mimir{}, and \hitman{}, for the downward pointing 30~mm $^{90}$Sr directional source. The values are taken from Gaussian fits performed over the histogram of the relevant reconstructed quantity. For \mimir{} and \hitman{} the source $z$ positions are shifted slightly from the true value on the x-axis to make the figure more legible.}
    \label{fig:dir-vertex-fitter-30mm-downward-comparison-all}
\end{figure}

\begin{figure}[ht!]
    \centering
    \includegraphics[width=0.45\linewidth]{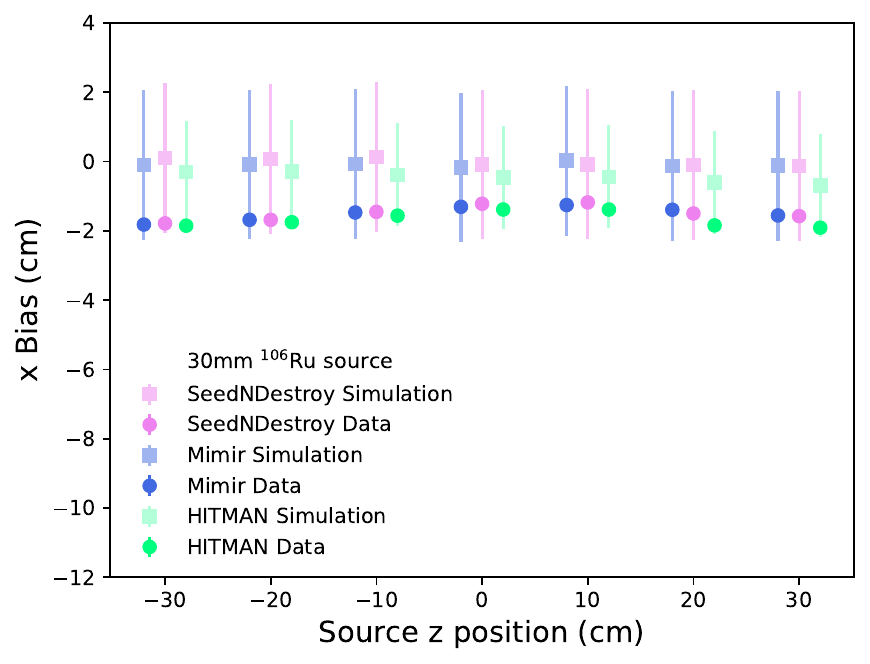}
    \includegraphics[width=0.45\linewidth]{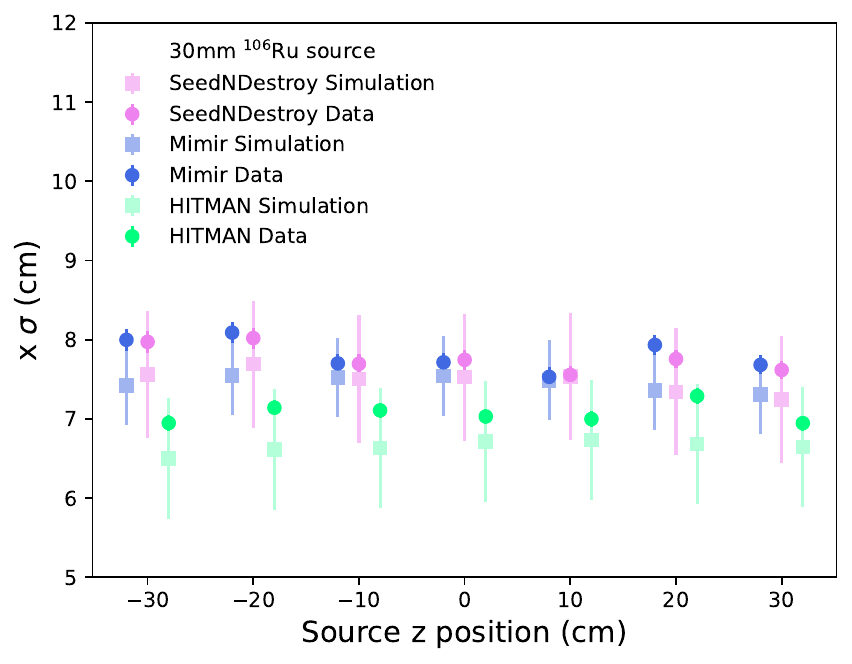}
    \includegraphics[width=0.45\linewidth]{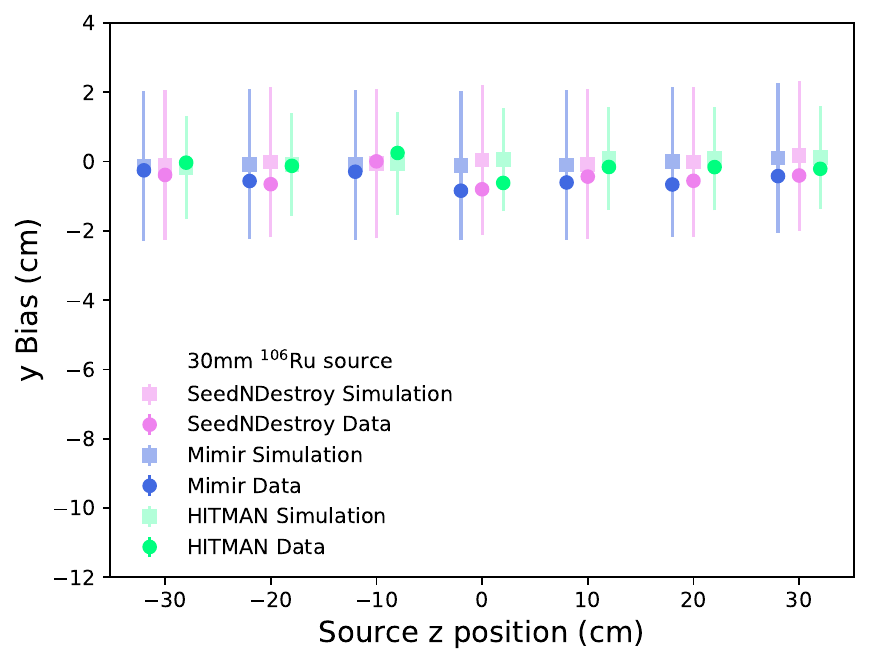}
    \includegraphics[width=0.45\linewidth]{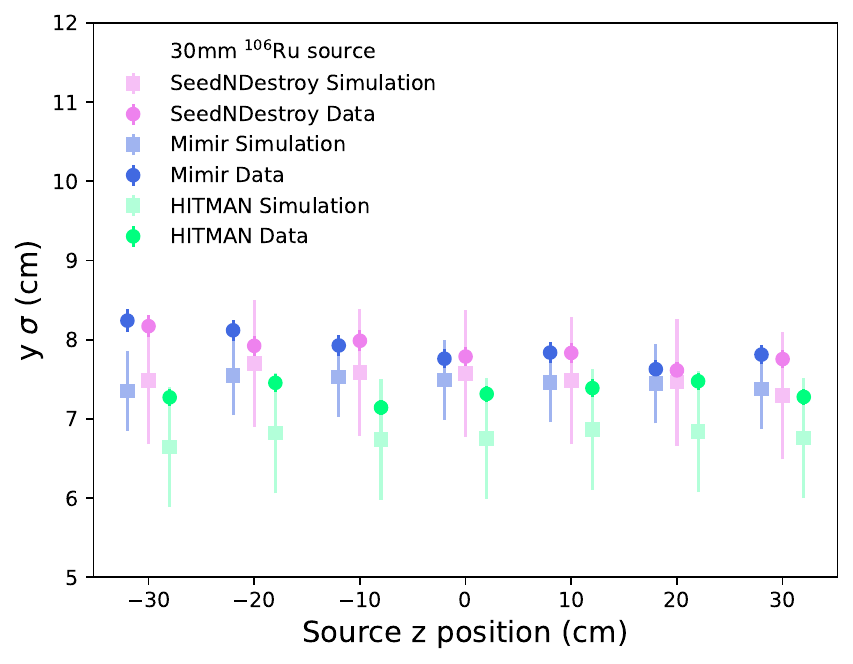}
    \includegraphics[width=0.45\linewidth]{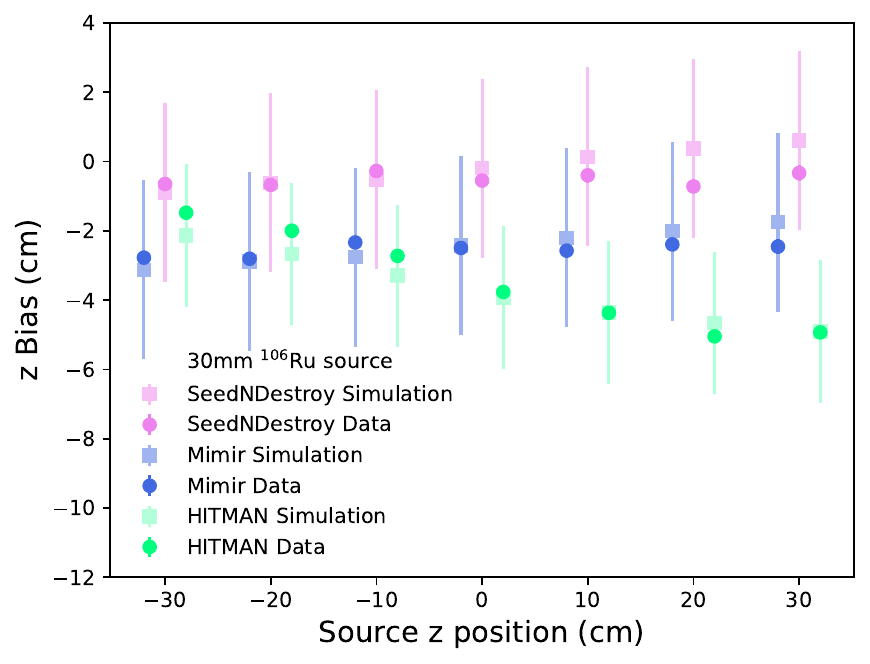}
    \includegraphics[width=0.45\linewidth]{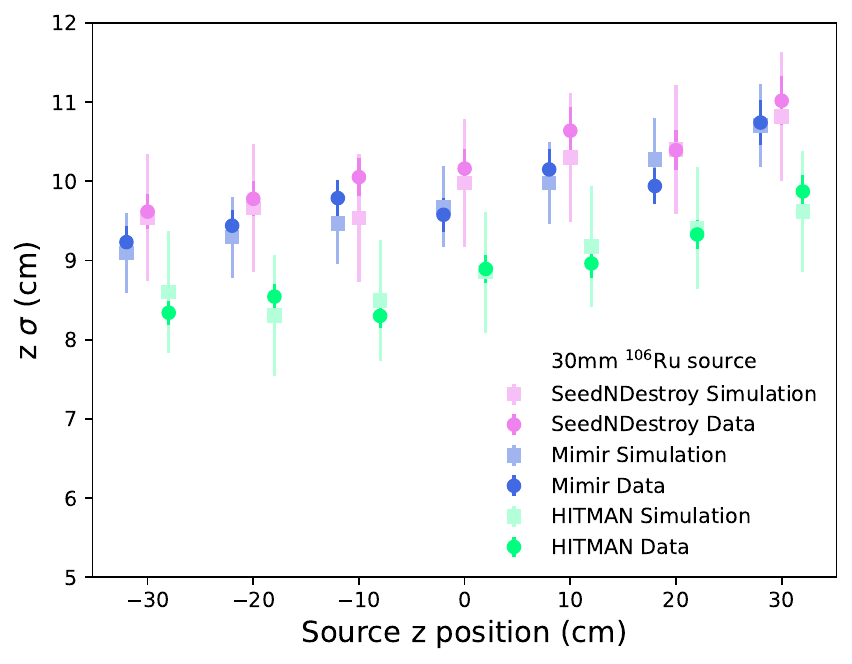}
    \caption{The bias (left) and resolution (right) for the $x$ (top), $y$ (middle), and $z$ (bottom) positions, for \snd{}, \mimir{}, and \hitman{}, for the downward pointing 30~mm $^{106}$Ru directional source. The values are taken from Gaussian fits performed over the histogram of the relevant reconstructed quantity. For \mimir{} and \hitman{} the source $z$ positions are shifted slightly from the true value on the x-axis to make the figure more legible.}
    \label{fig:dir-vertex-fitter-30mm-comparison-all}
\end{figure}

\begin{figure}[ht!]
    \centering
    \includegraphics[width=0.45\linewidth]{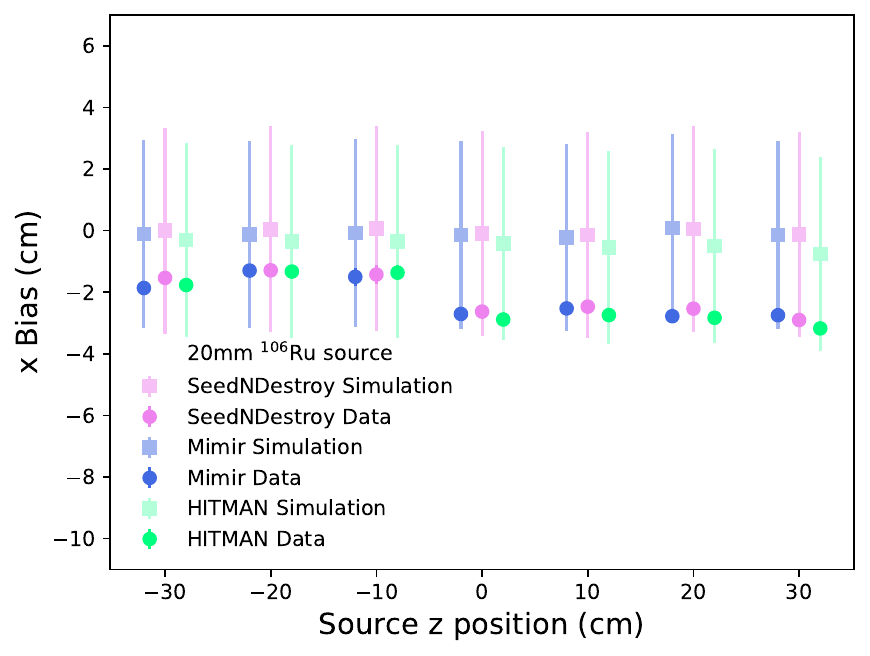}
    \includegraphics[width=0.45\linewidth]{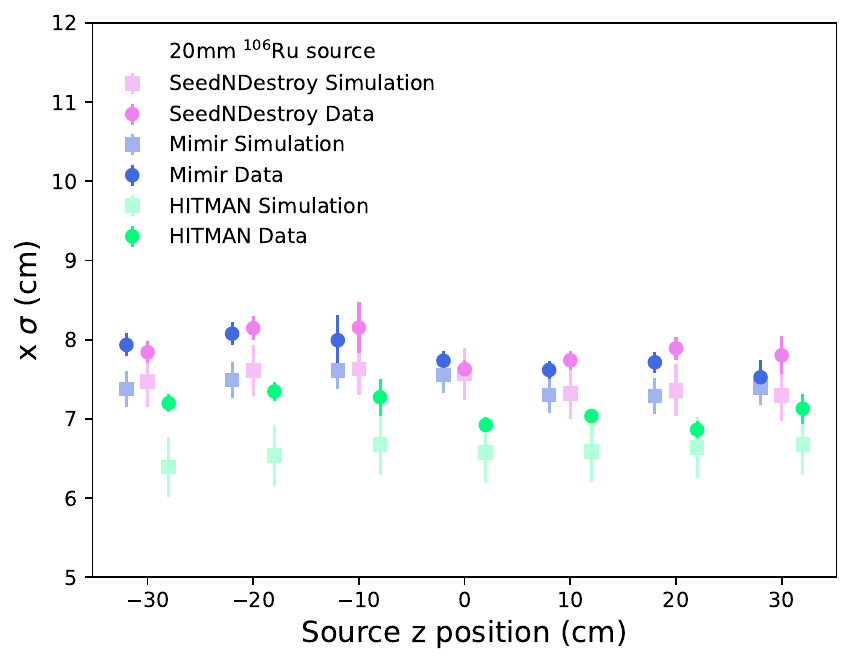}
    \includegraphics[width=0.45\linewidth]{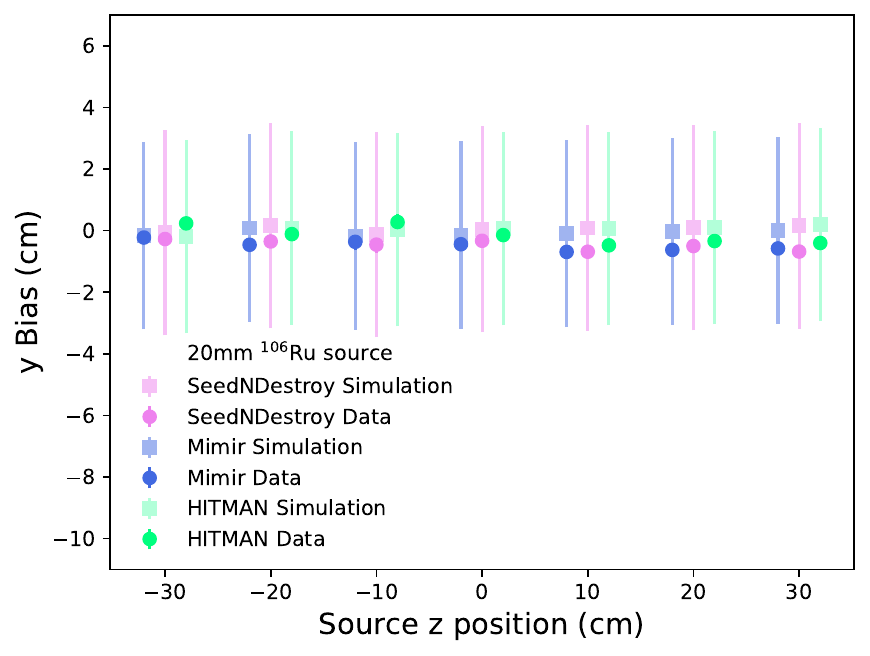}
    \includegraphics[width=0.45\linewidth]{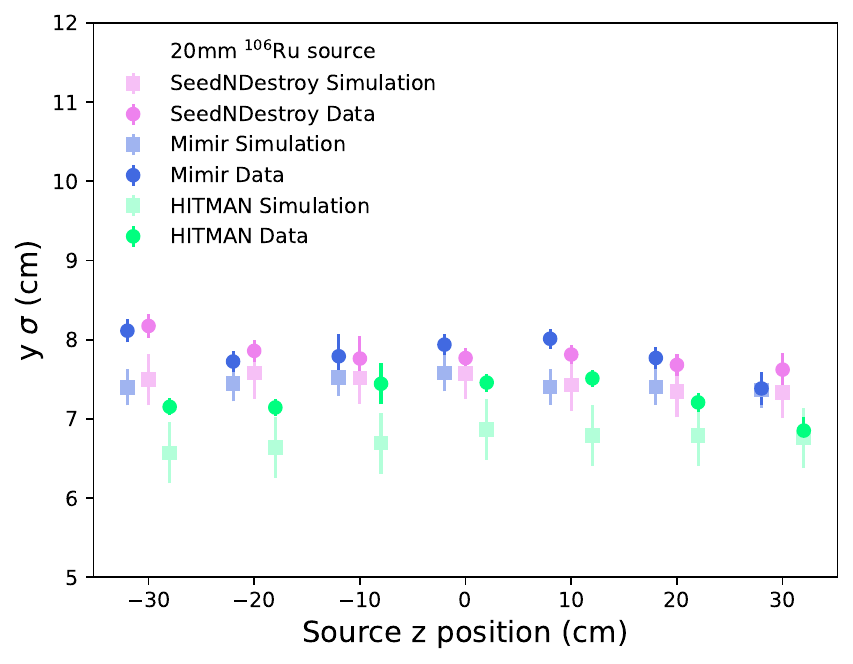}
    \includegraphics[width=0.45\linewidth]{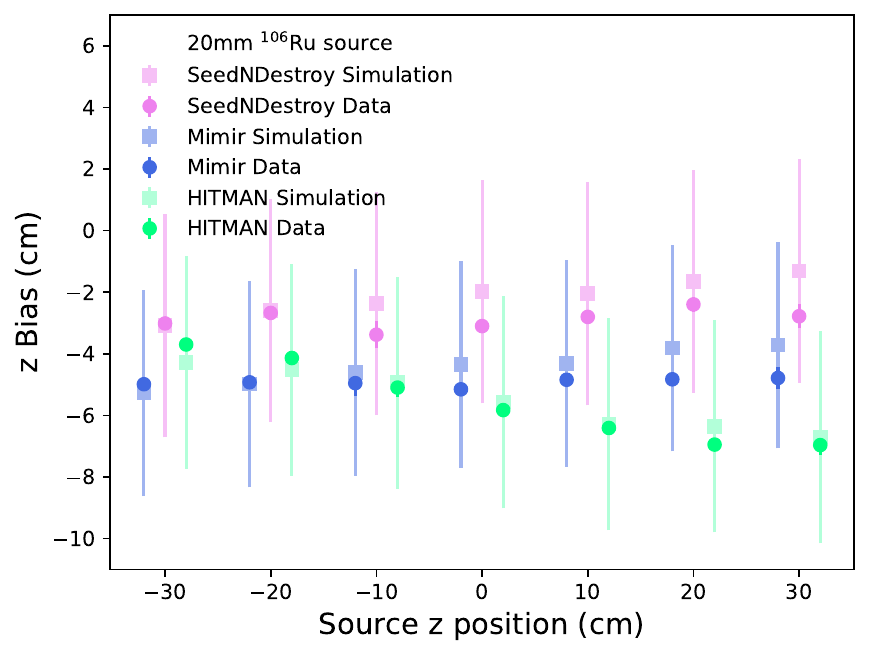}
    \includegraphics[width=0.45\linewidth]{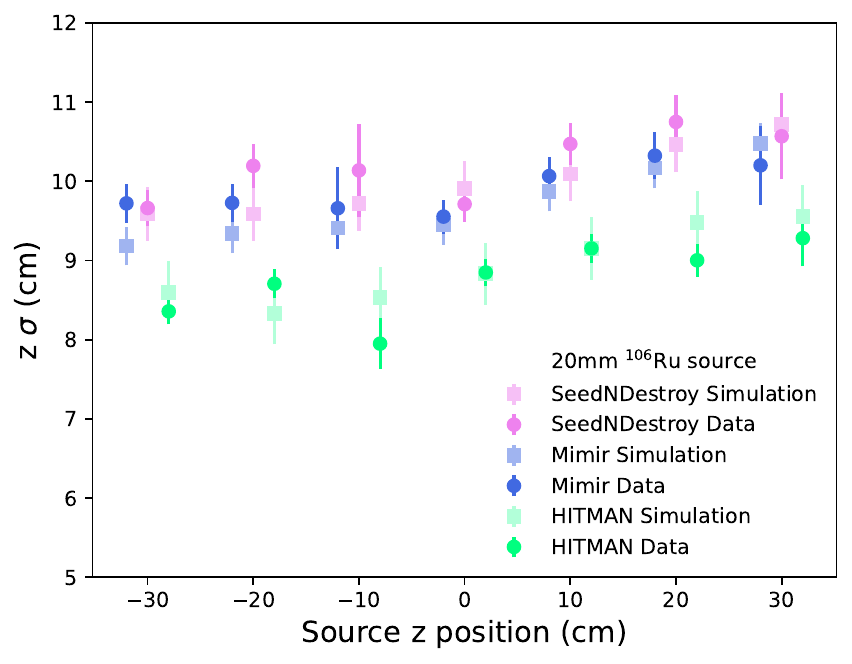}
    \caption{The bias (left) and resolution (right) for the $x$ (top), $y$ (middle), and $z$ (bottom) positions, for \snd{}, \mimir{}, and \hitman{}, for the downward pointing 20~mm $^{106}$Ru directional source. The values are taken from Gaussian fits performed over the histogram of the relevant reconstructed quantity. For \mimir{} and \hitman{} the source $z$ positions are shifted slightly from the true value on the x-axis to make the figure more legible.}
    \label{fig:dir-vertex-fitter-20mm-comparison-all}
\end{figure}

\begin{figure}[ht!]
    \centering
    \includegraphics[width=0.45\linewidth]{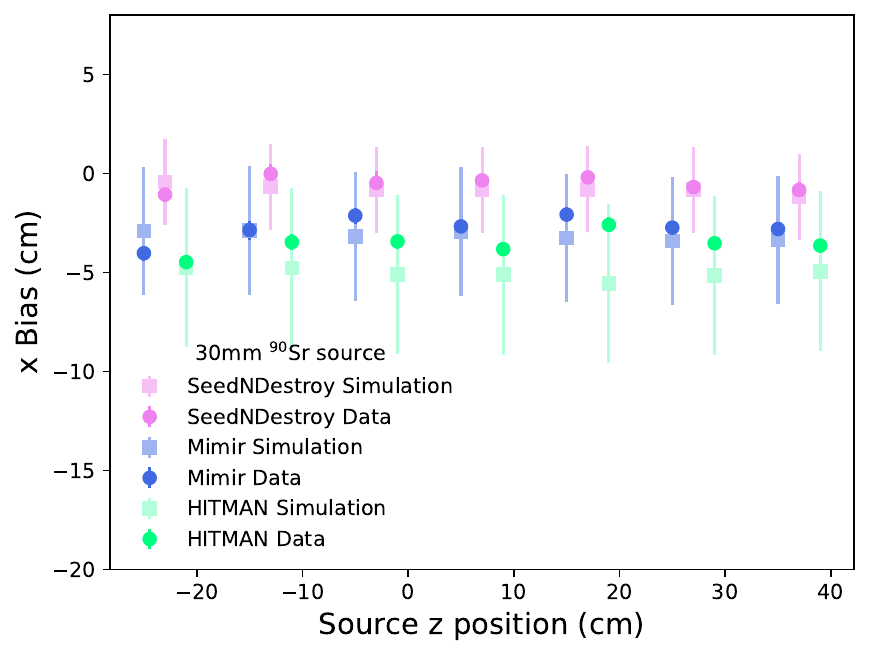}
    \includegraphics[width=0.45\linewidth]{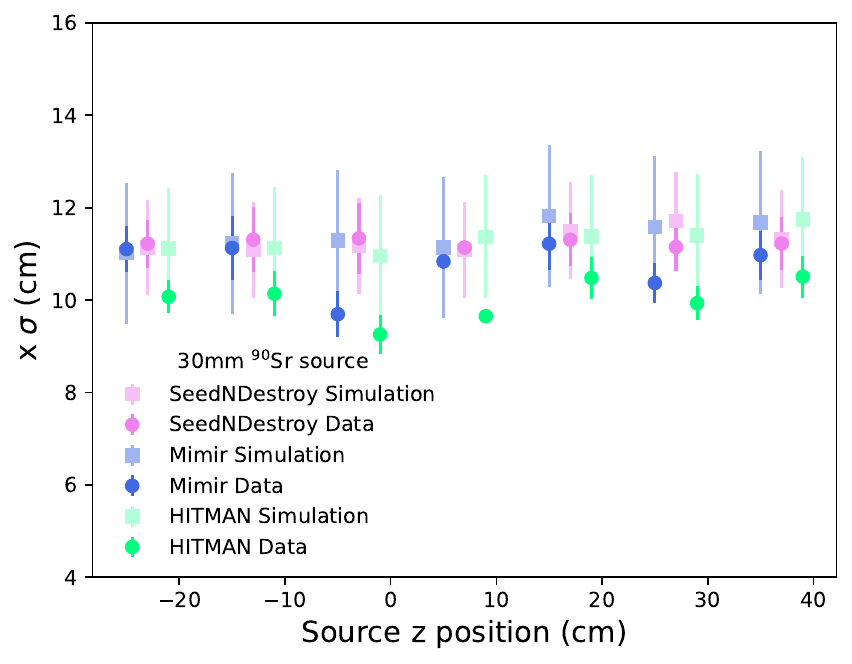}
    \includegraphics[width=0.45\linewidth]{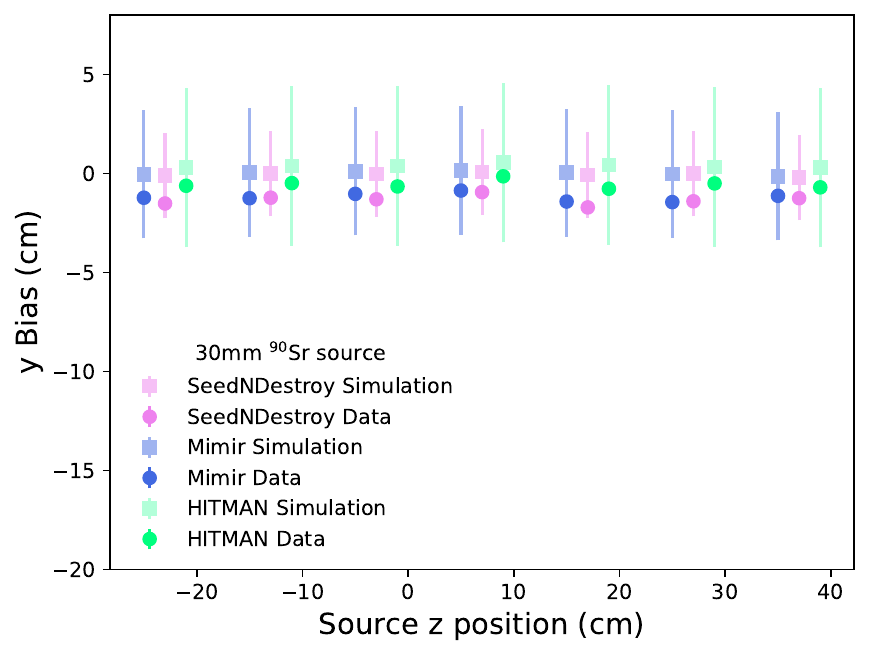}
    \includegraphics[width=0.45\linewidth]{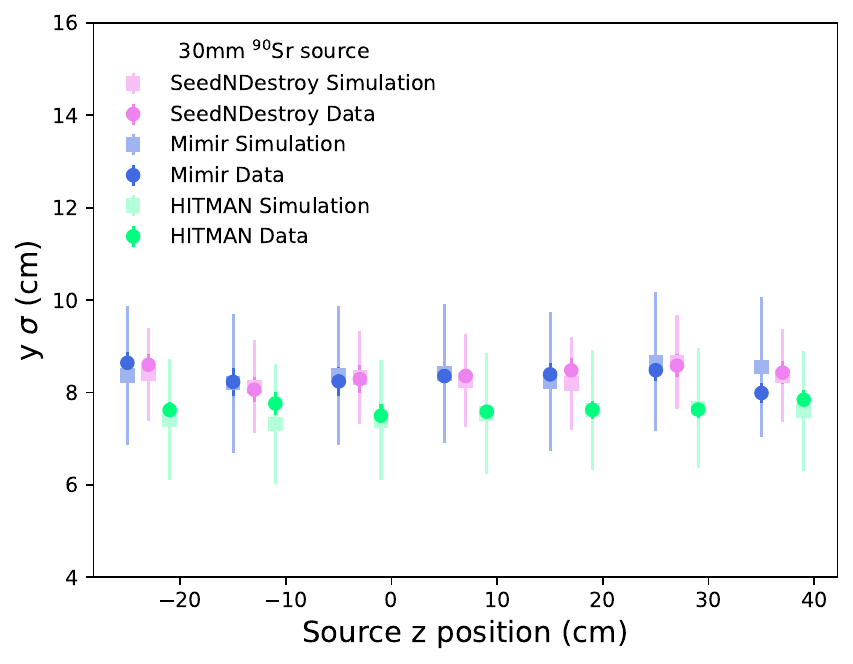}
    \includegraphics[width=0.45\linewidth]{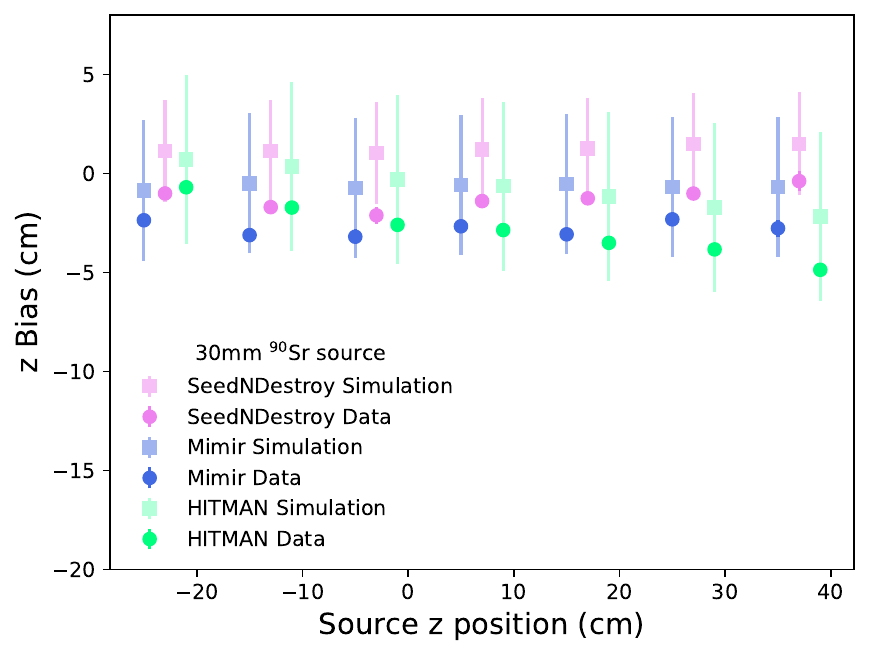}
    \includegraphics[width=0.45\linewidth]{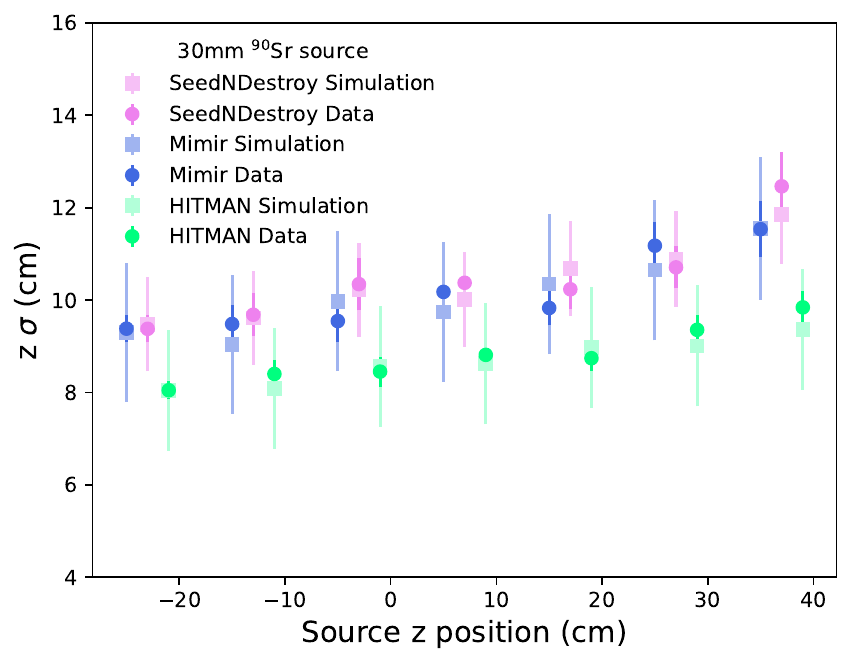}
    \caption{The bias (left) and resolution (right) for the $x$ (top), $y$ (middle), and $z$ (bottom) positions, for \snd{}, \mimir{}, and \hitman{}, for the 90$^{\circ}$  sideways pointing 30~mm $^{90}$Sr directional source. The values are taken from Gaussian fits performed over the histogram of the relevant reconstructed quantity. For \mimir{} and \hitman{} the source $z$ positions are shifted slightly from the true value on the x-axis to make the figure more legible.}
    \label{fig:dir-vertex-fitter-30mm-sideways-comparison-all}
\end{figure}

\begin{figure}[ht!]
    \centering
    \includegraphics[width=0.45\linewidth]{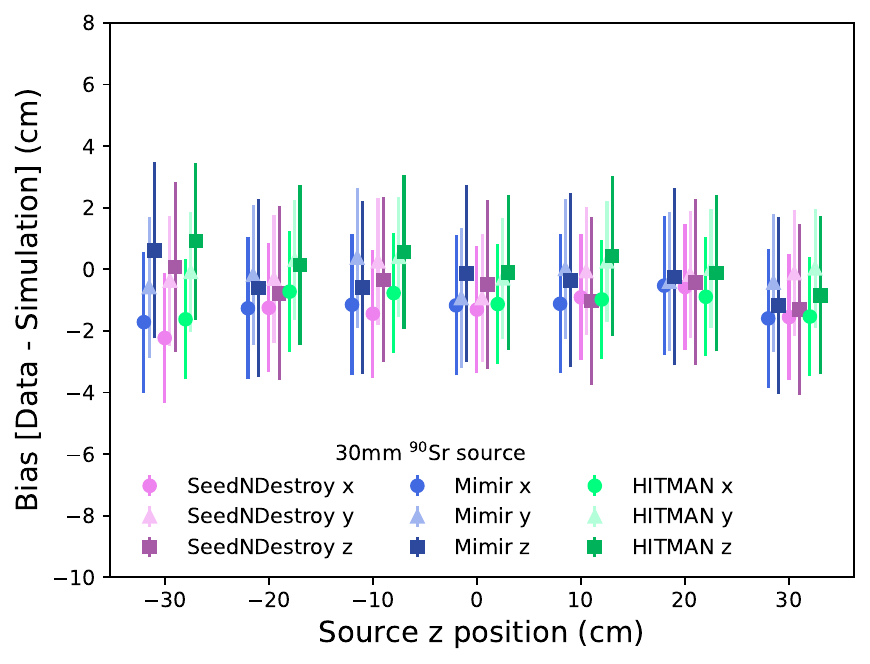}
    \includegraphics[width=0.45\linewidth]{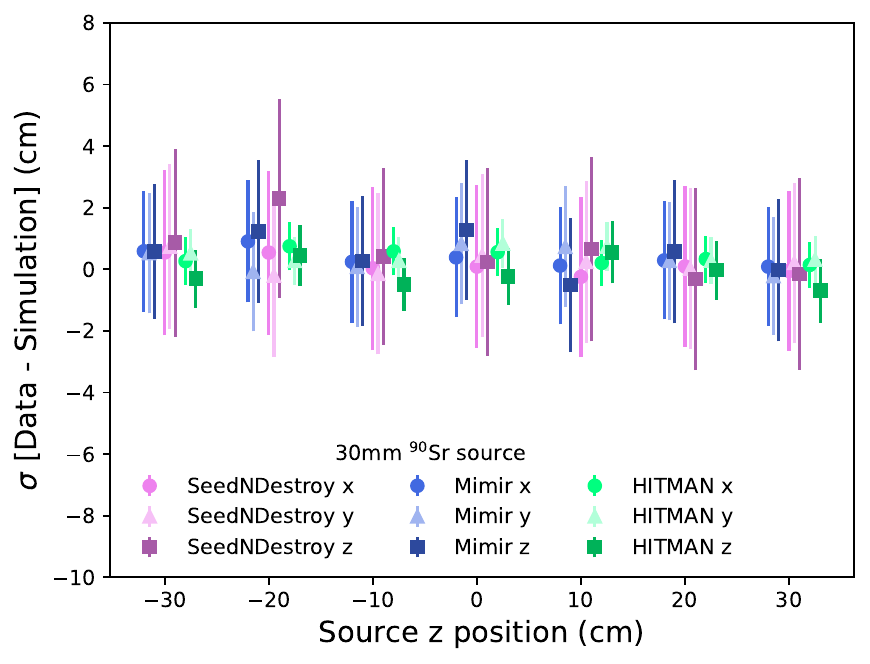}
t    \caption{The data minus the simulation for  the $x$, $y$, and $z$ bias and resolutions from the three vertex fitters, for the downward pointing 30~mm $^{90}$Sr directional source. For \mimir{} and \hitman{} the source $z$ positions are shifted slightly from the true value on the x-axis to make the figure more legible.}
    \label{fig:dir-vertex-diff-30mm-sr}
\end{figure}

\begin{figure}[ht!]
    \centering
    \includegraphics[width=0.45\linewidth]{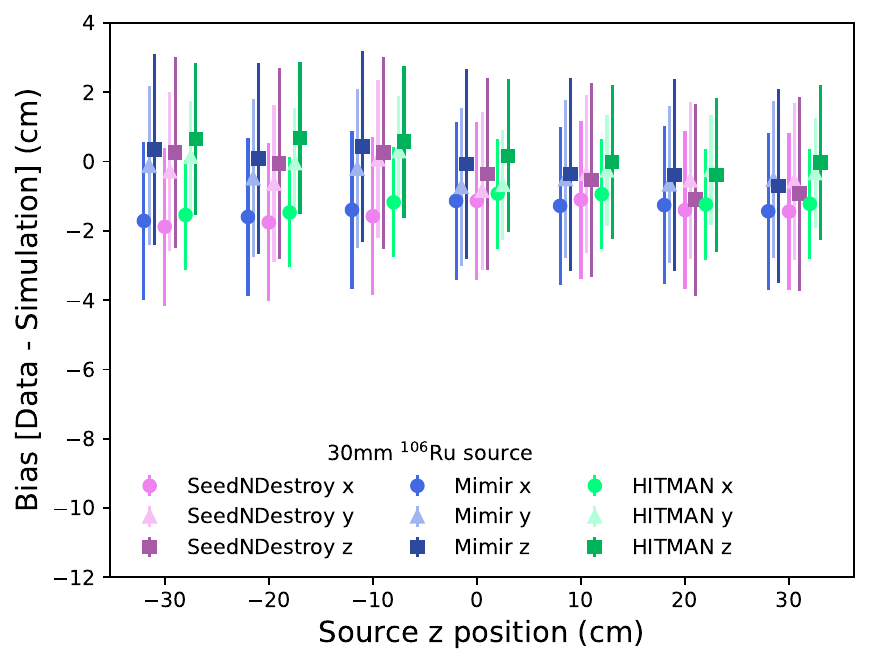}
    \includegraphics[width=0.45\linewidth]{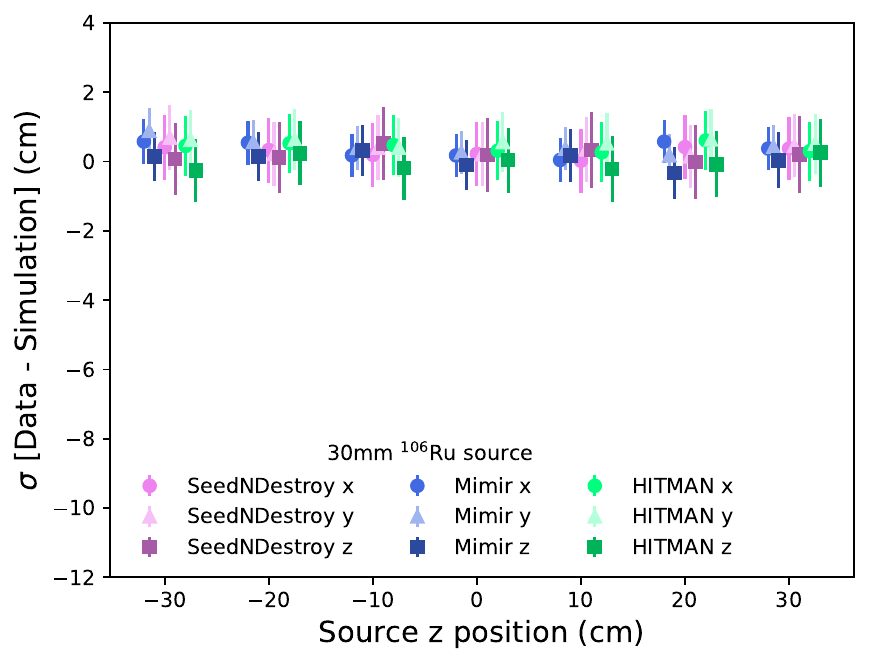}
    \caption{The data minus the simulation for  the $x$, $y$, and $z$ bias and resolutions from the three vertex fitters, for the downward pointing 30~mm $^{106}$Ru directional source. For \mimir{} and \hitman{} the source $z$ positions are shifted slightly from the true value on the x-axis to make the figure more legible.}
    \label{fig:dir-vertex-diff-30mm}
\end{figure}

\begin{figure}[ht!]
    \centering
    \includegraphics[width=0.45\linewidth]{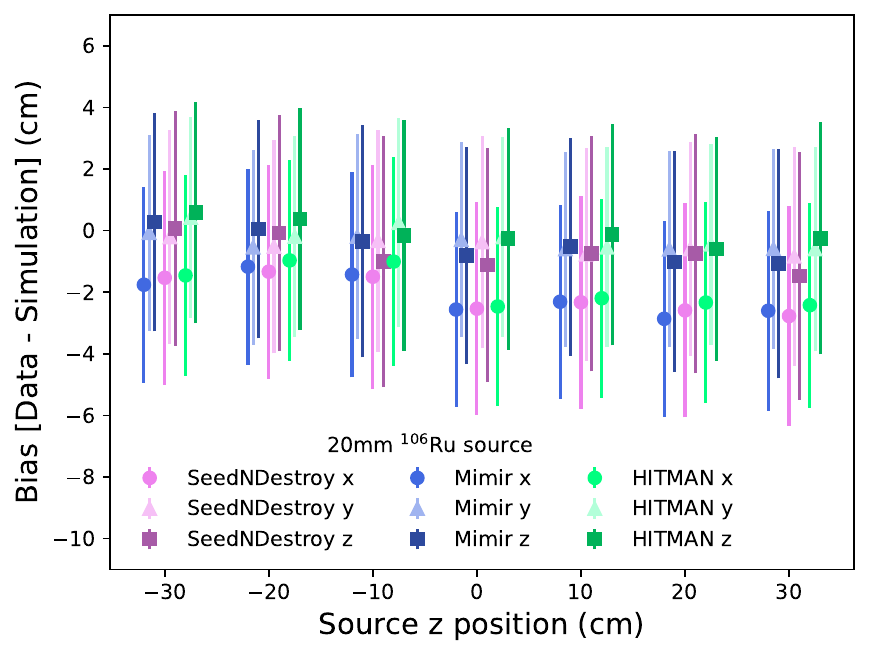}
    \includegraphics[width=0.45\linewidth]{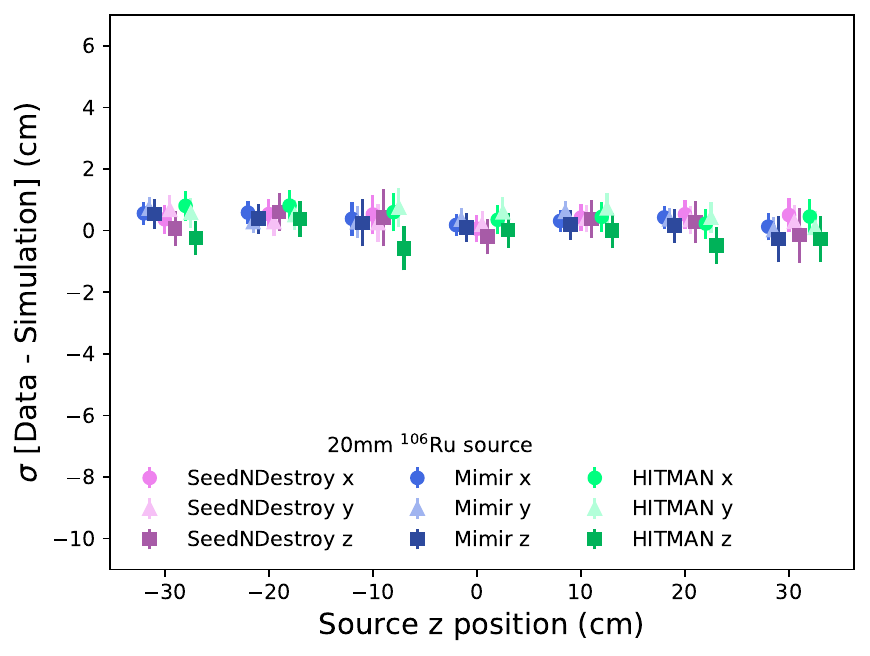}
    \caption{The data minus the simulation for  the $x$, $y$, and $z$ bias and resolutions from the three vertex fitters, for the downward pointing 20~mm $^{106}$Ru directional source. For \mimir{} and \hitman{} the source $z$ positions are shifted slightly from the true value on the x-axis to make the figure more legible.}
    \label{fig:dir-vertex-diff-20mm}
\end{figure}

\begin{figure}[ht!]
    \centering
    \includegraphics[width=0.45\linewidth]{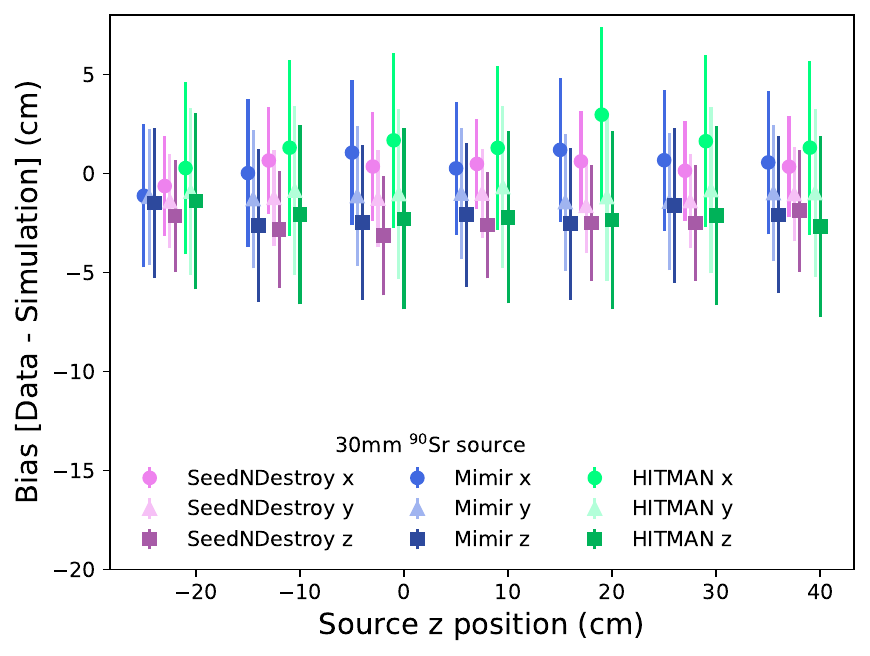}
    \includegraphics[width=0.45\linewidth]{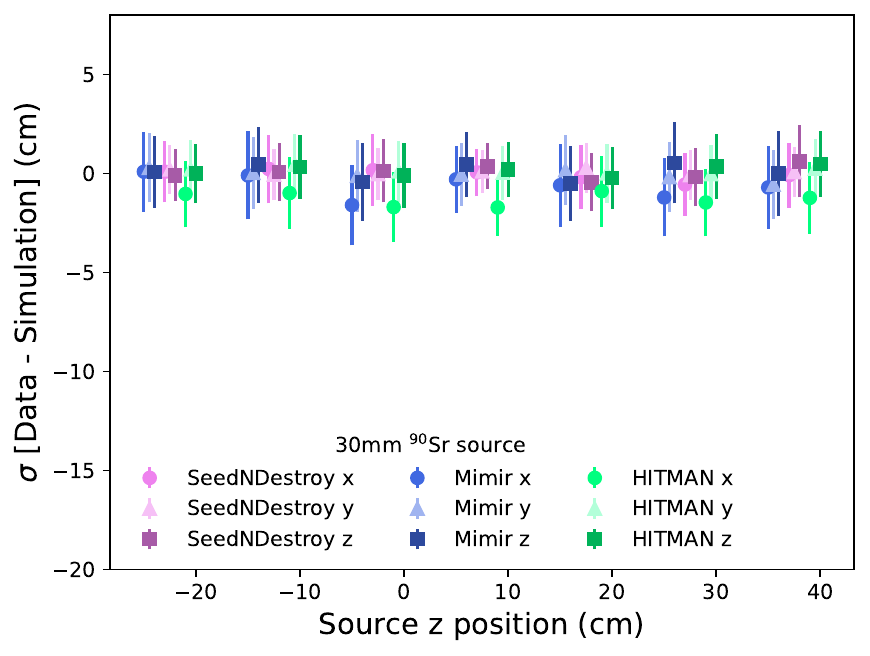}
    \caption{The data minus the simulation for  the $x$, $y$, and $z$ bias and resolutions from the three vertex fitters, for the 90$^{\circ}$ sideways pointing 30~mm $^{90}$Sr directional source. For \mimir{} and \hitman{} the source $z$ positions are shifted slightly from the true value on the x-axis to make the figure more legible.}
    \label{fig:dir-vertex-diff-30mm-sideways}
\end{figure}

In this section we show the vertex reconstruction results for \snd{}, \mimir{}, and \hitman{} for various directional source data. The reconstructed $x$, $y$, and $z$ bias and resolutions are shown in Fig.~\ref{fig:dir-vertex-fitter-30mm-downward-comparison-all}, \ref{fig:dir-vertex-fitter-30mm-comparison-all}, \ref{fig:dir-vertex-fitter-20mm-comparison-all}, and \ref{fig:dir-vertex-fitter-30mm-sideways-comparison-all} for the downward pointing 30~mm $^{90}$Sr, the downward pointing 30~mm $^{106}$Ru, the downward pointing 20~mm $^{106}$Ru, and the sideways pointing 30~mm $^{90}$Sr respectively. The differences between the data and simulations are shown in Figs.~\ref{fig:dir-vertex-diff-30mm-sr}, \ref{fig:dir-vertex-diff-30mm}, \ref{fig:dir-vertex-diff-20mm}, and \ref{fig:dir-vertex-diff-30mm-sideways} respectively. For the sideways pointing data, the azimuth of the source is fixed to 175 degrees in the (x, y) plane and the polar angle ($\theta$) is set to 90 degrees.

As with the thorium data described in Sec.~\ref{sec:vertex-results}, the data and simulation tend to agree to within 1-2~cm for both the bias and resolution. Overall, the $x$ and $y$ resolutions ($\sigma$) are less than 10~cm. The $z$ resolution is around 8-12~cm, and depends more strongly on the source $z$-position. The resolutions are slightly broader for the sideways pointing source ($\sim$8 - 12~cm) than for the downward pointing source ($\sim$8 - 10~cm) due to the better coverage at the bottom of the detector than for the barrel. Additionally, the resolution is slightly broader for the $^{90}$Sr data than the $^{106}$Ru data, due to the slightly lower endpoint energy.

\clearpage

\section{Muon Lifetime}\label{sec:appendix-B}

This section calculates the expected muon lifetime in water. Negative muons in water can undergo nuclear capture on oxygen, reducing their effective lifetime relative to the vacuum value. The corresponding decay rates of positive and negative muons are
\begin{equation}
\frac{1}{\tau_{\mu^+}} = R_{\text{vac}},  
\qquad
\frac{1}{\tau_{\mu^-}} = R_{\text{vac}} + R_{\text{oxy}},
\end{equation}
where $R_{\text{vac}}$ is the vacuum muon decay rate and $R_{\text{oxy}}$ is the capture rate on oxygen. With a $\mu^+$/$\mu^-$ ratio, $r$, the total probability that a muon at rest in water decays is
\begin{equation}
P(\text{decay}) = P(\text{decay}\mid\mu^+)P(\mu^+) + P(\text{decay}\mid\mu^-)P(\mu^-) =  \frac{r}{1+r} + \frac{R_{\text{vac}}}{R_{\text{vac}}+R_{\text{oxy}}}
\frac{1}{1+r}.
\end{equation}

The effective muon lifetime measured by tagging decays from both $\mu^+$ and $\mu^-$ is the posterior-weighted average of the $\mu^+$ and $\mu^-$ lifetimes: 
\begin{equation}
\tau_{tot} = \tau_{\mu^+}P(\mu^+ \mid \text{decay})\,+ \tau_{\mu^-}P(\mu^- \mid \text{decay})\,.
\end{equation}

Using the experimental results in \cite{muon-capture-rates,muon-ratio}, the expected muon lifetime in water is 2.04 $\pm$ 0.01 $\mu$s.

\clearpage

\bibliographystyle{apsrev4-2}
\bibliography{bibliography.bib}

\end{document}